\documentclass[useAMS,usenatbib,usegraphicx]{mn2e}

\newcommand{\der}{\mathrm{d}}
\newcommand{\reff}{r_\mathrm{eff}}
\newcommand{\rorb}{\langle r_\mathrm{orb} \rangle}
\newcommand{\vcirc}{v_\mathrm{circ}}

\newcommand{\kpc}{\mathrm{kpc}}
\newcommand{\fav}{f_h}
\newcommand{\mldyn}{\Upsilon_\mathrm{dyn}}
\newcommand{\mlnfw}{\Upsilon_\mathrm{NFW}}
\newcommand{\mlsc}{\Upsilon_\mathrm{SC}}
\newcommand{\mllog}{\Upsilon_\mathrm{LOG}}
\newcommand{\chigh}{\chi^2_\mathrm{GH}}
\newcommand{\chimin}{\chi^2_\mathrm{min}}
\newcommand{\chisc}{\chi^2_\mathrm{SC}}
\newcommand{\chinfw}{\chi^2_\mathrm{NFW}}
\newcommand{\chilog}{\chi^2_\mathrm{LOG}}

\newcommand{\betatheta}{\beta_\vartheta}
\newcommand{\betaphi}{\beta_\varphi}
\newcommand{\sigmatheta}{\sigma_\vartheta}
\newcommand{\sigmaphi}{\sigma_\varphi}

\title[Dynamical modelling of 17 Coma early-types]{Dynamical modelling of luminous and dark matter in 17 Coma early-type galaxies}
\author[J. Thomas et al.]{J. Thomas$^{1,2}$\thanks{E-mail: jthomas@mpe.mpg.de}, R. P. Saglia$^{2}$, R. Bender$^{1,2}$, 
D. Thomas$^{3}$, K. Gebhardt$^{4}$, 
\newauthor J. Magorrian$^{5}$, E.~M. Corsini$^{6}$ and G. Wegner$^{7}$\\
$^{1}$Universit\"atssternwarte M\"unchen, Scheinerstra\ss e 1, D-81679 M\"unchen, Germany\\
$^{2}$Max-Planck-Institut f\"ur Extraterrestrische Physik, Giessenbachstra\ss e, D-85748 Garching, Germany\\
$^{3}$Institute of Cosmology and Gravitation, Mercantile House, University of Portsmouth, Portsmouth, PO1 2EG, UK\\
$^{4}$Department of Astronomy, University of Texas at Austin, C1400, Austin, TX78712, USA\\
$^{5}$Theoretical Physics, Department of Physics, University of Oxford, 1 Keble Road, Oxford U.K., OX1 3NP\\
$^{6}$Dipartimento di Astronomia, Universit\`a di Padova, vicolo dell'Osservatorio 3, I-35122 Padova, Italy\\
$^{7}$Department of Physics and Astronomy, 6127 Wilder Laboratory, Dartmouth College, Hanover, NH 03755-3528, USA}
\begin{document}

\date{Accepted 1988 December 15. Received 1988 December 14; in original form 1988 October 11}

\pagerange{\pageref{firstpage}--\pageref{lastpage}} \pubyear{2002}

\maketitle

\label{firstpage}

\begin{abstract}
Dynamical models for 17 early-type galaxies in the 
Coma cluster are presented. The galaxy sample consists of flattened, rotating as well as
non-rotating early-types including cD and S0 galaxies with luminosities between 
$M_B = -18.79$ and $M_B = -22.56$. Kinematical long-slit observations cover at least the 
major and minor axis and extend to $1-4 \, \,\reff$. Axisymmetric Schwarzschild models 
are used to derive stellar mass-to-light ratios and dark halo parameters. 
In every galaxy the best fit with dark matter matches the data better
than the best fit without. 
The statistical significance is over 95 percent for 8 galaxies, around 90 percent for
5 galaxies and for four galaxies it is not significant. For the highly significant
cases systematic deviations between observed and modelled kinematics are clearly seen;
for the remaining galaxies differences are more statistical in nature. 
Best-fit models contain 10-50 percent dark matter inside
the half-light radius. The central dark matter 
density is at least one order of magnitude lower than the luminous mass density, 
independent of the assumed dark matter density profile. 
The central phase-space density of dark matter is often orders of magnitude lower than
in the luminous component, especially when the halo core radius is large. 
The orbital system of the stars along the 
major-axis is slightly dominated by radial motions. Some galaxies show
tangential anisotropy along the minor-axis, which is correlated with the 
minor-axis Gauss-Hermite coefficient $H_4$. Changing the balance between 
data-fit and regularisation constraints does not change the reconstructed mass structure 
significantly: model anisotropies tend to strengthen if the
weight on regularisation is reduced, but the general property of a galaxy to be radially or
tangentially anisotropic, respectively, does not change. This paper is aimed to
set the basis for a subsequent detailed analysis of luminous and dark matter scaling 
relations, orbital dynamics and stellar populations.
\end{abstract}



\section{Introduction}
\label{mass:outline}
Elliptical galaxies are numerous among the brightest galaxies and they harbour a significant
fraction of the present-day stellar mass in the universe \citep{Fuk98,Ren06}. 
Key parameters for the understanding 
of elliptical galaxy formation and evolution are, among others, the central dark matter 
density, the scaling radius of dark matter, the stellar mass-to-light
ratio and the distribution of stellar orbits. While the concentration of the dark matter halo puts constraints on the assembly epoch 
\citep{nfw96,J00,W02}, the orbital state contains imprints of the
assembly mechanism of ellipticals 
\citep[e.g.][]{vanAl82,Her92,Her93,Wei96,Dub98,Nab03,Jes05}.

Information about elliptical galaxy masses are in principle offered through various
channels. The analysis of X-ray halo temperatures, the kinematics of occasional gas discs 
and galaxy-galaxy lensing provide evidence for extended dark matter halos around 
early-type galaxies (e.g. \citealt{Ber93,Piz97,Loe99,Oos02,Hoe04,Fuk06,Hum06,Kle06,Man06}). These methods do not
constrain the inner halo-profiles strongly, however. At non-local redshifts 
strong lensing configurations allow a detailed reconstruction of the 
mass enclosed inside, say, $\reff$ (e.g. \citealt{Kee01,Kop06}). None of the above mentioned
observational channels is sensitive to dynamical galaxy parameters, such as the 
distribution of stellar orbits.

Dynamical modelling of stellar kinematics has the unique advantage that it
allows reconstruction of both the mass structure and the orbital state
of a galaxy. High-quality observations of the line-of-sight velocity 
distributions (LOSVDs) out to several $\reff$ are needed for this purpose.
To overcome the problems of measuring absorption line kinematics in the faint
outskirts of ellipticals, discrete kinematical tracers such as planetary nebulae
or globular clusters can be used to additionally constrain the mass distribution
(e.g. \citealt{Sag00,R03,Pie06}).

Since stars in galaxies behave collisionlessly to first order, 
the distribution of stellar orbits is not known a priori and very general 
dynamical methods are required to probe all the degrees of freedom in the 
orbital system.
So far only one large sample of 21 round, non-rotating giant ellipticals has been
probed for dark matter considering at least the full range of {\it spherical} models 
\citep{Kr00}. These models predict circular velocity curves constant
to about 10 per cent and equal luminous and dark matter somewhere inside $1 - 3 \, \reff$. 
Reconstructed halos of these models are $\sim 25$ times denser than in comparably bright 
spirals, which indicates a $\sim 3$ times higher formation redshift \citep{G01}. Not all
apparently round objects need to be intrinsically spherical; some may be face-on flattened
systems. 

Apparently flattened ellipticals have not yet been
addressed in much generality. Primarily, because {\it axisymmetric}
modelling is required to account for intrinsic flattening, inclination effects and rotation.
Fully general axisymmetric models involve three integrals of motion, one of
which -- the non-classical so-called third integral -- is not given explicitly 
in most astrophysically relevant potentials. Only recently, sophisticated numerical 
methods such as Schwarzschild's orbit superposition technique \citep{S79} have
provided fully general models involving all relevant integrals of
motion. Dynamical studies of samples of elliptical galaxies using
this technique are, however, based on kinematical data inside $r \la \reff$ 
\citep{Geb03,Cap05} and dark matter is not considered. 

The present paper is part of a project aimed to analyse the luminous and dark matter
distributions as well as the orbital structure in a sample
of flattened Coma ellipticals. The data for this project has been collected over the
last years and consists of ground-based as well as (archival and new) HST imaging and
measurements of line-of-sight velocity distributions (LOSVDs) along various position 
angles out to $1-4 \, \reff$ (\citealt{Meh00}; 
\citealt{Weg02}; \citealt{Cor07}). The implementation of our modelling 
machinery, which is an advanced version of the axisymmetric Schwarzschild code of
\citet{Ric88} and \citet{Geb00} 
has been described in detail in \citet{Tho04,Tho05}. In the present paper we
survey the models of the whole sample. This sets the basis
for subsequent investigations of luminous and dark matter scaling relations and stellar
populations in elliptical galaxies (Thomas et al. 2007a, in preparation).

In Sec.~\ref{sec:obs} the observations are summarised and
the modelling is outlined in Secs.~\ref{sec:setup} and \ref{sec:res}. 
The mass structure of our models and
the orbital anisotropies are described in Secs.~\ref{sec:mass} and \ref{sec:aniso}, 
respectively. Phase-space distribution functions for luminous and dark matter are 
the subject of Secs.~\ref{sec:dfstars} and \ref{sec:dfhalo}. We discuss the influence of regularisation on our results
in Sec.~\ref{sec:regula}. The paper closes with a short discussion and summary in 
Sec.~\ref{sec:sum}. A detailed comparison of models and data for each galaxy can be found in
App.~\ref{sec:fits}.

\section{Summary of observations}
\label{sec:obs}
The Coma sample consists of
seventeen early-type galaxies: two cD galaxies, nine ordinary giant ellipticals and six 
lenticulars or galaxies of intermediate type. 
They cover the luminosity interval
$-20.30<M_B<-22.56$, typical for luminous giant ellipticals/cDs. One single fainter galaxy
with $M_B=-18.8$ is also included (cf. Tab.~\ref{dattable}; 
magnitudes are from Hyperleda for a distance of
$d = 100 \, \mathrm{Mpc}$ to Coma; this corresponds to 
$H_0 = 69 \, \mathrm{km/s/Mpc}$). 
Effective radii are mostly between $3\farcs3<\reff<18\farcs4$. Only the 
four brightest galaxies have formally very large $\reff\approx30\arcsec-70\arcsec$ (cf. 
Tab.~\ref{dattable}; $\reff$ are from \citet{Meh00} and based on de-Vaucouleurs fits). 
All galaxies share the same distance
and the spatial resolution in the photometric as well as the kinematical observations 
is roughly comparable for all galaxies.

The photometric input for the modelling is constructed as a composite of ground-based
(outer parts) and HST imaging (inner parts). 
The two surface brightness profiles $\mu_\mathrm{grd}$ and $\mu_\mathrm{HST}$ 
are joined by shifting the HST profile according to 
the average $\langle \mu_\mathrm{grd}-\mu_\mathrm{HST} \rangle$ 
over a region where both data sets overlap and seeing effects are negligible
($\approx 5\arcsec - 16\arcsec$). The shift 
$\langle \mu_\mathrm{grd}-\mu_\mathrm{HST} \rangle$ is usually well defined (cf. Tab.~\ref{dattable}). 

\begin{table*}\centering
\begin{minipage}{170mm}
\begin{tabular}{lccc||ccccc||cccc}
\multicolumn{3}{c}{galaxy} & \multicolumn{6}{c}{photometry} & \multicolumn{4}{c}{kinematics}\\
\multicolumn{2}{c}{id} & type & \multicolumn{2}{c}{source} & $M_B$ & $\reff$ & 
$\epsilon_e$ & $\mathrm{rms}\langle \mu_\mathrm{grd} - \mu_\mathrm{HST} \rangle$ & maj & min & off & dia\\
GMP & NGC & & HST & grd & [mag] & [arcsec] & & 
[mag] & [$\reff$] & [$\reff$] & [$\reff$] & [$\reff$]\\
\multicolumn{1}{c}{(1)} & (2) & (3) & (4) & (5) & (6) & (7) & (8) & (9) & (10) & (11) & (12) & (13) \\
\hline
0144 & 4957    & E        & L97 & M00 & $-21.07$ & $18.4$ & $0.256$ & $0.011$ & $1.4$ & $0.7$ & --    & -- \\
0282 & 4952    & E        & L97 & M00 & $-20.69$ & $14.1$ & $0.315$ & $0.009$ & $1.7$ & $0.5$ & --    & -- \\
0756 & 4944    & S0       & W07 & M00 & $-21.77$ & $11.7$ & $0.657$ & $0.010$ & $3.0$ & $0.4$ & $2.5$ & -- \\
1176 & 4931    & S0       & W07 & M00 & $-20.32$ & $7.4$  & $0.552$ & $0.080$ & $4.7$ & $0.8$ & $3.7$ & -- \\
1750 & 4926    & E        & L97 & J94 & $-21.42$ & $11.0$ & $0.132$ & $0.058$ & $0.9$ & $0.9$ & --    & -- \\
1990 & IC 843  & E/S0     & W07 & M00 & $-20.52$ & $9.45$ & $0.485$ & $0.066$ & $3.3$ & $0.5$ & $1.8$ & -- \\
2417 & 4908    & E/S0     & L97 & J94 & $-21.06$ & $7.1$  & $0.322$ & $0.042$ & $2.2$ & $0.9$ & $0.9$ & -- \\
2440 & IC 4045 & E        & W07 & J94 & $-20.30$ & $4.37$ & $0.330$ & $0.038$ & $3.2$ & $1.0$ & --    & $1.0$ \\
2921 & 4889    & D        & L97 & J94 & $-22.56$ & $33.9$ & $0.360$ & $0.028$ & $0.7$ & $0.3$ & --    & -- \\
3329 & 4874    & D        & H98 & J94 & $-22.50$ & $70.8$ & $0.141$ & $0.057$ & $0.4$ & $0.1$ & --    & -- \\
3510 & 4869    & E        & L97 & J94 & $-20.40$ & $7.6$  & $0.112$ & $0.033$ & $2.0$ & $1.1$ & --    & -- \\
3792 & 4860    & E        & L97 & J94 & $-20.99$ & $8.5$  & $0.161$ & $0.071$ & $1.1$ & $1.0$ & --    & -- \\
3958 & IC 3947 & E        & L97 & J94 & $-18.79$ & $3.3$  & $0.323$ & $0.024$ & $1.7$ & $0.9$ & --    & -- \\
4928 & 4839    & E/S0 (D) & L97 & J94 & $-22.26$ & $29.5$ & $0.426$ & $0.104$ & $1.1$ & $0.1$ & --    & $0.2$\\
5279 & 4827    & E        & L97 & M00 & $-21.36$ & $13.6$ & $0.205$ & $0.019$ & $1.6$ & $0.7$ & --    & -- \\
5568 & 4816    & S0       & L97 & M00 & $-21.53$ & $55.7$ & $0.284$ & $0.075$ & $0.5$ & $0.1$ & $0.1$ & -- \\
5975 & 4807    & E        & L97 & M00 & $-20.73$ & $6.7$  & $0.170$ & $0.015$ & $2.9$ & $0.5$ & --    & $1.2$\\
\end{tabular}
\caption{Summary of observational data. (1, 2) galaxy id (GMP from \citealt{GMP}); (3) 
morphological type (from \citealt{Meh00}); (4, 5) HST and ground-based
photometry (L97 = HST/WFPC2 R-band data, Principal Investigator: 
John Lucey, Proposal ID: 5997; H98 = HST/WFPC2 R-band data, Principal Investigator: 
William Harris, Proposal ID: 6104; W07 = HST/WFPC2 R-band data, Principal Investigator: 
Gary Wegner, Proposal ID: 10884; M00 = Kron-Cousins $R_C$-band photometry from 
\citealt{Meh00}; J94 = Gunn $r$ photometry from \citealt{Jor94}); (6) 
absolute B-band magnitude (from Hyperleda; $H_0 = 69 \, \mathrm{km/s/Mpc}$); (7, 8)
effective radius $\reff$ and ellipticity $\epsilon_e$ at $\reff$ from \citet{Meh00}; (9)
$\mathrm{rms}\langle \mu_\mathrm{grd} - \mu_\mathrm{HST} \rangle$ between shifted HST surface
brightness $\mu_\mathrm{HST}$ and corresponding ground-based $\mu_\mathrm{grd}$; 
(10-13) radius
of the outermost kinematic data point along various
slit positions: maj/min/dia = position angle of $0\degr$/$90\degr$/$45\degr$
relative to major axis; off = parallel to major axis (in case of GMP5568: two offset-slits). The
offsets are quoted in the captions of Figs.~\ref{isoplotgh3329} - \ref{isoplotgh3958}.}
\label{dattable}
\end{minipage}
\end{table*}

The kinematic data to be fit by the models 
come from long-slit observations
along at least two position angles: the apparent major and minor axis, respectively.
Kinematical parameters from different sides of a galaxy are averaged.
As error-bars we use the maximum from the two sides or half of the scatter between
them, whatever is larger. This assumes that uncertainties in the observations are
mostly systematic (see also the discussion in Sec.~\ref{subsec:conf}). In case of pure statistical errors and an 
exactly axisymmetric system this would overestimate the errors by a factor 
$\sqrt{2}$. Thus, we are conservative.
 
The data are described in full detail in \citet{Meh00}, \citet{Weg02} and
\citet{Cor07}. Basic
parameters of the photometric and kinematic data set are summarised in Tab.~\ref{dattable}. 

Three galaxies deserve further comments:
\paragraph*{GMP5568/NGC4816:}
GMP5568 has been observed along four position angles.
In addition to major and minor-axis spectra, two observations were made with
the slits parallel to the major-axis: one with an offset of 
$\reff/4$, the other with $\reff/20$.

\paragraph*{GMP0144/NGC4957:}
The velocity dispersion peak of
GMP0144 is significantly off the photometric centre. Furthermore, GMP0144 
is the only galaxy in our sample that exhibits
a significant isophotal twist towards the centre. Thus, GMP0144 is likely 
triaxial near its centre. To reduce the influence of potentially
non-axisymmetric regions on our
modelling, kinematic measurements inside $r < 4 \arcsec$ are omitted.

\paragraph*{GMP5975/NGC4807:}
Dynamical models for GMP5975, based on major and minor-axis kinematics, have already been 
presented in \citet{Tho05}. There, a striking depopulation of retrograde orbits was
found. To check its significance we also determined kinematics along a diagonal slit. 
Here we present new models that include this additional kinematical data. Both, 
the mass structure and the distribution of stellar orbits did not change significantly.

\section{Dynamical modelling}
\label{sec:setup}
We model the kinematic and photometric observations with
Schwarzschild's orbit superposition method \citep{S79}. 
Details about our implementation are given
in \citet{Tho04,Tho05}. Basic steps of the method are briefly recalled in this section.

\subsection{Deprojection and inclination}
\label{subsec:depro}
The surface photometry is deprojected to obtain the 3d luminosity distribution
$\nu$ for each galaxy (using
the program of \citealt{mag99}). We consider radial profiles of surface-brightness, ellipticity
and isophotal shape parameters $a_4$ and $a_6$ \citep{bm87} for the deprojections\footnote{In
case of GMP1176 isophotal shape parameters up to $a_{12}$ are used to 
represent the isophotes appropriately (see also \citealt{Cor07}).} (cf. 
App.~\ref{sec:fits}). For each galaxy, we probe three different inclinations in
the deprojection, and subsequent dynamical modelling, respectively:
(1) $i=90\degr$ (edge-on), where the deprojection is intrinsically 
least flattened; (2) a minimum inclination that is found by requiring the 
deprojection to be intrinsically as flattened as an E7 galaxy; (3) an intermediate
inclination, for which the deprojection looks like an E5 galaxy from the side.
This inclination scheme emerges as a compromise between limited computation time on 
the one side and the strategy to get the most conservative
estimate of uncertainties on intrinsic properties on the other. In many galaxies the
inclination is only poorly constrained (cf. Sec.~\ref{subsec:inclin}). In other words,
when we quote 68 percent confidence uncertainties on intrinsic properties below this
includes in many cases models from all three probed inclinations, including those
connected with the rather extreme intrinsic E5 and E7 shapes.

In case of GMP0756, GMP1176, GMP1990 and GMP2417 
only the edge-on orientation is considered. These galaxies are all highly flattened. 
In addition, they appear either discy (e.g. GMP1176)
or have thin dust features (GMP1990, GMP2417; cf. \citealt{Cor07}), implying 
that they are seen close to edge-on.

The dynamical modelling of GMP5975 in \citet{Tho05} revealed that only models at $i=90\degr$ were within the  
one sigma confidence region. It was also found that the deprojection of GMP5975 becomes 
implausibly boxy in the outer parts, if it is assumed that the galaxy is significantly 
inclined. We therefore reanalysed the extended data set for GMP5975 only
with $i=90\degr$.

\subsection{Mass model}
\label{subsec:mass}
With the luminosity density $\nu$ given, a trial mass density distribution can be defined by
\begin{equation}
\label{rhorho}
\rho = \Upsilon \, \nu + \rho_\mathrm{DM}.
\end{equation}
The stellar mass-to-light ratio $\Upsilon$ is assumed constant throughout the
galaxy. Concerning the dark matter density $\rho_\mathrm{DM}$ we probe the 
two following parametric prescriptions. Firstly the NFW-distribution \citep{nfw96}
\begin{equation}
\label{nfw}
\rho_\mathrm{NFW}(r,r_s,c) \propto \frac{1}{(r/r_s)(1+r/r_s)^2},
\end{equation}
where $r_s=r_{200}/c$ is a scaling radius, $r_{200}$ is a measure of the virial radius 
and $c$ is the concentration of the halo. Simulations predict
the two halo parameters $r_s$ and $c$ to be correlated, such that
the distribution (\ref{nfw}) can be read as a one-parameter family
of dark matter halos \citep{nfw96}. To be explicit, we use
\begin{equation}
\label{family}
r_s^3 \propto 10^{(A-\log c)/B} \left( 200 \, \frac{4 \pi}{3} \, c^3 \right)^{-1}
\end{equation}
with $A=1.05$ and $B=0.15$ \citep{nfw96,R97}. We consider spherical as well as flattened
NFW halos, where the latter are derived from equation (\ref{nfw}) by the
substitution $r \rightarrow r \sqrt{\cos^2 (\vartheta) + \sin^2 (\vartheta) / q^2}$
($q$ is the constant flattening of the isodensity contours, $\vartheta$ is the
latitude in spherical coordinates).

The second halo family used is the logarithmic potential
\begin{equation}
\label{nis}
\rho_\mathrm{LOG}(r) \propto v_c^2
\frac{3r_c^2+r^2}
{(r_c^2+r^2)^2},
\end{equation}
that gives rise to an asymptotically constant circular velocity $v_c$ and a
flat central density core inside $r \la r_c$ (\citealt{Bin81}).

In the gravitational potential generated by the mass distribution (\ref{rhorho}) 
we compute typically about $18000$ orbits as described in \citet{Tho04}. 

\subsection{Orbit superposition}
\label{subsec:orbitfit}
The final orbit superposition model is constructed according to the 
maximum entropy technique of \citet{Ric88}. It consists in solving 
\begin{equation}
\label{maxs}
\hat{S} \equiv S - \alpha \, \chi^2_\mathrm{LOSVD} \rightarrow \mathrm{max},
\end{equation}
with $S$ denoting the Boltzmann entropy
\begin{equation}
\label{bentropy}
S \equiv - \int f \ln \left( f \right) \,
\der^3r \, \der^3v
\, = - \sum_i w_i \ln \left( \frac{w_i}{V_i} \right)
\end{equation}
and $f$ being the phase-space distribution function (DF) of the model.
In Schwarzschild models -- by construction -- the DF is constant along individual orbits.
The corresponding phase-space density $f_i$ along orbit $i$ is the ratio
\begin{equation}
\label{eq:fi}
f_i \equiv \frac{w_i}{V_i}
\end{equation}
of the total amount of light $w_i$ on the orbit (the so-called orbital weight)
and the orbital phase-space volume $V_i$. The $w_i$ that solve 
equation (\ref{maxs}) are obtained 
iteratively: starting with a low $\alpha = 10^{-10}$ we solve equation (\ref{maxs}) for a
fixed set of $\alpha_i$, using the
orbital weights obtained at $\alpha_{i-1}$ as initial guess for the solution
at $\alpha_i$. 

\begin{table*}\centering
\begin{minipage}{170mm}
\begin{tabular}{l||cc||cccc||cccc||cccc}
 & \multicolumn{2}{c}{no DM} & \multicolumn{4}{c}{LOG halos} & 
\multicolumn{4}{c}{NFW halos}\\
GMP & 
$\mlsc$ & $\chisc$ & 
$\mllog$ & $r_c$ & $v_c$ & $\chilog$ & 
$\mlnfw$ & $c$ & $q$ & $\chinfw$ & 
halo & $\Delta \chi^2_\mathrm{halo}$ & $\Delta \chi^2_\mathrm{DM}$ & $i$\\
\multicolumn{1}{c}{(1)} & (2) & (3) & (4) & (5) & (6) & (7) & (8) & (9) & (10) & (11) & (12) & (13) & (14) & (15) \\
\hline
0144 & 7.0 & 0.400& 5.0 & 4.4 & 212  & 0.383   & 4.5 & 17.17 & 0.7 & 0.336& NFW & -2.45 & 3.3  &$50^{50}_{50}$\\
0282 & 6.5 & 0.436& 5.0 & 17.0 & 502 & 0.244  & 4.5 & 11.24 & 0.7 & 0.256& LOG & 1.01   & 16.9 &$60 ^{70}_{60}$\\
0756 & 3.0 & 1.253& 2.6 & 12.7 & 215 & 0.930  & 2.2 & 20.2 & 0.7 & 0.942 & LOG & 1.57   & 41.3 & 90\\
1176 & 2.5 & 1.353& 2.0 & 3.4 & 200  & 0.724   & 2.0 & 18.0 & 1.0 & 0.707 & NFW & -1.8  & 67.2 &90\\
1750 & 7.0 & 0.540& 6.0 & 18.7 & 500 & 0.452  & 6.0 & 12.5 & 1.0 & 0.469 & LOG & 0.81   & 4.2  &$65^{90}_{65}$\\
1990 & 10.0& 0.301& 10.0 & 13.1 & 105 & 0.291 & 9.0 & 24.0 & 1.0 & 0.298 & LOG & 0.72   & 1.0  & 90\\
2417 & 8.5 & 0.244& 8.0 & 23.8 & 500 & 0.206  & 7.0 & 14.76 & 0.7 & 0.216& LOG & 0.46   & 1.8  &90\\
2440 & 7.0 & 0.579& 6.5 & 10.9 & 482 & 0.453  & 6.5 & 16.47 & 0.7 & 0.475& LOG & 1.69   & 9.6  &$60^{60}_{60}$\\
2921 & 9.0 & 0.112& 6.5 & 8.2 & 425 & 0.073   & 6.5 & 9.2 & 0.7 & 0.067  & NFW & -0.47  & 3.3  &$90^{90}_{60}$\\
3329 & 12.0& 0.325& 7.0 & 3.6 & 400 & 0.307   & 9.0 & 10.85 & 0.7 & 0.309& LOG & 0.22   & 1.4  &$90^{90}_{45}$\\
3510 & 6.0 & 0.425& 5.5 & 11.6 & 287 & 0.398  & 5.0 & 16.12 & 0.7 & 0.398& LOG & 0.67   & 2.5  &$90^{90}_{60}$\\
3792 & 9.0 & 0.370& 8.0 & 15.3 & 550 & 0.339  & 8.0 & 10.0 & 1.0 & 0.349 & LOG & 0.54   & 1.7  &$60^{90}_{40}$\\
3958 & 6.0 & 0.229& 5.0 & 6.8 & 274 & 0.162   & 4.0 & 14.7 & 1.0 & 0.174 & LOG & 0.42   & 2.4  &$90^{90}_{70}$\\
4928 & 10.0& 0.232& 8.5 & 29.1 & 507 & 0.109  & 7.0 & 12.7 & 1.0 & 0.122 & LOG & 0.66   & 6.4  &$90^{90}_{70}$\\
5279 & 7.0 & 0.132& 6.5 & 28.4 & 482 & 0.099  & 6.0 & 15.9 & 0.7 & 0.109 & LOG & 0.71   & 2.3  &$90^{90}_{90}$\\
5568 & 7.0 & 0.162& 6.0 & 66.7 & 650 & 0.103  & 5.0 & 17.2 & 0.7 & 0.104 & LOG & 0.12   & 5.2  &$90^{90}_{50}$\\
5975 & 4.0 & 0.580& 3.0 & 1.7 & 200 & 0.333   & 3.0 & 15.0 & 0.7 & 0.314 & NFW & -1.37  & 19.1 &90        \\
\end{tabular}
\caption{Summary of modelling results. (1) galaxy id (cf. Tab.~\ref{dattable}); 
(2-3) best-fit stellar $\mlsc$ [$M_\odot/L_\odot$] ($R_C$-band) and  achieved goodness-of-fit 
$\chisc$ (per data point) without dark matter; (4-7) as (2-3), but for logarithmic halos 
with parameters $r_c$ [kpc] and $v_c$ [km/s]; (8-11) same as (2-3), but for NFW halos with 
concentration $c$ and flattening $q$; (12-13) best-fit halo-profile with significance 
$\Delta \chi^2_\mathrm{halo} = (\chinfw-\chilog)\times N_\mathrm{data}$; (14) evidence for
dark matter $\Delta \chi^2_\mathrm{DM} = (\chisc - \chimin) \times N_\mathrm{data}$;
(15) inclination of best-fit with minimum and maximum in the 68 percent confidence
region of calculated models (where no range is quoted, only edge-on models were calculated).}
\label{modtable}
\end{minipage}
\end{table*}

The $\chi^2$-term in equation (\ref{maxs}) quantifies deviations between model and data.
We do not include the photometry in the $\chi^2$, but treat the deprojected luminosity 
distribution as a boundary condition for the solution of equation (\ref{maxs}).
To fit the measured LOSVDs, which are parameterised in terms of the Gauss-Hermite parameters
$v$, $\sigma$, $H_3$ and $H_4$ \citep{Ger93,vdMF93} we proceed as follows: 
the Gauss-Hermite parameters are used to generate
binned data LOSVDs ${\cal L}^{jk}_\mathrm{dat}$. Errors are
propagated via Monte-Carlo simulations. These data LOSVDs and the corresponding
model quantities ${\cal L}^{jk}_\mathrm{mod}$ are used to get the
$\chi^2_\mathrm{LOSVD}$ of equation (\ref{maxs}):
\begin{equation}
\label{chilosvd}
\chi^2_\mathrm{LOSVD} \equiv \sum_{j=1}^{N_{\cal L}} \, \sum_{k=1}^{N_\mathrm{vel}} 
\left(
\frac{{\cal L}^{jk}_\mathrm{mod}-{\cal L}^{jk}_\mathrm{dat}}
{\Delta {\cal L}^{jk}_\mathrm{dat}}
\right)^2.
\end{equation}
The above sum includes all $N_{\cal L}$ data points and each LOSVD is represented
by $N_\mathrm{vel}$ bins in projected (line-of-sight) velocity.

With the orbital weights $w_i$ determined, the dynamical state of the model is
completely specified, i.e. the phase-space distribution function is
known (cf. equation \ref{eq:fi}). In the course of this paper we will not only consider
the DF, but also the orbital anisotropy. It can be
quantified by the so-called anisotropy parameters
\begin{equation}
\label{eq:bt}
\betatheta \equiv 1 - \frac{\sigma_\vartheta^2}{\sigma_r^2}
\end{equation}
(meridional anisotropy) and
\begin{equation}
\label{eq:bp}
\betaphi \equiv 1 - \frac{\sigma_\varphi^2}{\sigma_r^2}
\end{equation}
(azimuthal anisotropy). Internal velocity dispersions $\sigma$ in the above
equations are computed in spherical coordinates
$r$, $\vartheta$ and $\varphi$, oriented such that $\varphi$ is the azimuth in the 
equatorial plane and $\vartheta$ is the latitude. 

\subsection{Regularisation}
\label{subsec:regula}
The parameter $\alpha$ in equation (\ref{maxs}) controls the relative weight of
data-fit and entropy maximisation. The higher $\alpha$ the better the fit and the
larger the noise in the derived distribution function, or orbital weights, respectively. 
Ideally, regularisation has to be optimised case-by-case for each galaxy. This
holds in principle for both, the value of the regularisation parameter $\alpha$, as well as for the
functional form of $S$. Firstly, because the
spatial resolution and/or coverage as well as the signal-to-noise of the observations
vary from galaxy to galaxy and regularisation should be adapted to 
that. This primarily concerns the choice of $\alpha$.
Secondly, different galaxies have different intrinsic structures. Specifically, the
degree to which the entropy of a stellar system is maximised may vary in phase-space.
Consider, for example, a cold disc inside a hot spheroid. The disc has low entropy and
to fit its rotation, $\alpha$ needs to be large (models with the maximum entropy according
to equation \ref{bentropy} have no rotation). On the other hand, the spheroid-dominated region in phase-space can have
higher entropy and using a large $\alpha$ in the fit amplifies the noise in the corresponding
parts of the phase-space distribution function (DF). 
The dilemma as to the choice of $\alpha$ in such cases could be solved by 
adjusting the function $S$ appropriately.

For the Coma galaxy modelling we use the same regularisation for all
galaxies: $\alpha = 0.02$ and the entropy of equation (\ref{bentropy}). The value
for $\alpha$ has been obtained by Monte-Carlo simulations of isotropic rotator test
galaxies with realistic, noisy mock data \citep{Tho05}. Applying it to the
whole sample is motivated by the similar spatial coverage and resolution of all our
Coma observations (cf. Sec.~\ref{sec:obs}). 
Furthermore, since it has been obtained from fitting isotropic rotators it has proven
to be sufficiently large to fit non-maximum entropy, rotating galaxies. It might
be slightly too large for non-rotating galaxies. Thus we expect models of non-rotating 
galaxies to possibly be noisier than those of rotating galaxies, but we do not
expect that our imposed regularisation is too restrictive. 
In any case, we will explicitly investigate the
dependency of model results on the choice of $\alpha$ in Sec.~\ref{sec:regula}.

\subsection{Best-fit model and uncertainties}
\label{subsec:bfit}
To obtain the best-fit mass model for a given $\alpha$ 
we calculate orbit models as described in Secs.~\ref{subsec:mass} and \ref{subsec:orbitfit}
for various combinations of the relevant parameters:
($r_c,v_c,\Upsilon$) in case of LOG-halos, ($c,\Upsilon,q$) in case of NFW-halos or just 
$\Upsilon$ for models without dark matter, respectively. 
Logarithmic halos are probed on
a grid with $\Delta r_c \approx \reff/2$ and $\Delta v_c = 50 \, \mathrm{km/s}$ 
(in some cases the grid is refined around the location of the lowest $\chi^2$). 
Typically we explore $N_r \times N_v \approx 100$ halos. The analogous numbers for the
NFW halos read $\Delta c = 2.5$, $N_c \approx 12$ and $N_q = 2$ ($q \in \{0.7,1.0\}$). 
The step-size $\Delta \Upsilon$ 
for the mass-to-light ratio equals 10-20 percent of the best-fit $\mldyn$, independent 
of the halo type. Around the best-fit
model the resolution in $\Upsilon$ is doubled, resulting in $N_\Upsilon \approx 6-10$ 
models with different
mass-to-light ratios for each halo. This sums up to about $600-1000$ models with logarithmic 
halos and $120-240$ models with NFW halos. The final number of models is up to a 
factor of three larger, depending on the number $N_i \le 3$ of probed inclinations $i$. 
The total number of models per galaxy is around $1000-3000$.

\begin{figure*}\centering
\begin{minipage}{170mm}
\includegraphics[width=83mm,angle=0]{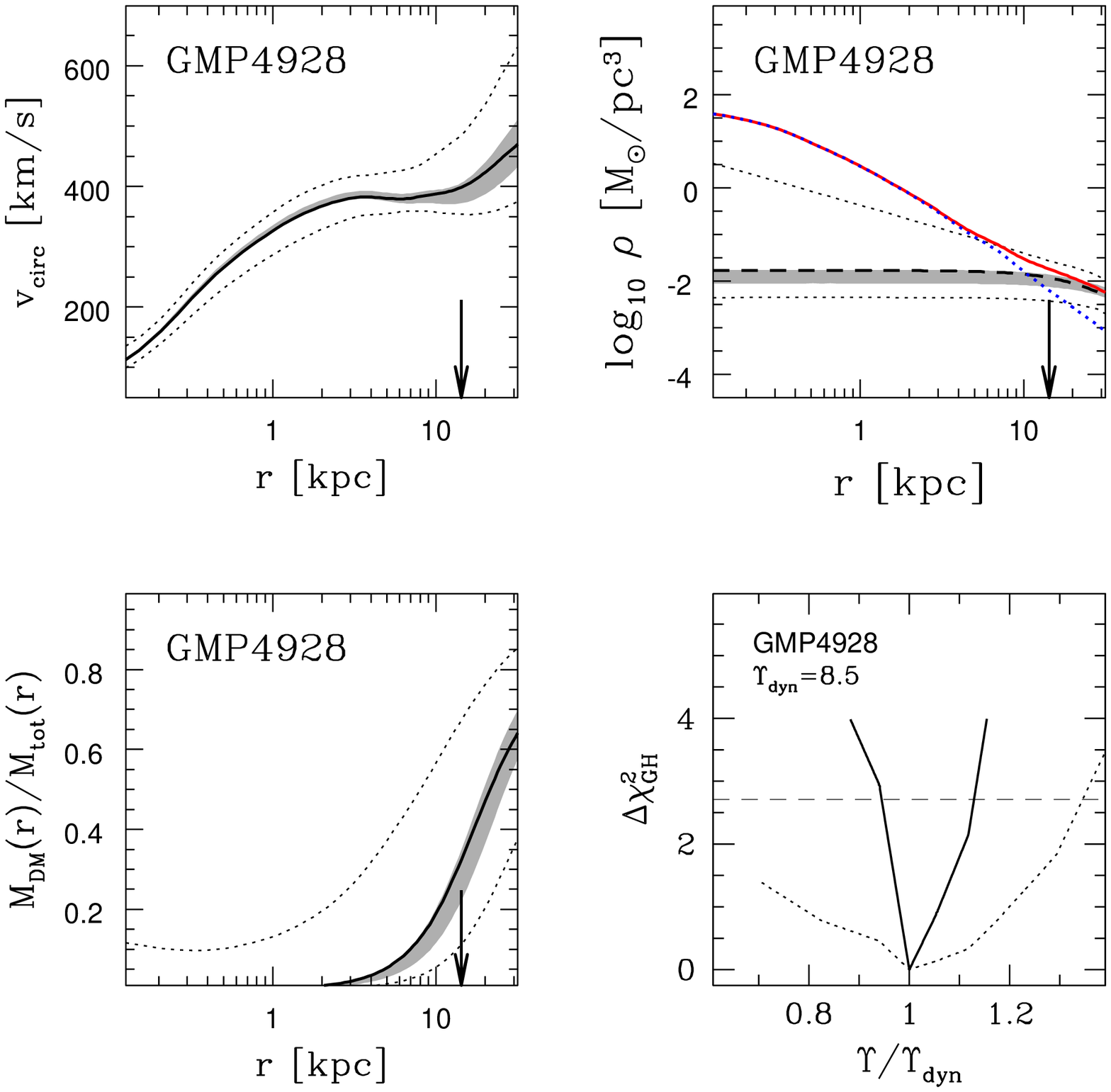}
\includegraphics[width=83mm,angle=0]{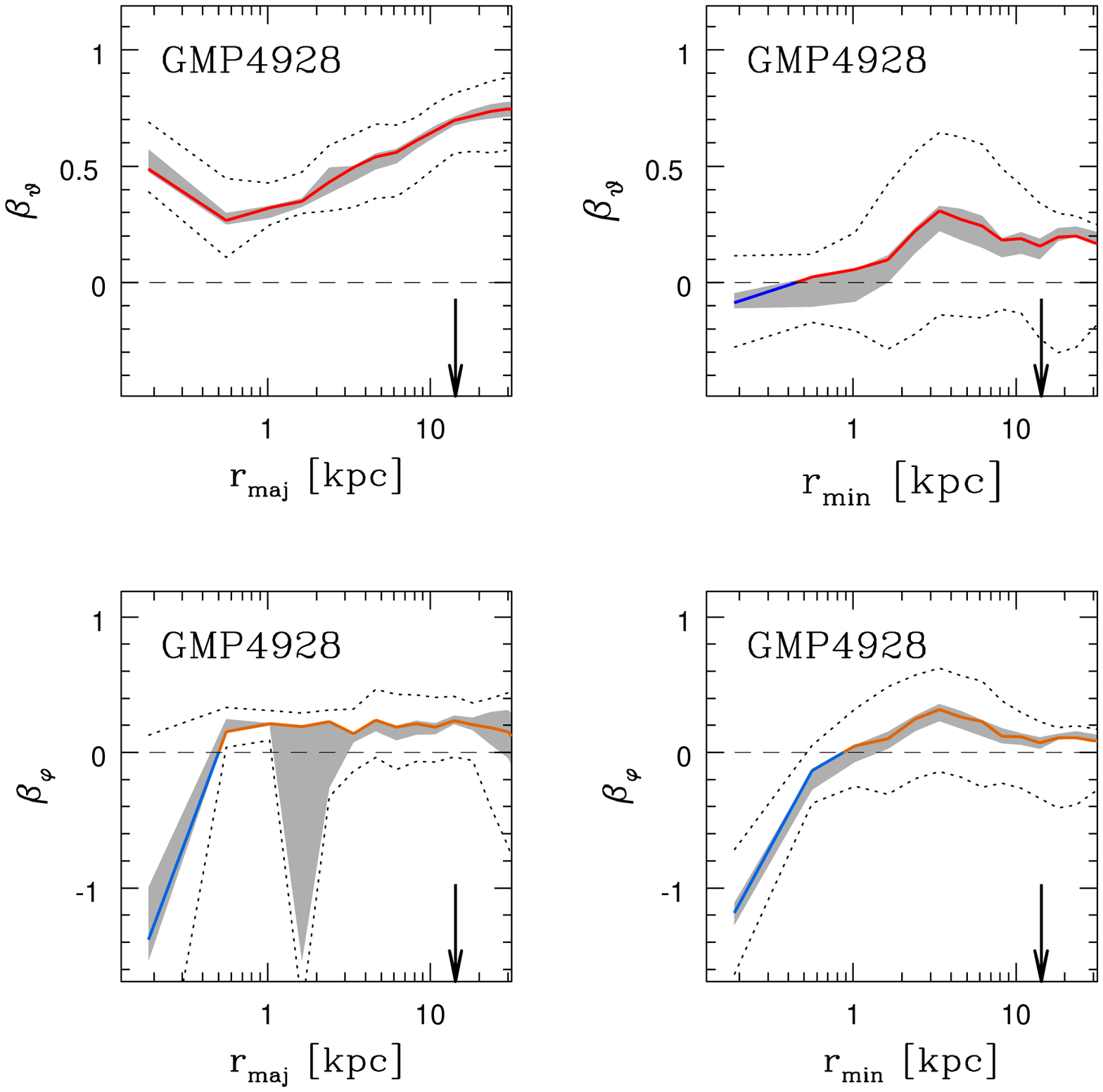}
\caption{Comparison of confidence intervals with (shaded) and without (dotted) rescaling the 
$\chi^2$ (details in the text). Individual panels show circular velocity $\vcirc$, mass 
densities $\rho$ (stellar: blue/dotted, dark matter: black/dashed and total: red/solid), 
dark matter fraction $M_\mathrm{DM}/M$, $\chi^2_\mathrm{GH}$ (dotted without rescaling, solid
with rescaling) and velocity anisotropies $\beta_\vartheta$ and 
$\beta_\varphi$ along major and minor-axis, respectively.}
\label{4928:rescaled}
\end{minipage}
\end{figure*}

Among these models we determine the best-fit according to the lowest 
$\chi^2_\mathrm{GH}$, defined as
\begin{eqnarray}
\label{chigheq}
\chi^2_\mathrm{GH} \equiv \sum_{j=1}^{N_{\cal L}} 
\left[
\left( 
\frac{v^j_\mathrm{mod} - v^j_\mathrm{dat}}{\Delta v^j_\mathrm{dat}} 
\right)^2 +
\left( 
\frac{\sigma^j_\mathrm{mod} - \sigma^j_\mathrm{dat}}{\Delta \sigma^j_\mathrm{dat}} 
\right)^2 + \right. \nonumber \\
\left.
\left( 
\frac{H^j_{3,\mathrm{mod}} - H^j_{3,\mathrm{dat}}}{\Delta H^j_{3,\mathrm{dat}}} 
\right)^2 +
\left( 
\frac{H^j_{4,\mathrm{mod}} - H^j_{4,\mathrm{dat}}}{\Delta H^j_{4,\mathrm{dat}}} 
\right)^2
\right].
\end{eqnarray}
Here, $v^j_\mathrm{dat}$ is the rotation according to the
Gauss-Hermite parameterisation of the LOSVDs (other parameters analogously). A detailed
discussion about the difference between $\chi^2_\mathrm{GH}$ and $\chi^2_\mathrm{LOSVD}$
can be found in \citet{Tho05}.

Confidence intervals on model 
quantities are set by the corresponding minimum and maximum values obtained over all 
models within $\Delta \chi^2_\mathrm{GH} = 1.1$ from the
minimum $\chi^2_\mathrm{GH}$. The value of $\Delta \chi^2_\mathrm{GH} = 1.1$ is slightly
more conservative than the classical $\Delta \chi^2_\mathrm{GH} = 1.0$ and has been
derived by means of Monte-Carlo simulations \citep{Tho05}.

As a byproduct of the iterative technique to solve 
equation (\ref{maxs}), we get -- for each set of ($r_c,v_c,\Upsilon,i$), 
($c,q,\Upsilon,i$) and ($\Upsilon,i$), respectively -- 
orbit models for about $N_\alpha \approx 50$ different values
of the regularisation parameter. This allows us to 
derive a best-fit model for each $\alpha_i$ and to explore the dependency of best-fit
model parameters on $\alpha$ (cf. Sec.~\ref{sec:regula}).

\section{Modelling results}
\label{sec:res}
Modelling results are summarised in Tab.~\ref{modtable}.
In the remainder of this Section
we collect some general notes on these results.

\subsection{Goodness-of-fit}
\label{subsec:conf}
The goodness-of-fit obtained under the different assumptions about the overall mass 
distribution are given in columns (3), (7) and (11) of Tab.~\ref{modtable}. Thereby
\begin{equation}
\chisc \equiv \min \{ \chigh(\Upsilon,\, i)/N_\mathrm{data}\},
\end{equation}
\begin{equation}
\chilog \equiv \min \{ \chigh(r_c,\, v_c,\, \Upsilon,\, i)/N_\mathrm{data}\}
\end{equation}
and
\begin{equation}
\chinfw \equiv \min \{ \chigh(c,\, q,\, \Upsilon,\, i)/N_\mathrm{data}\}
\end{equation}
are minimised over all relevant mass parameters. Differences between models with and 
without dark matter are further discussed in Sec.~\ref{subsec:ml}. Here we only refer to the
fact that our models are able to reproduce 
the observations with a $\chigh$ per data point which is in many cases significantly 
smaller than unity. The largest
deviations between model and data occur for the S0 GMP0756, possibly related to the low
$H_4$ along the offset-axis, which are not followed by our models.
Fits to some galaxies are as good as $\chimin \la 0.1$, where
\begin{equation}
\label{eq:chimm}
\chimin \equiv \min \{\chilog, \, \chinfw , \chisc\}
\end{equation}
describes the overall minimum of $\chigh$ for a given galaxy.
In many cases where $\chimin$ is particularly low, error bars of
the observations are much larger than the point-to-point scatter of the data points.
This concerns GMP5279, GMP2921, GMP4928, GMP5568 and GMP3958, where
the observational errors are likely overestimated (see also the fits in App.~\ref{sec:fits}). 
In some other systems, like for example GMP0144, the error bars used in the modelling are 
rather large, because they also include side-to-side variations of the kinematics, which
are often also larger than the point-to-point scatter on a given side of the galaxy. 
Thus, both effects might partly explain the low $\chimin$ of these galaxies.

Very low $\chimin$ raise the question whether confidence intervals of 
model properties calculated as described in Sec.~\ref{subsec:bfit} (and shown in 
Figs.~\ref{vcirccomparison}, \ref{denscomparison}, 
\ref{anisominor} and \ref{anisomaj} below) are overestimated. In cases where the
observational errors are obviously too large it is reasonable to rescale them
until $\chimin \approx 1$. In fact, this has been done for GMP5975 in
\citet{Tho05}, where error bars were scaled such that
$\chimin \approx 0.7$. This value was determined from Monte-Carlo simulations
of isotropic rotator models. To quantify the effect of rescaling, 
Fig.~\ref{4928:rescaled} exemplifies confidence intervals for one galaxy of the
sample (GMP4928) once without rescaling the $\chi^2_\mathrm{GH}$ and once with
rescaling all observational error bars to $\chimin = 0.7$. As it can be seen, 
uncertainty regions shrink a lot after rescaling.

\begin{figure*}\centering
\begin{minipage}{166mm}
\includegraphics[width=164mm,angle=0]{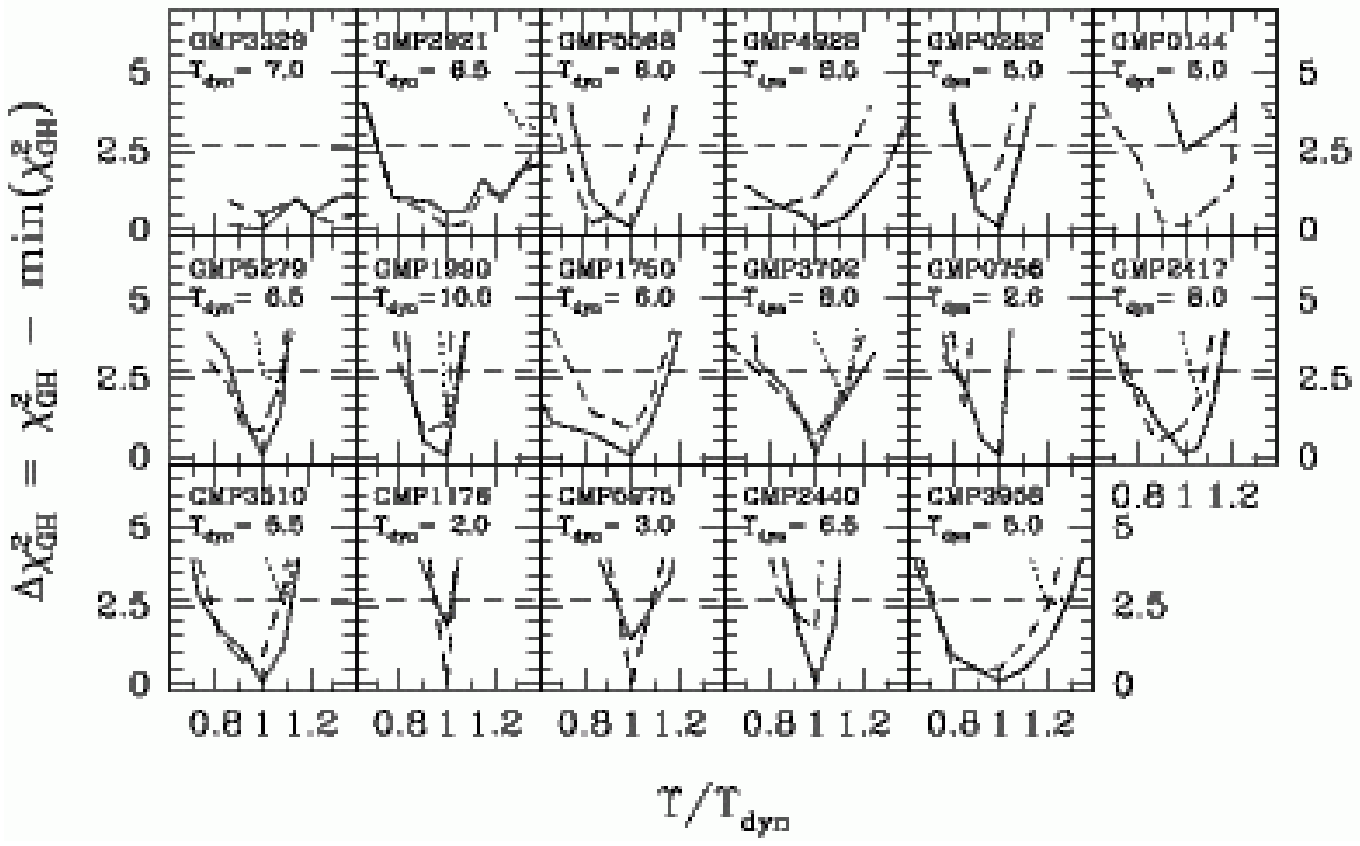}
\caption{Confidence levels $\Delta \chi^2_\mathrm{GH}$ versus $\Upsilon$ 
(normalised to the best-fit $\mldyn$). Solid: logarithmic halos; dashed: NFW halos; dotted: mass follows
light; horizontal dashed: 90 percent confidence limit. Where a dotted line is missing
the self-consistent case is ruled out with more than 95 percent confidence (exception: GMP3329,
where the best-fit $\mlsc \approx 1.7 \, \mldyn$ is outside the plotted region). From 
top-left to bottom-right galaxies are arranged in order of decreasing total mass inside 
$\reff$.}
\label{chicomp}
\end{minipage}
\end{figure*}

Globally rescaling the error-bars is not appropriate for all galaxies, however. Errors
along the major-axis of GMP3792, 
for example, might be overestimated, but
those along the minor-axis seem not. Moreover, in a case like GMP1750
slight minor-axis rotation -- which cannot be fit with axisymmetric models -- 
adds a constant to $\chimin$. Just rescaling to $\chimin \approx 0.7$
in all galaxies would introduce an artificial dependency of uncertainty regions on minor-axis
rotation, which is not appropriate. In order to treat all galaxies of the sample
homogeneously we do not rescale the
$\chi^2_\mathrm{GH}$, but give the most conservative error-bars for our models.
The corresponding
confidence intervals may be interpreted as the maximal uncertainty on derived model
quantities, while the shaded regions of Fig.~\ref{4928:rescaled} may be interpreted
as lower limits for these uncertainties.

Fig.~\ref{chicomp} shows the dependency of
\begin{equation}
\Delta \chi^2_\mathrm{GH}(\Upsilon) \equiv \chi^2_\mathrm{GH}(\Upsilon) - \mathrm{min} \, (\chi^2_\mathrm{GH}),
\end{equation}
where $\mathrm{min} \, \chigh \equiv \chi^2_\mathrm{min} \times N_\mathrm{data}$ 
(cf. equation \ref{eq:chimm}) and 
$\chi^2_\mathrm{GH}(\Upsilon)$ is minimised over all NFW-fits, logarithmic-halo fits and
self-consistent fits with the given $\Upsilon$. For all but one galaxy, we find a clear minimum in
$\chigh(\Upsilon)$. The exceptional case, GMP3329, is peculiar in many respects: 
(1) It is among the brightest galaxies of the sample and has
a very large $\reff$. The data only cover the region inside $r \la \reff/2$. (2) The
surface-brightness profile shows a break near $0.2 - 0.3 \, \reff$ 
(cf. upper panel of Fig.~\ref{isoplotgh3329}). (3) At about the same
projected distance from the centre the velocity dispersion dips and rises again at larger
radii (cf. lower panel of Fig.~\ref{isoplotgh3329}).
The poor constraints on the mass-to-light ratio in this system could be related to the 
poor data coverage. It might also be that GMP3329 is actually composed of two subcomponents. 
If these have different mass-to-light ratios $\Upsilon$, then the $\chigh$-curve may have 
two corresponding local minima and the poor data coverage may smooth out these into a 
flat plateau. Finally, our modelling of GMP3329 may suffer from the Coma core being
possibly not in dynamical equilibrium, as indicated by the kinematics of intra-cluster 
planetary nebulae \citep{Ger07}.

\subsection{Model inclinations}
\label{subsec:inclin}
Most of the best-fit models are edge-on according to the last column of Tab.~\ref{modtable}. This is surprising
at first sight because if galaxies are oriented randomly then one would expect only about 
6 out of 17 objects to have inclinations larger than $i\ga70\degr$. Omitting
the five systems for which only edge-on models were calculated (GMP0756, GMP1176, GMP1990,
GMP2417 and GMP5975; cf. Sec.~\ref{subsec:depro}) and taking
into account the uncertainties quoted in Tab.~\ref{modtable} there are 3 galaxies out
of 11 where inclinations $i<70\degr$ are ruled out by our modelling (at the 68 
percent confidence level). This is in good agreement with the expectation for random
inclinations. Nevertheless, we now discuss a little more whether 
our modelling might be subject to a slight inclination bias.

First, one possible issue is that 
using the same regularisation for all galaxies might introduce a subtle 
bias towards edge-on configurations. Consider a rotating system: the lower the assumed
inclination of the model the larger its intrinsic rotation needs to be in order to match the 
observations after projection. Thus, the system will be dynamically colder and its
entropy will be lower. Turning the argument around: usage of a constant $\alpha$ enforces the
same weight on entropy in the inclined model as in the edge-on model. Since the inclined
model has to have lower entropy, however, its fit may be less good. This might drive
models of rotating galaxies towards $i=90\degr$. We do not expect this to affect
conclusions drawn from our modelling
results strongly, because, as it has been argued in Sec.~\ref{subsec:depro}, error bars on
intrinsic properties include in many cases models at different inclinations, even
extreme cases. But it might drive the {\it best-fit} model to occur preferentially 
around $i=90\degr$.

Second, for face-on galaxies noise in the kinematics may be a source of bias towards
edge-on models as well \citep{Tho07}. It can cause
rotation measurements $v \ne 0$ even for exactly face-on, axisymmetric galaxies. An edge-on
model can in principle fit these $v\ne0$, whereas face-on models necessarily obey 
$v \equiv 0$. Thus, everything else fitting equally
well, the contribution of noise in $v$ to the $\chi^2$ would be smaller in edge-on than in
face-on models. Since we have no clear candidate face-on galaxy in our sample we
do not expect this issue to be relevant to the Coma sample, however.

\begin{figure}\centering
\includegraphics[width=84mm,angle=0]{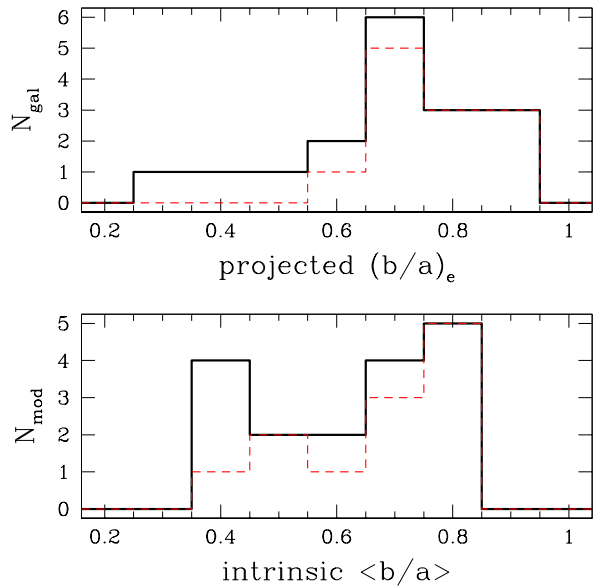}
\caption{Top: histogram of apparent short-to-long axis ratios at $\reff$. Bottom: intrinsic best-fit 
short-to-long axis ratio $\langle b/a \rangle$ (averaged over $r/\reff \in [0.5,2.5]$).
Black/solid: whole sample; red/dashed: without S0s.}
\label{fig:shape}
\end{figure}

The third thing to note is that galaxies for our sample may be not at random
inclinations. The sample is
designed to complement earlier studies on round galaxies and we explicitly selected
flattened, rotating ellipticals and S0s to be fitted. The distribution of apparent
axis-ratios $(b/a)_e \equiv 1 - \epsilon_e$ (with $\epsilon_e$ from Tab.~\ref{dattable})
is shown in the upper panel of Fig.~\ref{fig:shape}. The sample exhibits a tail of 
highly flattened systems, which lacks, for example, in the distribution of bright galaxies 
with de-Vaucouleurs profiles in the Sloan Digital Sky Survey \citep{Vin05}. This tail
is produced by the S0 galaxies in our sample and clearly shows that the sample as a whole 
is biased towards flattened systems. The ellipticity distribution of those galaxies that are 
classified as ellipticals in Tab.~\ref{dattable} (dashed line in Fig.~\ref{fig:shape}) is
still shifted slightly towards higher ellipticities with respect to the bright ellipticals 
of \citet{Vin05}. In combination with the lack of round objects in our sample, this 
indicates that even our 11 ordinary ellipticals are slightly biased, 
but the sample is too small for a definite appraisal.

Fourth, even assuming our sample galaxies are at random inclinations and that
the first two just discussed points are irrelevant (regularisation and noise) 
and that inclinations can be reconstructed uniquely from ideal data with ideal models then 
we still could be faced with a slight bias in our models: as it has been described in
Sec.~\ref{subsec:depro} we do not probe a fine grid in inclinations but look for
extreme cases. Our models provide for each galaxy only the choice between
edge-on or intrinsically E5/E7, respectively. Since an intrinsic E5/E7 shape is a
rather extreme assumption this might drive the modelling towards the edge-on option as 
well. 

\begin{figure*}\centering
\begin{minipage}{166mm}
\includegraphics[width=164mm,angle=0]{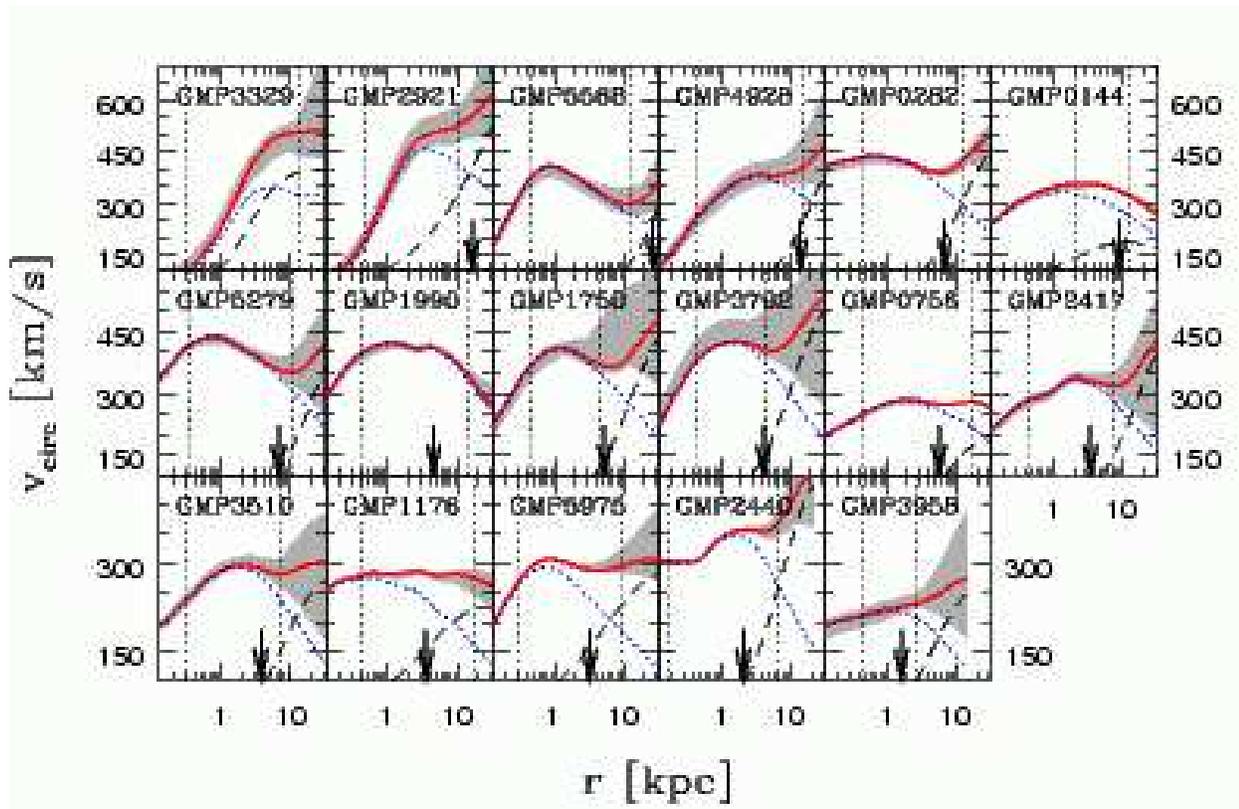}
\caption{Compilation of circular velocity curves. Thick/red: luminous+dark (68 per cent confidence
region shaded); dotted/blue: luminous only; dashed: dark matter only; vertical dotted
lines: boundaries of kinematical data; arrows denote the effective radius $\reff$. From 
top-left to bottom-right galaxies are arranged in order of decreasing total mass inside 
$\reff$.}
\label{vcirccomparison}
\end{minipage}
\end{figure*}

The distribution of intrinsic axis ratios of the Coma models is shown in the lower
panel of Fig.~\ref{fig:shape}. It peaks at $b/a = 0.8$, consistent with deprojections 
of the frequency function of elliptical galaxy apparent flattenings 
\citep{Tre96,Vin05}. Compared with these studies, the distribution in
the lower panel of Fig.~\ref{fig:shape} has relatively more galaxies on the flatter
side of the peak and relatively fewer galaxies on the rounder side. Now, if the
modelling would be subject to a strong bias towards $i\to90\degr$ then we would expect
the opposite: an overestimation of intrinsically roundish galaxies. Thus, the lower
panel of Fig.~\ref{fig:shape} argues against
a strong modelling bias towards high inclinations. However, the argument is not conclusive, 
because the sample itself maybe biased against apparently round galaxies. This 
could partly compensate for an inclination bias in the modelling.

In conclusion, although there might be a slight inclination bias in the modelling and/or the
sample galaxies, Fig.~\ref{fig:shape} reveals that either this bias is not very
strong, or that modelling and sample biases roughly counterbalance each other.

\section{Luminous and dark matter distribution}
\label{sec:mass}
Now we discuss the distribution of luminous and dark matter in the Coma models.

\begin{figure*}\centering
\begin{minipage}{166mm}
\includegraphics[width=164mm,angle=0]{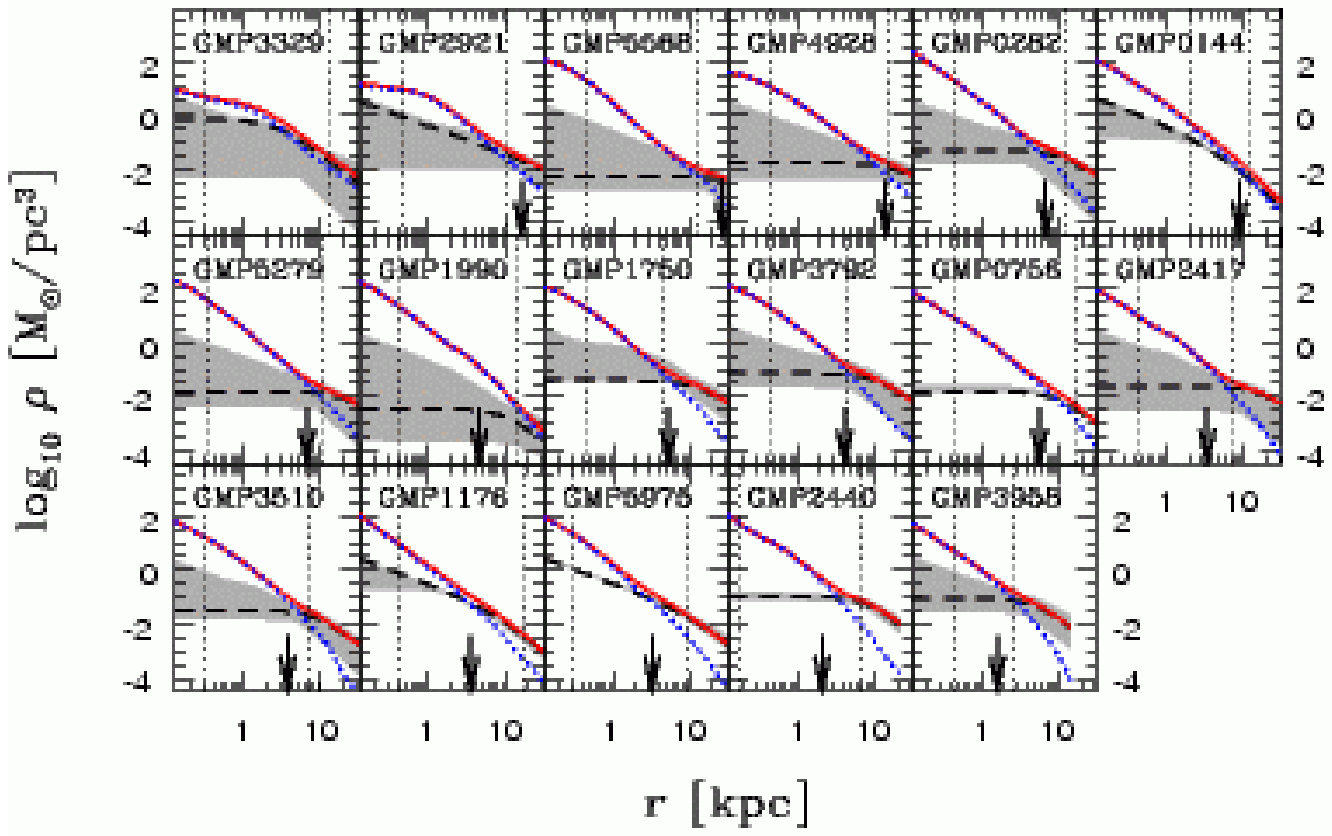}
\includegraphics[width=164mm,angle=0]{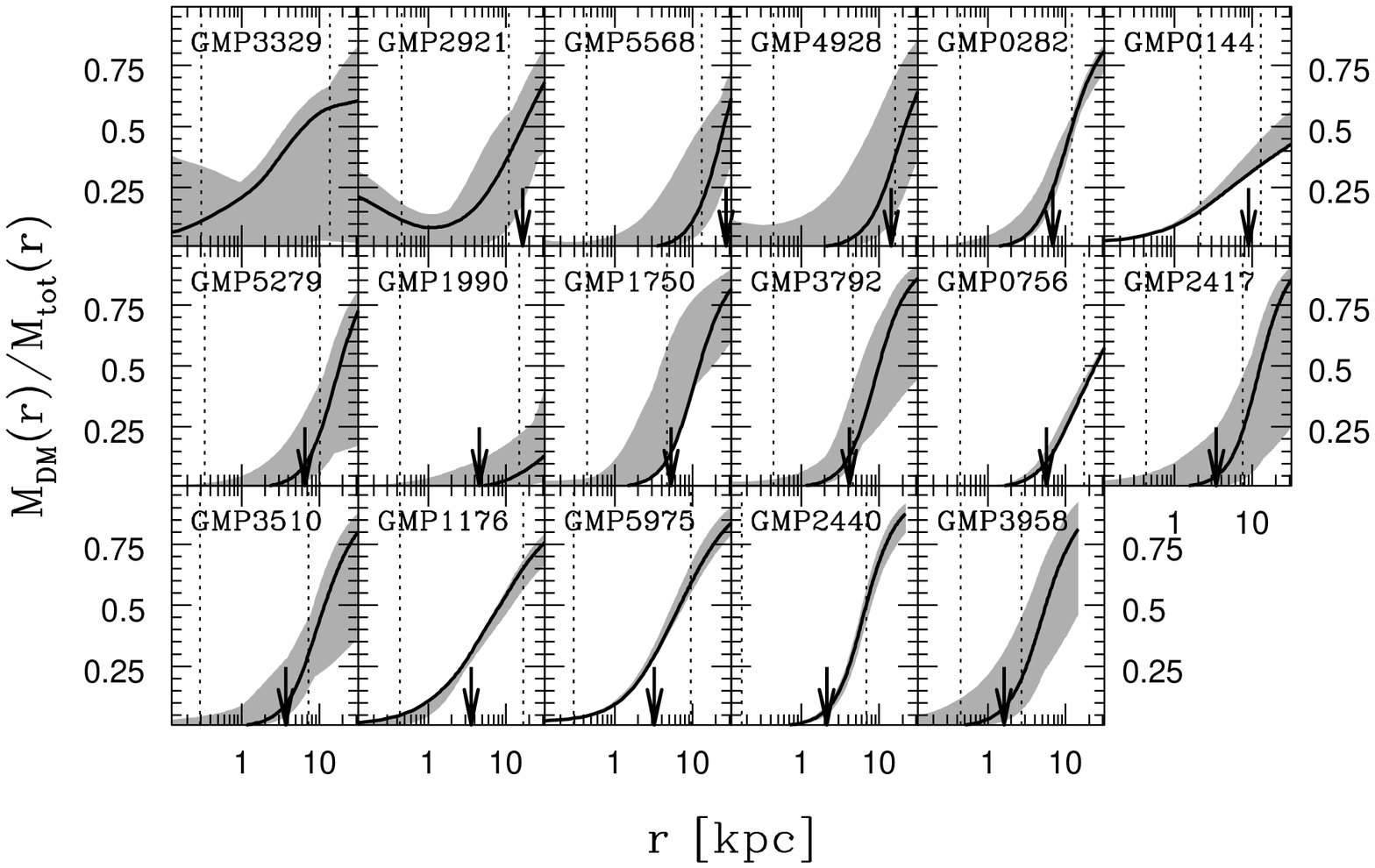}
\caption{Top: spherically averaged mass densities. Red (solid): total mass; blue (dotted):
luminous mass; black (dashed): dark matter with errors (shaded). Bottom: corresponding spherically averaged
dark matter fractions with 68 percent confidence regions. Vertical dotted lines
and arrows as in Fig.~\ref{vcirccomparison}.}
\label{denscomparison}
\end{minipage}
\end{figure*}

\subsection{Does mass follow light?}
\label{subsec:ml}

According to Tab.~\ref{modtable}, the best-fit model of each galaxy contains a dark matter
halo. Column (14) of the table states that eight galaxies have at least a two sigma 
detection of a dark matter halo (GMP0282,
GMP0756, GMP1176, GMP1750, GMP2440, GMP4928, GMP5568 and GMP5975).
The best fitting models with and without dark matter, respectively, are compared
to the kinematic data in App.~\ref{sec:fits}. From this comparison it follows
that models without a halo obviously
fail to reproduce the kinematic data for the above mentioned galaxies. The evidence
for dark matter thereby comes mostly from the innermost and outermost kinematic data points:
without dark matter, the energy of the models is too low, when compared 
to the data at large radii and too high, when compared to the central data (for example
illustrated by the dispersion profile of GMP5975). The reason for the differences at small
radii is likely that part of the missing outer mass in models without a halo 
is compensated for by a larger mass-to-light ratio (cf. Sec.~\ref{subsec:mtols}). 
This, in turn, causes an increase
of the central mass and central velocity dispersion, respectively.
In GMP1750 the dispersion profile without dark matter fits
systematically worse than the one with dark matter at all radii. 
Concerning GMP5568, the dispersion along one of the offset-slits
is particularly large, larger than in all other slits. It is not entirely clear if these
large dispersions are real. If not, then they erroneously increase
the evidence for dark matter. However, because the error-bars of the corresponding data points
are very large, these dispersions are not the dominant driver for the dark halo detection in
GMP5568.

In the rest of the sample the detection of dark matter -- if at all -- 
is more of statistical nature. 
Models with and without dark matter for GMP0144 and GMP2921 differ 
in a similar fashion as those of GMP1750. For GMP3510, GMP3958 and GMP5279 
small differences between models with and without dark matter can be seen at the last kinematic
data points, but the formal significance for dark matter is less than 90 percent. 
We believe that the statistical significance for dark matter in these five cases is
underestimated due to our very conservative error estimates.

In the four remaining objects GMP1990, GMP2417, GMP3329 and GMP3792 the evidence for 
dark matter is generally low. Poor evidence for dark matter in GMP3329 maybe
related to the overall poor constraints that the measured kinematics put on its 
mass-to-light ratio (cf. Sec.~\ref{subsec:conf}). GMP1990 is consistent with the assumption 
that mass follows light.

Our sample thus roughly divides into three categories: (1) galaxies that are clearly 
inconsistent with a constant mass-to-light ratio (8 galaxies out of 17). 
(2) Cases where models with
and without a dark halo differ systematically, but where the formal evidence for a dark 
halo is less than two sigma (5 galaxies). In these cases we expect that our very conservative
error bars lead to an underestimation of the dark matter detection. (3) Systems in which 
the evidence for dark matter is generally weak (4 galaxies).

Models and data of some galaxies with a clear dark halo detection still 
differ systematically in the outer parts (e.g. GMP0756, GMP1176 and
GMP5975). Decreasing the weight on regularisation reduces these differences. 
However, according to the discussion
of Sec.~\ref{subsec:regula} we do not expect that we have significantly over-regularised our
models. Even in case we would have, the derived halos of these systems 
do not depend much on the choice of the regularisation 
parameter, such that conclusions upon the masses of these galaxies are robust
(cf. Sec.~\ref{sec:regula}). It might be possible that differences between models and data
are related to changes in the stellar population, that other our adopted halo profiles are 
not appropriate for these systems, or that the corresponding outer regions
are not in equilibrium or not axisymmetric. We plan further investigations
of these topics for future publications.

\subsection{Circular velocity curves}
\label{rotkurven}
Fig.~\ref{vcirccomparison} shows the best-fit circular velocity curves for the
Coma galaxies. The shapes of the curves vary from cases with two local extrema 
(e.g. GMP5279) to a case of monotonic increase (GMP3958). The sample provides examples 
of rising as well as falling outer circular-velocity curves. Flattened,
rotating galaxies have fairly flat circular velocity curves beyond the central rise 
(less than 10 percent variation up to the last kinematic data point in
GMP0282, GMP0756, GMP2417, GMP3510, GMP1176 and GMP5975).

\subsection{Mass densities and dark matter fractions}
\label{subsec:rho}
Spherically averaged density profiles of all Coma galaxies are surveyed in the top
panel of
Fig.~\ref{denscomparison}. The luminosity distribution of most galaxies shows a power-law core that
smoothly joins with the outer light-distribution. Towards the most luminous galaxies
the central slope of the luminosity distribution flattens out (GMP3329 and GMP2921). 
The inner breaks in the light profiles of GMP1990 ($r \approx 2 \, \kpc$), 
GMP2417 ($r \approx 1 \, \kpc$) and GMP2440 ($r \approx 0.3 \, \kpc$) originate from 
prominent dust features. 

The central regions are dominated by luminous
matter. Halo densities in the centre are at least one order of magnitude lower than
stellar mass densities -- independently of the halo profile being either of the logarithmic or 
of the NFW type. The radius where 
dark and luminous mass densities equalise is inside the kinematically sampled region of 
each galaxy.
In some galaxies the transition from the luminous inner parts to the dark matter dominated
outskirts is very smooth (for example GMP0756, GMP1176, GMP5975). The corresponding dark halo
components are relatively concentrated (NFW halos) and the circular velocity curves 
are fairly flat. In other galaxies the transition is marked by a break in the total mass 
profile and a dip in $\vcirc$ (for example GMP0282). We will come back to these different 
circular velocity curve shapes in Thomas et al. (2007a, in preparation).

Dark matter fractions of the best-fit Coma models are shown in the bottom
panel of Fig.~\ref{denscomparison}. In most galaxies
10 to 50 percent of the mass inside $\reff$ is dark.

\subsection{NFW or logarithmic halos?}
\label{subsec:haloprof}
The evidence for or against logarithmic and NFW halos is summarised in column 
(13) of Tab.~\ref{modtable}.
The majority of best-fit models (13 out of 17) is obtained with
logarithmic halos. However, the significance for one or the
other halo profile providing the better fit is in most cases low.
As already implied by the approximate flatness of the circular velocity curves, the overall
effect of the halo component is to keep the outer logarithmic slope of the total mass density
around $-2$ (i.e. the case of an exactly constant $\vcirc$). This can be achieved 
either with logarithmic halos (asymptotically) or with suitably scaled NFW halos (over
a finite radial range around the scaling radius). Differences in the profiles' inner slopes
seem to play a minor role, perhaps because the inner mass profile turns out to closely follow
the light profile. If elliptical galaxy circular velocity curves are roughly flat over a 
very extended radial range, then NFW fits will break down at some point. With the data at 
hand no clear decision in favour of one of the two profiles can be made.

Most of the best-fit NFW-models are obtained with a flattened halo. Since we cannot
significantly discriminate between NFW and LOG-halos, firm statements about the flattening
of the halos are not possible.

\begin{figure*}\centering
\begin{minipage}{166mm}
\includegraphics[width=164mm,angle=0]{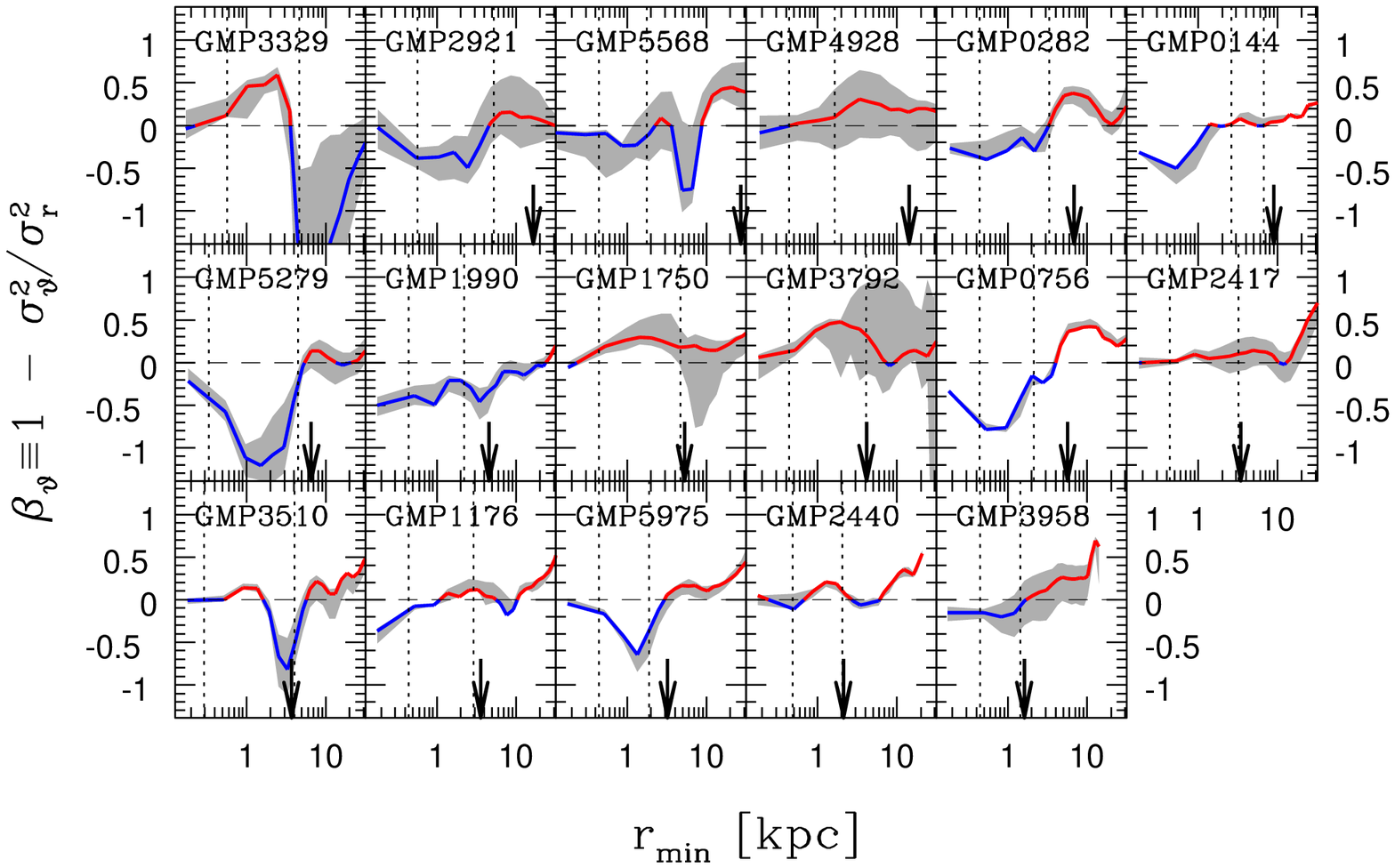}
\includegraphics[width=164mm,angle=0]{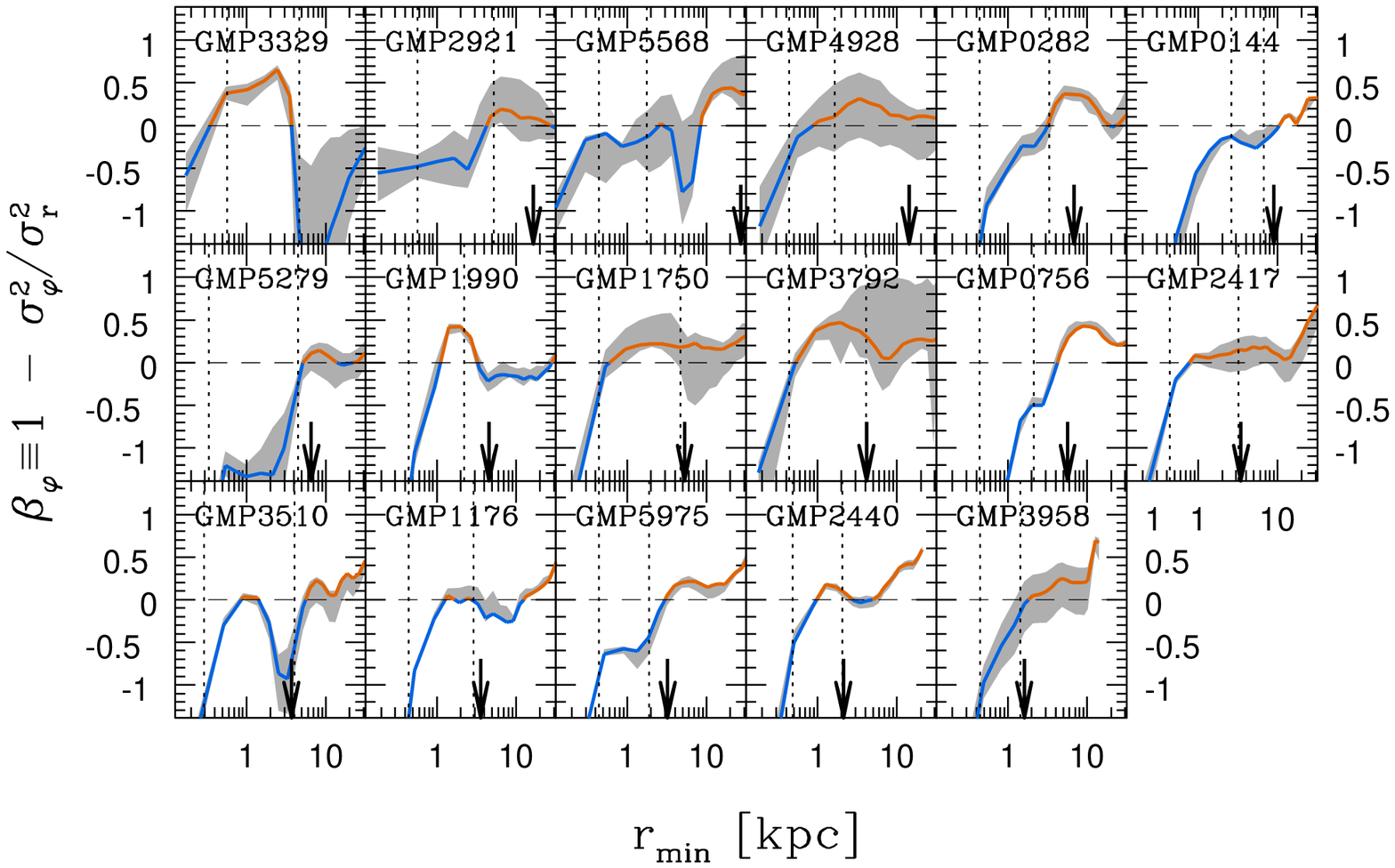}
\caption{Minor-axis anisotropy profiles. Top: meridional anisotropy
$\betatheta$; bottom: azimuthal anisotropy $\betaphi$; solid: best-fit models
(in the colour-version radial anisotropy is highlighted in red, 
tangential anisotropy in blue); shaded: 68 percent
confidence region; dotted: region with kinematic data; arrows: $\reff$.}
\label{anisominor}
\end{minipage}
\end{figure*}

\subsection{The fraction of mass that follows the light}
\label{subsec:mtols}
From Tab.~\ref{modtable} it can be taken that
mass-to-light ratios $\mlsc$ of self-consistent models are on average ($17 \pm 10$) 
percent larger
than those of models with a dark matter halo. Concerning the difference between
logarithmic and NFW halos, the best-fit $\mlnfw$ is generally 
equal or lower than the corresponding
$\mllog$.

\section{Velocity Anisotropy}
\label{sec:aniso}
Having explored the mass structure of the models we next focus on their
dynamics. In this section we will consider the velocity anisotropies defined
in equations (\ref{eq:bt}) and (\ref{eq:bp}), respectively.

The discussion will be restricted to the minor-axis and major-axis bins of the
Schwarzschild models, respectively. Each galaxy is covered by kinematical observations
along at least these axes and the internal orbital structure is
best constrained there.

\subsection{The polar region}
\label{subsec:kinpolar}
Fig.~\ref{anisominor} surveys 
velocity anisotropy profiles along the intrinsic
minor-axis of the Coma galaxy sample. According to axial symmetry 
$\betatheta \equiv \betaphi$ holds directly on the symmetry axis. 
Hence, the upper and lower panel of Fig.~\ref{anisominor} are overall very similar.
The minor-axis bins of the models, however, form a cone with opening angle $\Delta \vartheta=25 \degr$ 
around the $z$-axis. Thus, they include regions off the symmetry axis, 
where the equivalence between azimuthal and 
meridional dispersions does not hold, such that $\betatheta$ and $\betaphi$ 
in Fig.~\ref{anisominor} are not identical.

In Fig.~\ref{anisominor}
the very central anisotropies should not be regarded as reliable. Firstly, because
the central bins are affected from incomplete orbit sampling, resulting in
artificially large azimuthal dispersions \citep{Tho04}. Secondly, 
for numerical reasons the innermost bin is not resolved in $\vartheta$, but
averaged over all $\vartheta \in [0\degr, \, 90\degr]$.

In the spatial region with kinematical data the Coma galaxies offer
different degrees of minor-axis
anisotropy, from strongly tangential (GMP5279) to moderately radial (GMP3792).
Towards the centre $\betatheta \rightarrow 0$, while $\betaphi$ becomes negative
(most likely due to the incomplete orbit sampling). Going outward, many but not all galaxies 
exhibit a gradual change in dynamical structure, often in form of a minimum or maximum 
in $\beta$. Around the last data point most models are isotropic. 
The most radial system is GMP3792.

\begin{figure*}\centering
\begin{minipage}{166mm}
\includegraphics[width=164mm,angle=0]{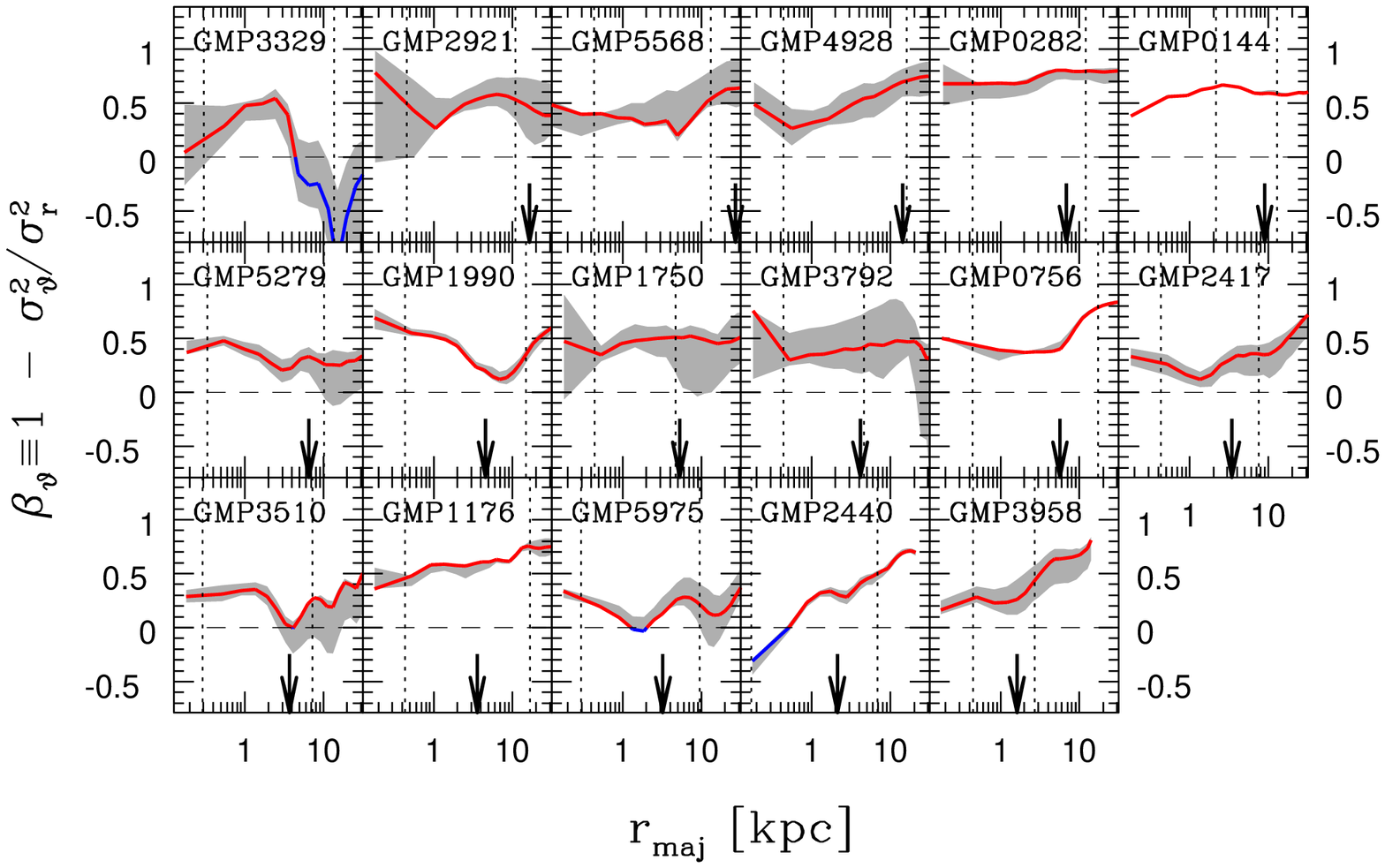}
\includegraphics[width=164mm,angle=0]{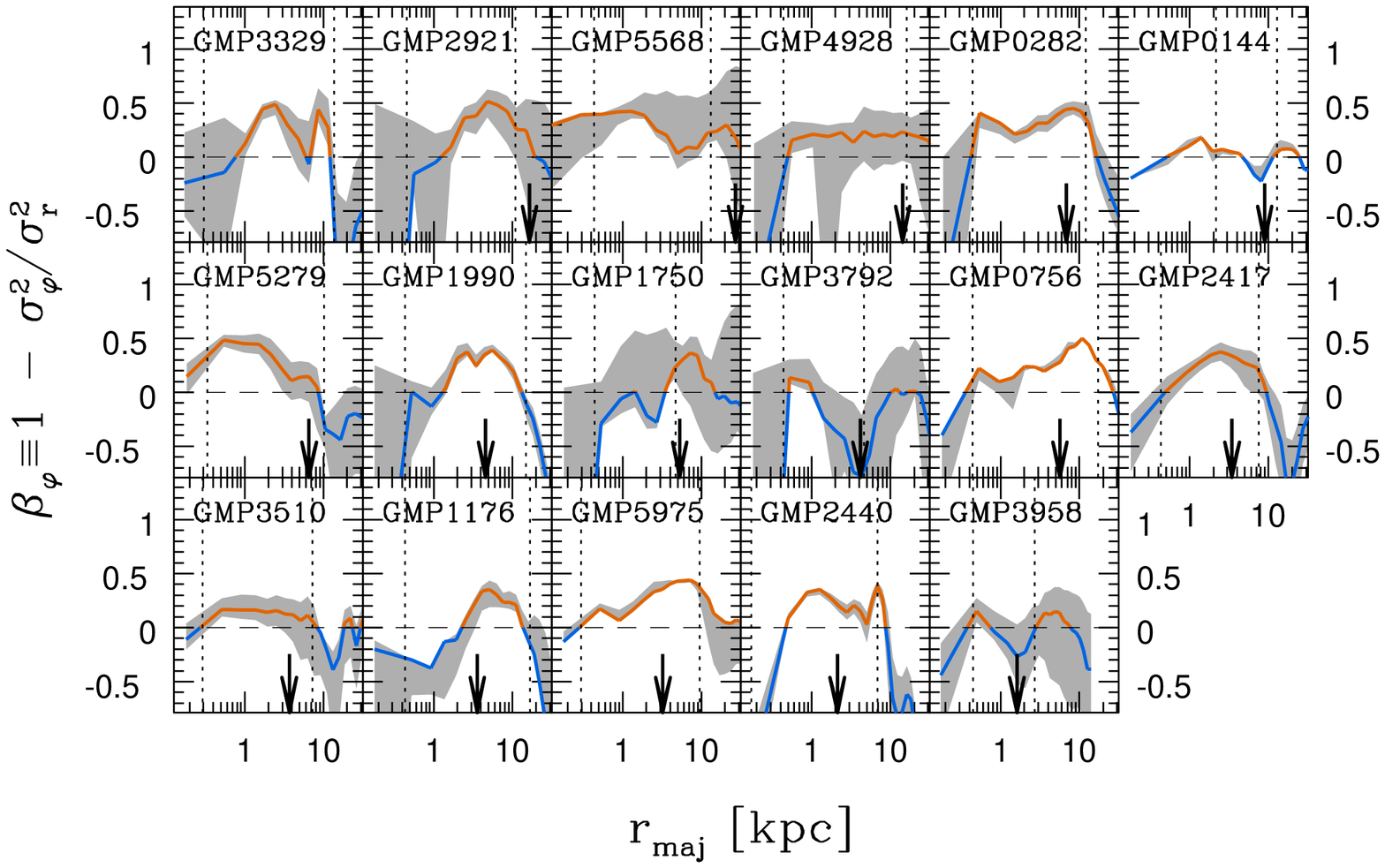}
\caption{As Fig.~\ref{anisominor}, but for the major axis.}
\label{anisomaj}
\end{minipage}
\end{figure*}

\subsection{The equatorial plane}
Velocity anisotropy profiles in the equatorial plane are shown in Fig.~\ref{anisomaj}.
Note that unlike along the minor-axis axial 
symmetry does not imply any relationship 
between $\betaphi$ and $\betatheta$ at low latitudes. 

\subparagraph*{\bf Meridional anisotropy.}
Contrasting the situation around the poles, almost no galaxy exhibits tangential anisotropy 
$\betatheta<0$. Apart from the peculiar object GMP3329 (cf. Sec.~\ref{subsec:conf}) 
all galaxies have $\betatheta>0$ over the kinematically sampled 
radial range. The average $\betatheta$ turns out to be related to the intrinsic flattening
of the galaxies (Thomas et al. 2007a, in preparation). Uncertainties on the intrinsic shape therefore
propagate into uncertainties on $\betatheta$. As it has been stated in Sec.~\ref{subsec:depro},
in many cases it is not possible to distinguish between different inclinations with
high significance. Hence, intrinsic shapes are likewise poorly constrained and the 
uncertainties on $\betatheta$ become large. A typical example 
is GMP3792: the best-fit inclination is $i=60\degr$ and requires a relatively flattened
intrinsic configuration with large $\betatheta$. However, models at higher as well as lower 
$i$ are within the 68 percent confidence region. Consequently, the shaded area includes
also models with different flattening and $\betatheta$.

\begin{figure}\centering
\includegraphics[width=84mm,angle=0]{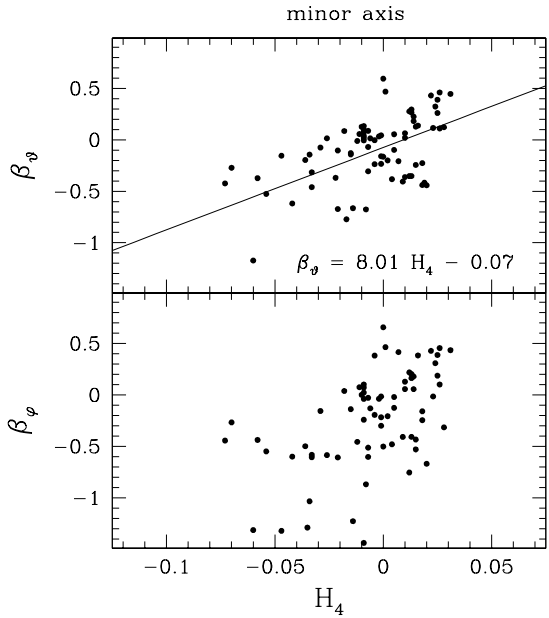}
\caption{Short-axis anisotropy against minor-axis $H_4$. The line in the upper
panel shows a linear fit (quoted in the panel).}
\label{betah4min}
\end{figure}

\subparagraph*{\bf Azimuthal anisotropy.}
More diversity than in $\betatheta$ is offered by azimuthal velocities. In GMP3792,
for example, $\betaphi<0$ suggests that the system
may be composed of two flattened subsystems with low net angular momentum, causing
large $\varphi$-motions. GMP3510, GMP3958 and GMP0144 are relatively isotropic 
($\sigma_r \approx \sigma_\varphi$) over the kinematically sampled spatial region.
GMP5279, instead, offers $\betatheta \approx \betaphi >0$, implying
$\sigma_r>\sigma_\varphi$ and $\sigma_r > \sigma_\vartheta$ over the region with data.

\subsection{Relations between anisotropy and observed kinematics}
The intrinsic short-axis velocity anisotropies are closely related to the observed (local) 
$H_4$. This can be taken from Fig.~\ref{betah4min}, where for each (projected) 
radius $R$ with a
measurement of $H_4$ the local $H_4(R)$ is plotted against the internal anisotropy 
$\beta(r=R)$ at the same radius. 
Internal radii $r$ have not been corrected for inclination since most models 
are edge-on (cf. Sec.~\ref{subsec:inclin}). 
From the figure a tight correlation of $\betatheta$ with $H_4$ follows (quoted
in the plot): the smaller $H_4$, the more tangentially anisotropic the model.
A similar trend occurs between $\betaphi$ and $H_4$ (lower panel). This reflects
that $\betatheta \approx \betaphi$ around the symmetry axis (see above). 

Comparable trends between $\beta$ and $H_4$ have also been found in spherical models
(e.g. \citealt{Ger93,Mag01}). The similarity between spherical models on the one
hand and the polar region of axisymmetric models on the other might be connected to the
fact that in both cases $\sigmaphi = \sigmatheta$. In other words, effectively there is 
only one degree of freedom in the stellar anisotropy ($\betaphi=\betatheta$) and, if the potential
is fixed, there must be a close relationship between $\beta$ and the shape of the
LOSVD (as measured by $H_4$). Experiments with spherical models indicate that the 
dependency of $H_4$ on the potential is weaker than its variation with $\beta$ 
(e.g. \citealt{Ger93,Mag01}). If the
same holds for axisymmetric potentials, then this would explain why along the polar axis
of axisymmetric models $\beta$ depends in about the same way on $H_4$ as in spherical
models. Note, however, that our Schwarzschild models provide many more internal degrees of 
freedom than the smooth spherical models considered by \citet{Ger93} and \citet{Mag01}. 
This becomes apparent when the influence of regularisation on the fit is lowered and the 
scatter around the relation shown in Fig.~\ref{betah4min} increases (cf. Sec.~\ref{kin:reg}).

\begin{figure}\centering
\includegraphics[width=84mm,angle=0]{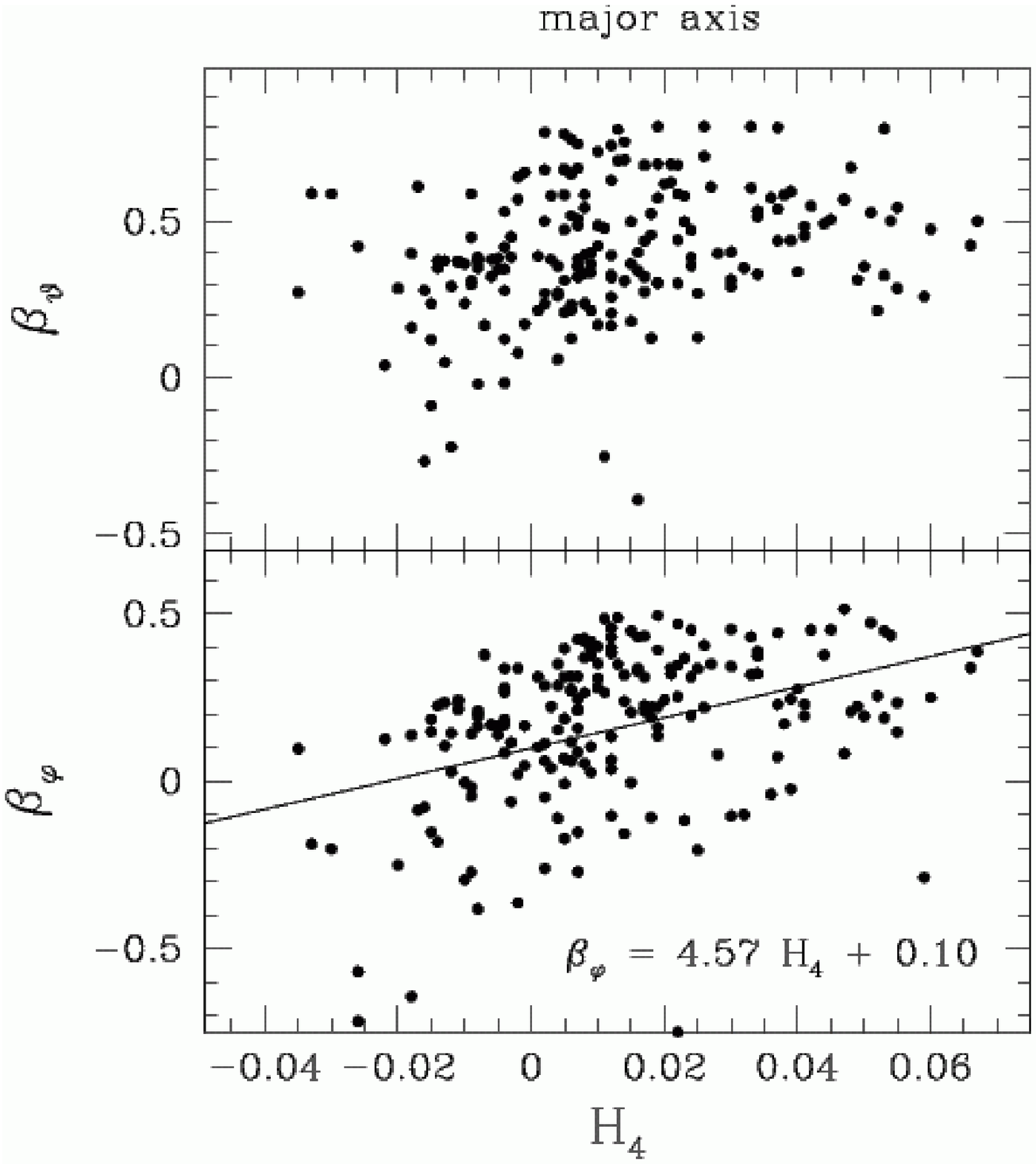}
\caption{As Fig.~\ref{betah4min}, but for the major-axis.}
\label{betah4maj}
\end{figure}

In contrast to the polar region, no tight correlation between $H_4$ and velocity
anisotropy is found around the equatorial plane (cf. Fig.~\ref{betah4maj}). This holds especially for
$\betatheta$, whereas there is a slight trend of $\betaphi$ to increase with
$H_4$. For comparison, a linear fit is shown in the lower panel. A detailed investigation of the orbital structure
will be presented in another paper (Thomas et al. 2007a, in preparation).

\begin{figure*}\centering
\begin{minipage}{166mm}
\includegraphics[width=164mm,angle=0]{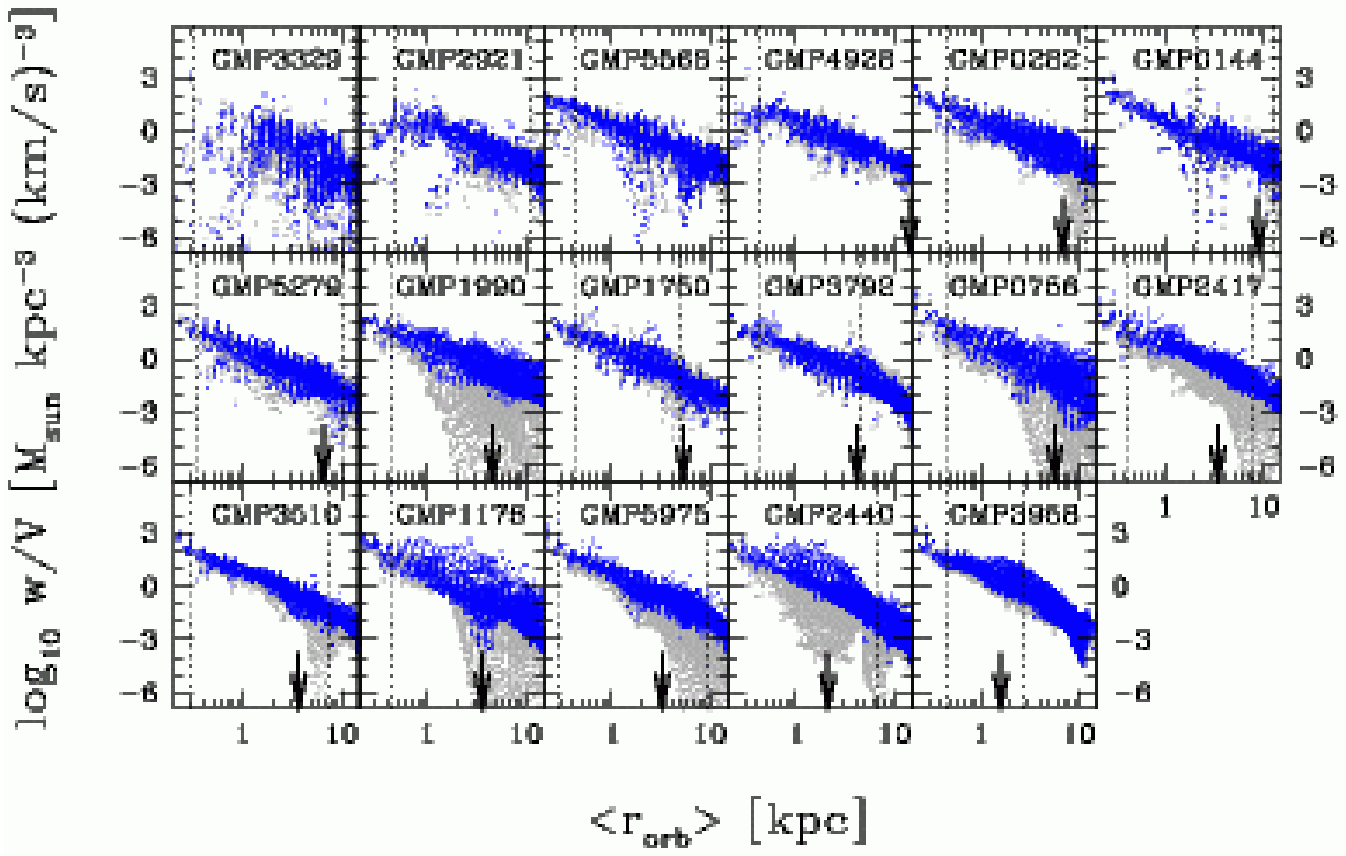}
\includegraphics[width=164mm,angle=0]{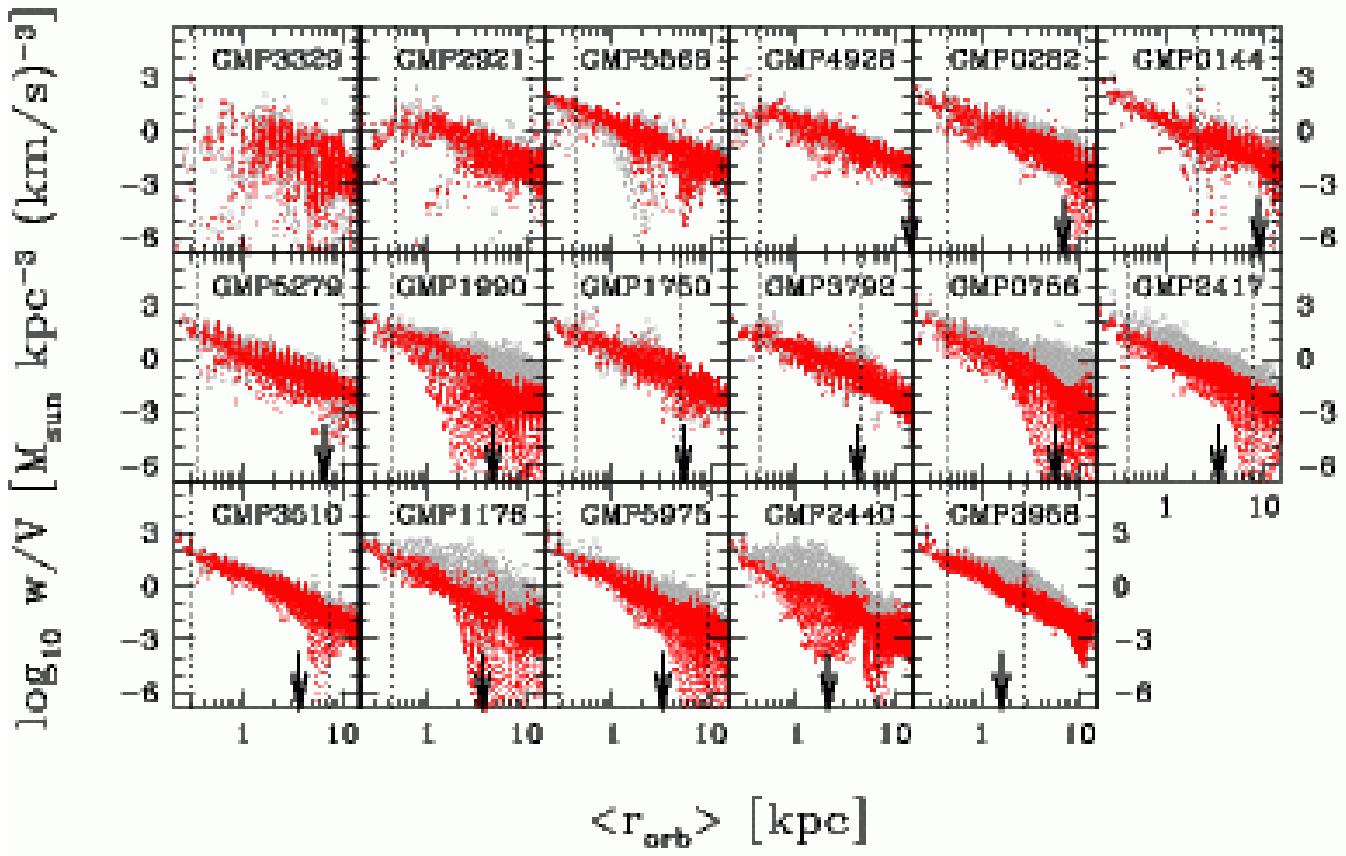}
\caption{Luminous matter phase-space densities. Top: prograde orbits are highlighted (blue);
bottom: retrograde orbits are highlighted (red); vertical dotted lines: boundaries of 
kinematical data; arrows: $\reff$.}
\label{dflum}
\end{minipage}
\end{figure*}

\section{Phase-space distribution function of the stars}
\label{sec:dfstars}
A more fundamental quantity related to a stellar dynamical system than its anisotropy 
is its phase-space
distribution function $f$. It describes the density of stars in phase-space and offers the
most detailed and comprehensive view on its dynamical state. For stationary systems 
the DF is a function of the (isolating) integrals of motion and, thus, constant along
individual orbits (Jeans theorem; e.g. \citealt{Bin87}). 
To be considered in axisymmetric potentials are the energy $E$, angular
momentum $L_z$ along the axis of symmetry and, in most astrophysically relevant potentials,
the so-called third integral $I_3$. To be physically meaningful the DF has to obey the
further condition that it is positive everywhere. In Schwarzschild models the constancy
of the DF along orbits is explicitly taken into account during the orbit integration.
Its positive definiteness is guaranteed as long as the orbital weights $w_i$ are positive
(cf. equation \ref{eq:fi}).
In other words, the very existence of our Schwarzschild models ensures that the 
luminous component of the model is
stationary and physically meaningful (positive density).

A detailed investigation of the full dependency of the
DF on all integrals of motion and its connection to stellar population properties 
will be the subject of another publication (Thomas et al. 2007a,
in preparation). Here we only consider some general properties of the DF. For this
purpose it is convenient to define a mean orbital radius
\begin{equation}
\rorb_i \equiv \sum_k \frac{\Delta t^k_i}{T_i} r^k_i,
\end{equation}
where $T_i$ is the total integration time of orbit
$i$ and $r^k_i$ is its radius at time-step $k$ (lasting $\Delta t^k_i$). 
In rough terms $\rorb$ can be interpreted as a measure of the orbital binding energy.

Fig.~\ref{dflum} surveys the DFs of all 17 Coma galaxies. Each dot represents the phase-space
density of a single orbit. To roughly trace the angular-momentum dependency of the DF
prograde orbits with $L_z>0$ are highlighted in the top panel and retrograde orbits 
($L_z<0$) are highlighted in the bottom one.

The figure shows that with decreasing galaxy mass differences between prograde and 
retrograde orbits in phase-space become more significant. 
This partly reflects an increasing importance of rotation in lower mass galaxies of
our sample. Often, the highest phase-space densities of prograde orbits are
nearly constant over some radial region (e.g. GMP3958 around $\reff$, GMP2440 inside
$r \la \reff$). 
In many, but not all, rotating galaxies
the dominance of prograde orbits comes along with a strong depression of retrograde orbits.
In the outer parts of GMP5975, for example, retrograde orbits have phase-space densities
up to 10 orders of magnitude smaller than prograde orbits. Such low-density
orbits can actually be regarded as being absent in the models \citep{Tho05}. Concerning
the significance of this depopulation it is interesting to note that in case of
GMP5975 it was originally found in models based on major and minor-axis data only
\citep{Tho05}, but remains almost unchanged in our new models including additional kinematical
information along a diagonal axis (cf. Sec.~\ref{sec:obs}). 
The depopulation of retrograde orbits cannot be a general modelling artifact
since it does
not appear in all rotating galaxies. A counter-example is the least-massive object,
GMP3958: it rotates but does not show a strong depression of retrograde orbits in its
outer parts. 

One galaxy of the sample, GMP5568, hosts a counter-rotating central
disk \citep{Meh98}, which shows up by a dominance of retrograde orbits around 
$\rorb \approx 1\,\kpc$ and 
$w/V \approx 10 \, M_\odot/\mathrm{pc}^3/(\mathrm{km}/\mathrm{s})^3$ in Fig.~\ref{dflum}.
In most rotating galaxies the majority of retrograde orbits follows approximately a power-law like
straight line (e.g. GMP3958, GMP2417). 
Retrograde orbits in GMP1990 follow such a power-law like
distribution only outside $\reff$. Near the centre of the galaxy retrograde orbits
exhibit some excess density, compared to a power-law extrapolation of the behaviour 
outside $\reff$. This may reflect a faint inner
counter-rotating sub-component, although unlike in GMP5568, prograde orbits always
dominate in GMP1990 (and the observed sense of rotation is the same at all radii).

In some galaxy models orbital phase-space densities are more spread than in others:
for example the DFs of GMP3329 and GMP5568 look particularly noisy. 
In case of GMP3329 its peculiar kinematics, already discussed in 
Sec.~\ref{subsec:ml}, may be responsible for the distorted phase-space 
distribution of orbits. 
\begin{figure*}\centering
\begin{minipage}{166mm}
\includegraphics[width=164mm,angle=0]{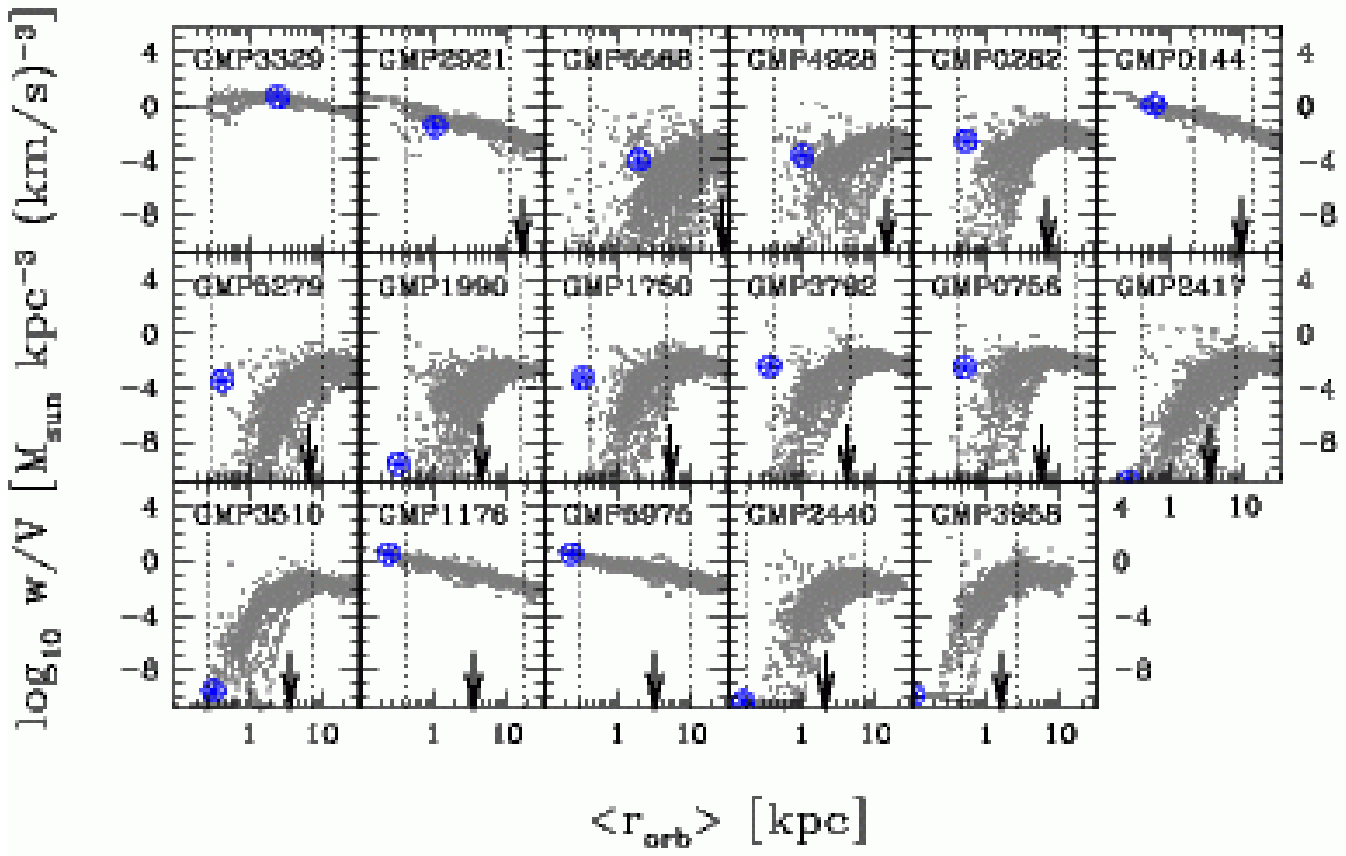}
\caption{As Fig.~\ref{dflum}, but for dark matter. No distinction between prograde and 
retrograde orbits is made (see text for details); large (blue) symbols: average phase-density of
all orbits with $\rorb<0.1 \, \reff$.}
\label{dfdark}
\end{minipage}
\end{figure*}

\section{Phase-space distribution function of dark matter}
\label{sec:dfhalo}

So far we have only considered the phase-space distribution function of the 
luminous component of our models. To ensure that these models are physically meaningful
we also need the dark halos to be supported by an everywhere positive
phase-space distribution function. Without the baryons present, the existence of DFs 
for our halo profiles is known. In case of NFW-halos it follows trivially from the fact that
they arise in $N$-body simulations and DFs for LOG-halos have been constructed 
explicitly by \citet{Eva93}. With a significant contribution of baryons (or any other component)
to the overall gravitational potential, the existence of these DFs is no longer
guaranteed, however. For example, if a cored halo
(central logarithmic density slope $\gamma =0$) is embedded in a cuspy baryonic
component ($\gamma = -1$) and if the core radius exceeds a critical limit
around $r_C \ga 3 \, \reff$, then central phase-space densities become negative
in isotropic or radially anisotropic systems \citep{Cio92,Cio99}. In contrast,
a cuspy halo can always be supported \citep{Cio96}. Thus, the existence of a 
plausible halo DF for our
LOG-halos, which often have core radii near or beyond the critical limit (cf. Tabs.~\ref{dattable}
and \ref{modtable}) is not obvious. The main goal of this section is to investigate whether
we can find a positive definite DF for all our best-fit models, or whether the phase-space
analysis rules out some of our halo profiles.

\subsection{Construction and existence of the halo distribution function}
Our modelling machinery allows 
to construct a DF for dark matter in an analogous way as for luminous matter: by solving 
equation (\ref{maxs}) with an orbit superposition. The only difference to the calculation 
of the luminous matter orbit superposition is that now the dark matter density profile 
is used as the boundary condition and not the deprojected light-profile $\nu$. In addition,
since we lack of any kinematic information about the hypothetical halo constituents we set
$\alpha = 0$ in equation (\ref{maxs}) and maximise the entropy of the orbits. 
For our goal of finding at least one positive
definite DF this does not imply any loss of generality.

Within the numerical resolution of our orbit models we find indeed orbit superpositions with
positive orbital weights $w_i>0$ that allow to reconstruct the halo density profile in each case.
The corresponding DFs are stationary by construction and positive 
everywhere. The fact that we even find positive definite DFs for those LOG-halos that
are beyond the above cited critical core-radius can have several reasons: our models are
slightly different from the ones used in \citet{Cio99} (different radial run in outer parts,
baryonic component flattened in our case). In addition, our orbit superpositions can well
produce
tangential anisotropy, which helps to maintain a positive DF. Finally, our orbit
models have a finite resolution. We cannot exclude that reconstructing the halo density
with higher resolution would force some orbital weights to become negative.

\subsection{Differences between NFW and LOG-halos}
Apart from the mere existence, there are significant differences in the
derived DFs, however. This can be taken from plots of the halo DFs in Fig.~\ref{dfdark}. 
NFW-halo DFs (GMP2921, GMP0144, GMP1176 and GMP5975) are
monotonic with respect to $\rorb$ and regular. The high degree of regularity (compared to
the corresponding luminous matter DFs) reflects the maximisation of
entropy, whereas noise in the stellar kinematics and sub-structuring of 
stars in phase-space tend 
to broaden the stellar DF (cf. Sec.~\ref{sec:dfstars}) .

DFs of LOG-halos exhibit a drop of central phase-space density, as predicted by
\citet{Cio99}. In GMP3329, where the halo is very concentrated, the 
drop is rather gentle. With increasing core-radius the drop becomes more substantial. In 
addition, the noise in the DF increases with increasing core radius. 
Such disturbances in the DF, even though we maximise the entropy, indicate
that a fine-tuning of the orbits is necessary for large core-radii to be
supported by a positive-definite DF. This, and the non-monotonic dependency of orbital
phase-space densities on $\rorb$ 
could imply that the corresponding DFs and, thus, also the spatial density profiles 
are unstable. If this is indeed the case, then the phase-space
analysis would provide a strong argument against large-cored halo profiles, independent from
the kinematic fits. Of course, the halo DFs shown in Fig.~\ref{dfdark} are not unique,
as stated above: the models have no access to the orbit distribution in the halo, apart
from those constraints coming from the shape of the density-profile alone. Details of the 
DFs in Fig.~\ref{dfdark} are therefore physically meaningless. However, that the entropy
maximisation does not yield smooth DFs for LOG-halos with large cores suggests that
-- independent from our ignorance about the details of the orbit distribution -- 
smooth dark matter DFs in the corresponding baryonic potential wells are unlikely.

In any
case, a systematic stability analysis is out of the scope of this
paper. What we can conclude here is, that based on the kinematic fits and based on
the mere existence of a positive definite halo DF, we cannot rule out one or the other
halo profile with our orbit models. Shape and structure of LOG-halo DFs make them
the less likely option, however.

\subsection{Central dark matter density}
According to Sec.~\ref{subsec:rho} the central density of dark matter is
often orders of magnitudes lower than the luminous mass density. This suggests that it
is in many cases weakly constrained. Could it be even lower than in LOG-halos? According
to the above phase-space analysis this seems unlikely, because lower central densities 
would most likely augment the disturbances of the halo DF. Thus, the central spatial dark matter densities
of our LOG-halos are likely lower limits to the true central dark matter densities.

\subsection{Central dark matter phase-space density}
From the dark matter DFs of Fig.~\ref{dfdark} we have also calculated 
a mean central phase-space density
\begin{equation}
\label{favdef}
\fav \equiv \left( \frac{\sum w_i}{\sum V_i} \right)_{0.1},
\end{equation}
where the sums on the right hand side are intended to comprise all orbits
with $\rorb<0.1 \, \reff$. These central phase-space densities are flagged in
Fig.~\ref{dfdark} by the large symbols. Comparison with 
Fig.~\ref{dflum} shows that central dark matter phase-space densities are
particularly low in systems with strong rotation. Exceptions are GMP1176 and GMP5975
with their NFW halos.

\begin{figure}\centering
\includegraphics[width=84mm,angle=0]{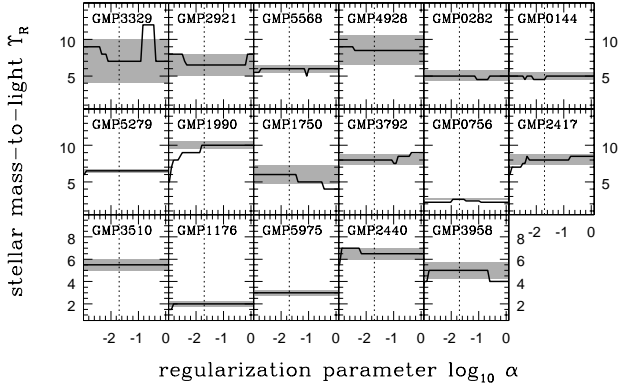}
\caption{Best-fit $\mldyn$ versus regularisation
parameter $\alpha$. Dotted line: $\alpha=0.02$, the regularisation adopted for the
best-fit models; shaded: one sigma confidence region for $\alpha=0.02$.}
\label{reg:star}
\end{figure}

The uncertainty in the dark-halo DF related to our ignorance about dark matter 
kinematics of course affects $\fav$. According to our above discussion the drop in LOG-halo
DFs seems a feature connected to the density profile, though, and we do not expect
that reasonably isotropic or radially anisotropic LOG-halo DFs exist for which $\fav$ increases by 
orders of magnitude. Thus, although $\fav$
is subject to many uncertainties, it is likely good as an order of magnitude estimation
for the central dark matter phase-space density connected with the mass decomposition
made in Sec.~\ref{subsec:mass}.

\section{Regularisation}
\label{sec:regula}
As it has been discussed in Sec.~\ref{subsec:regula} the same regularisation $\alpha = 0.02$ 
is adopted for all Coma galaxies. In the following we will discuss the dependency of 
our modelling results on the choice of $\alpha$.

\begin{figure}\centering
\includegraphics[width=84mm,angle=0]{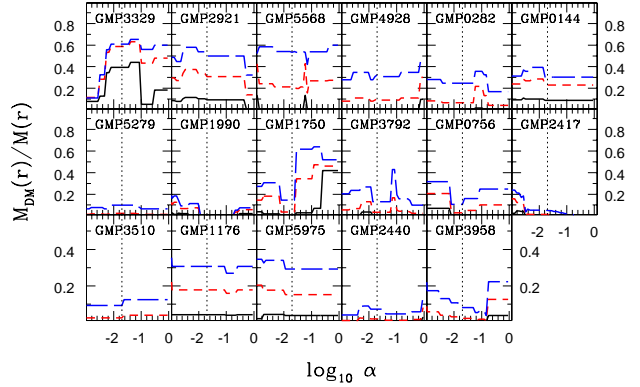}
\caption{Dark matter fractions at $0.1 \, \reff$ (black/solid), $0.5 \, \reff$
(red/short-dashed) and $1.0 \, \reff$ (blue/long-dashed) versus regularisation
parameter $\alpha$. Vertical dotted lines: $\alpha=0.02$.}
\label{reg:dmfrac}
\end{figure}

\begin{figure}\centering
\includegraphics[width=84mm,angle=0]{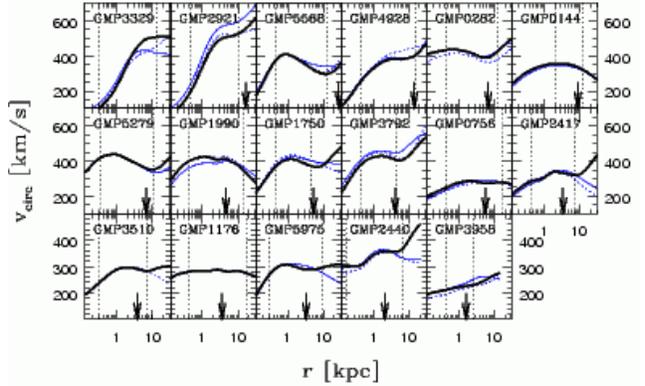}
\caption{Best-fit circular velocity curves for different values of the regularisation parameter: 
$\alpha = 0.001$ (blue, solid), $\alpha = 2.7$ (blue, dotted) 
and $\alpha=0.02$ (black, solid).}
\label{reg:vcirc}
\end{figure}

\begin{figure*}\centering
\begin{minipage}{164mm}
\includegraphics[width=81mm,angle=0]{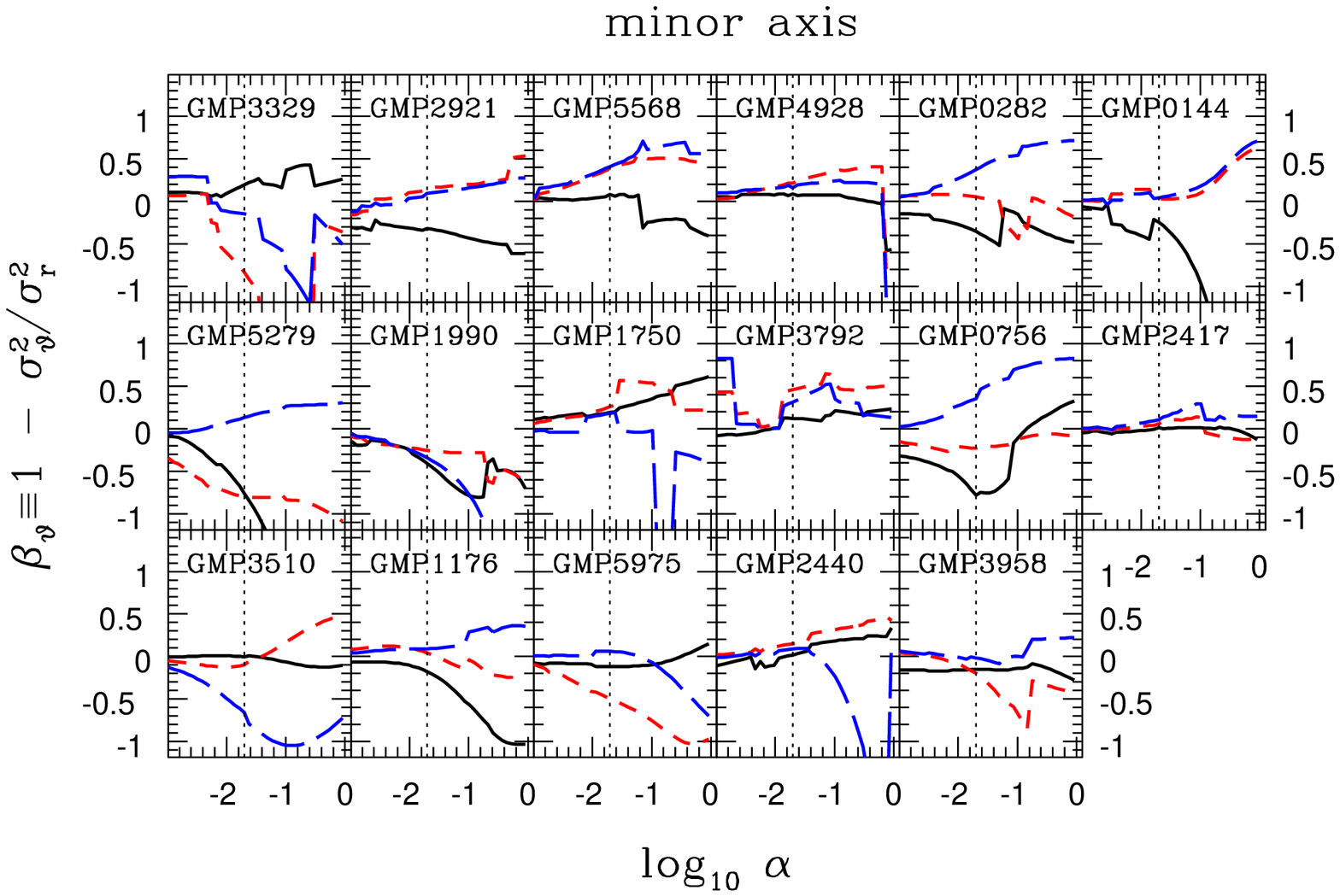}
\includegraphics[width=81mm,angle=0]{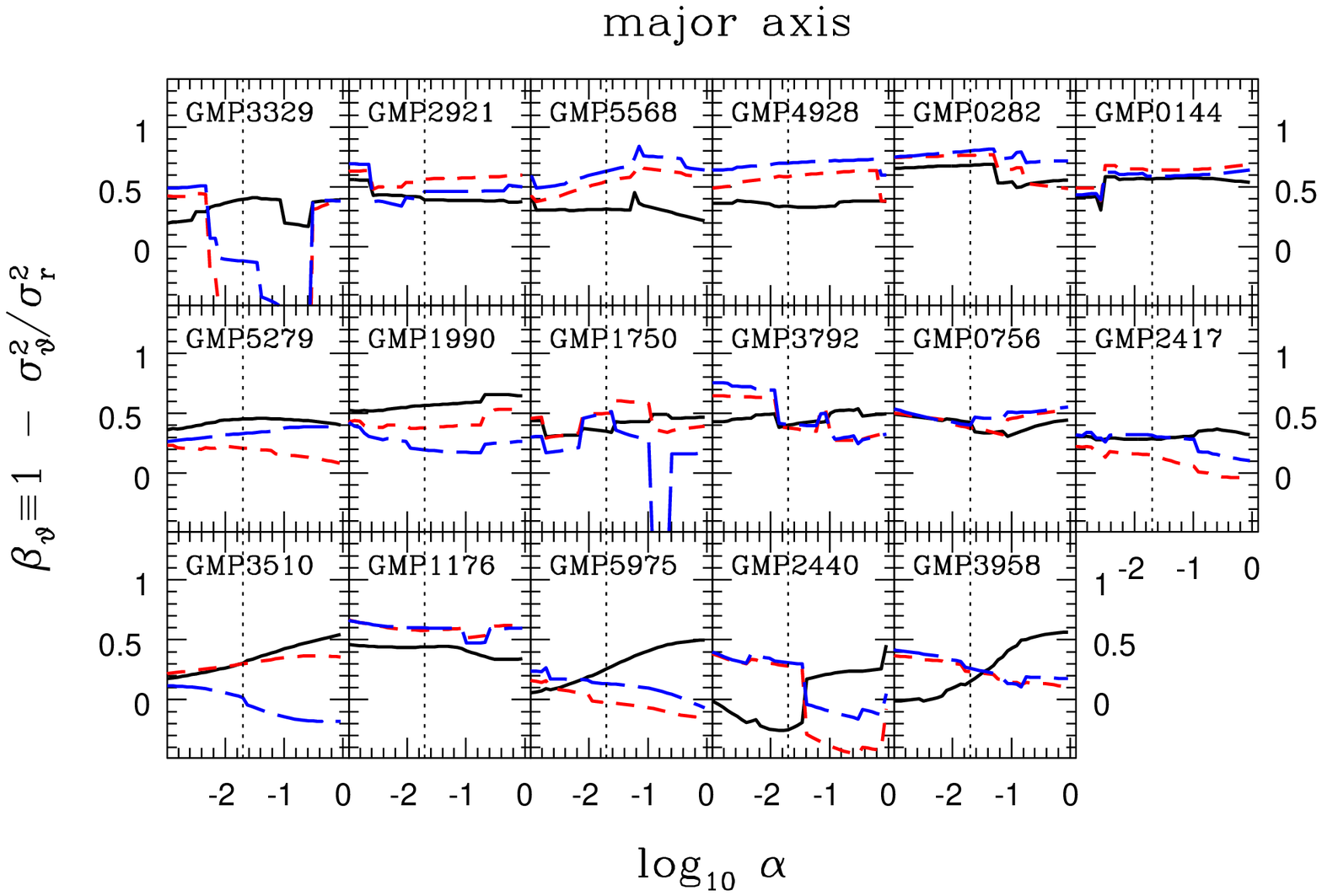}
\includegraphics[width=81mm,angle=0]{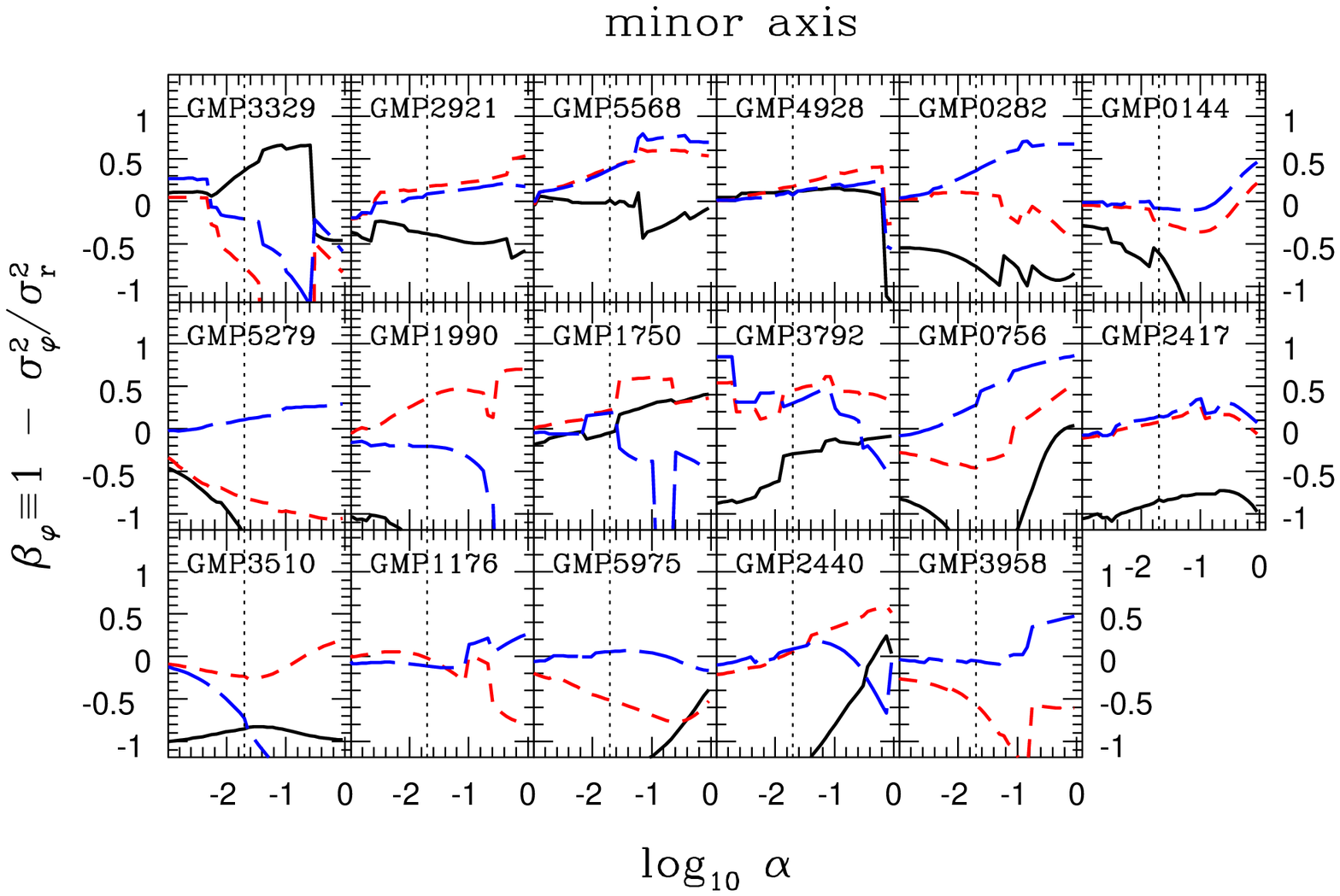}
\includegraphics[width=81mm,angle=0]{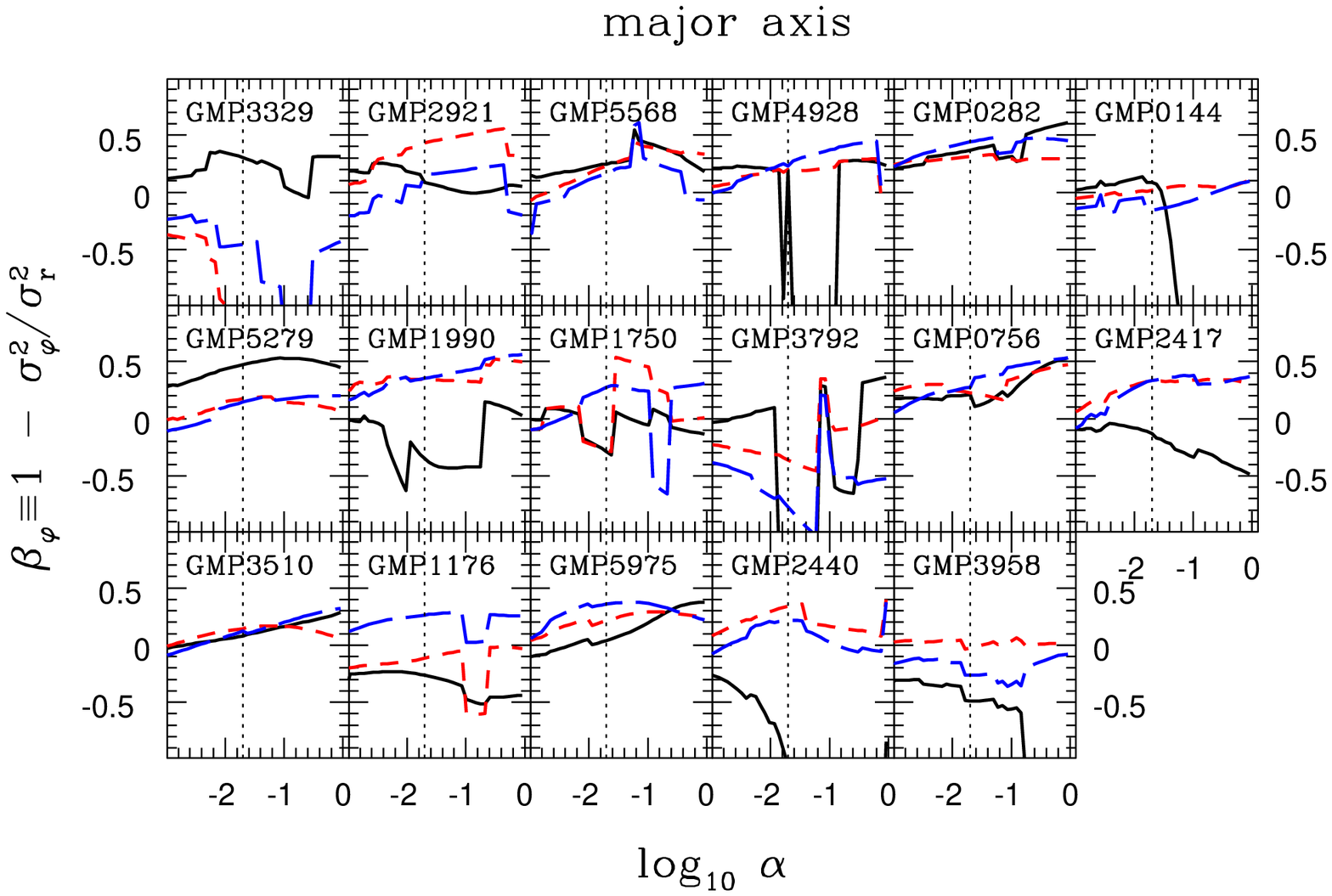}
\caption{Best-fit anisotropy at $0.1 \, \reff$ (black/solid), $0.5 \, \reff$
(red/short-dashed) and $1.0 \, \reff$ (blue/long-dashed); top: meridional anisotropy, bottom: azimuthal anisotropy;
left: minor-axis, right: major-axis; vertical dotted lines: $\alpha=0.02$.}
\label{reg:kin}
\end{minipage}
\end{figure*}

\subsection{The influence of regularisation on model masses}
\label{mass:reg}
Fig.~\ref{reg:star} surveys the best-fit stellar mass-to-light ratios $\mldyn$ over 
the regularisation interval $\alpha \in [10^{-5},3]$. Two conclusions can be drawn
from the figure. First, no systematic trend of $\mldyn$ with $\alpha$ is noticeable. In
GMP1990, for example, $\mldyn$ increases with $\alpha$, while in GMP1750 
it decreases. Second, in most of the
sample galaxies the weight on regularisation has barely any effect on $\mldyn$ (e.g. 
GMP5568, GMP0282, GMP0144, GMP5279, GMP2417, GMP3510, GMP1176, GMP5975, GMP2440).

The best-fit dark matter fractions at three representative radii are shown in
Fig.~\ref{reg:dmfrac} as a function of $\alpha$. As could have been expected, 
the dark matter fraction and $\mldyn$ are correlated: in most cases where $\mldyn$, say,
increases, the dark matter fraction decreases (and vice versa). Since there is no
systematic trend of $\mldyn$ with $\alpha$ it follows that there is also no systematic
trend of the dark matter fraction with $\alpha$. Moreover, the variation of dark matter 
fractions with $\alpha$ is within the quoted error budget of Fig.~\ref{denscomparison}.

Finally, Fig.~\ref{reg:vcirc} shows circular velocity curves for three different
values of $\alpha$. The influence of $\alpha$ on the shape of the circular
velocity curve is weak. Only in a few systems the general shape of the circular 
velocity curve changes with $\alpha$ (for example GMP3510). These changes occur
mostly outside the region covered by kinematic data, however.

\subsection{The influence of regularisation on model kinematics}
\label{kin:reg}
Now to the influence of $\alpha$ on the derived
velocity anisotropies: the left panels of Fig.~\ref{reg:kin}
show best-fit
meridional and azimuthal velocity anisotropies at three representative radii
as a function of $\alpha$. The figures indicate that maximum
entropy fits ($\alpha \rightarrow 0$) yield isotropy along the minor-axis.
Lowering the weight on regularisation generally increases the anisotropy --
the absolute value of $\beta$ -- in the models, as could have been expected. 
There is no specific trend of $\beta$ with $\alpha$: some systems
gain more tangential anisotropy with increasing $\alpha$ (for example GMP5279), while
others become more radial (for example GMP0756). In most cases the dependency of $\beta$ on
$\alpha$ is monotonic and $\beta(\alpha)$ does not change sign. 
In other words, whether or not a galaxy model is
tangentially or radially anisotropic does not depend on $\alpha$. Only the
exact degree of anisotropy changes with $\alpha$. 

Major-axis velocity anisotropies are plotted on the right hand side of 
Fig.~\ref{reg:kin}. In contrast to the minor-axis case there is no trend of 
$\betatheta \to 0$ for $\alpha \to 0$. Variations of intrinsic velocity anisotropies 
with $\alpha$ are slightly weaker along the equator than they are along the minor-axis.
Since the trend of $\beta$ with $\alpha$ is again
monotonic and sign preserving in most cases, the general property of a galaxy to be radially or
tangentially anisotropic is insensitive to the particular choice of $\alpha$.

\begin{figure}\centering
\includegraphics[width=84mm,angle=0]{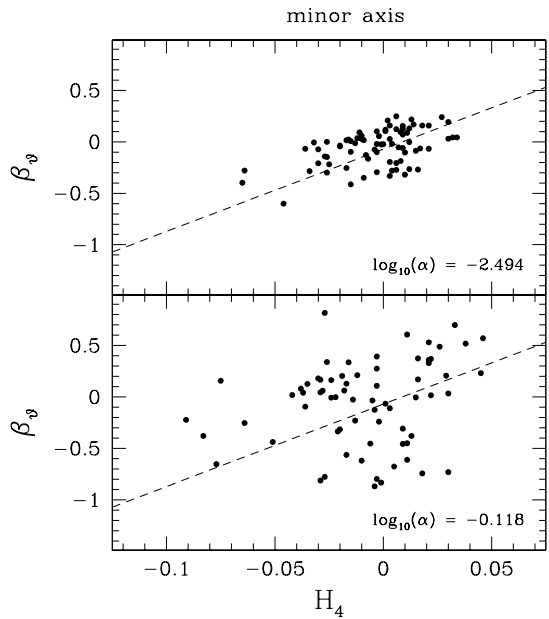}
\caption{As upper panel of Fig.~\ref{betah4min}, but for different values of the regularisation
parameter $\alpha$ (indicated in the panels). Dashed: linear fit for $\alpha=0.02$ (cf. 
Fig.~\ref{betah4min}).}
\label{betah4minNR}
\end{figure}

From the top-left panel of Fig.~\ref{reg:kin} it is clear that the relation between
(the $\alpha$-dependent) anisotropy
and (the $\alpha$-independent) $H_4$ described in Sec.~\ref{sec:aniso} 
must change with different
amounts of regularisation in the models. This is illustrated in Fig.~\ref{betah4minNR},
which repeats the upper panel of Fig.~\ref{betah4min} for two different values
of $\alpha$. The consequences of stronger regularisation are displayed in the top
panel, while weaker regularisation leads to the distribution shown in the bottom
panel.
For comparison the linear fit from Fig.~\ref{betah4min} is shown by the
dashed line. Increasing the weight on regularisation makes the correlation tighter, but
does not alter the slope. With less regularisation the scatter increases, because models 
start to fit the noise in the data. The mean relation is in any case robust
against different choices of $\alpha$. Note, that a correlation between an
{\it intrinsic} property (like $\betatheta$) and an {\it observed} one (like $H_4$) 
cannot be the result of the entropy maximisation. In fact, for $\alpha = 0$ the meridional
anisotropy along the minor-axis vanishes ($\betatheta \approx 0$) and the relation 
breaks down.

Concluding, the dynamical structure of the fits
depends more strongly on the choice of $\alpha$ than the mass distribution does 
(cf. Sec.~\ref{mass:reg}). Thereby no clear trend of velocity anisotropies with 
$\alpha$ is noticeable. The monotonic behaviour of $\beta$ with respect to $\alpha$ 
in most cases ensures that the general property of a model to be radially 
or tangentially anisotropic is independent of the choice of $\alpha$.

\begin{figure}\centering
\includegraphics[width=84mm,angle=0]{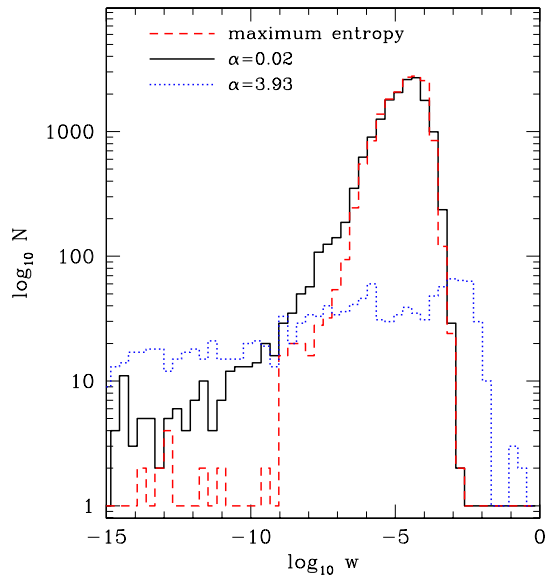}
\caption{Distribution of orbital weights for three differently regularised models 
(as indicated in the plot). All models are calculated in the best-fitting mass distribution
of GMP5975.}
\label{histo}
\end{figure}

To give an example of how $\alpha$ influences the orbit distribution
Fig.~\ref{histo} shows the histogram of orbital weights in the best-fit mass-model of
GMP5975 for three different $\alpha$. As can be seen the model at
large $\alpha$ (weak regularisation) 
is dominated by a few orbits that carry almost the entire 
light. All other orbits are essentially depopulated in the model
(only orbits with weights $\log w>-15$ are included in the plot). 
The model at $\alpha=0.02$ is still relatively close to the maximum
entropy distribution.

\section{Discussion and summary}
\label{sec:sum}
We have surveyed axisymmetric Schwarzschild models for a sample of
17 Coma early-type galaxies. The models are fitted to measurements 
of line-of-sight velocity distributions out to $1-4\,\reff$. Stellar mass-to-light
ratios and dark halo parameters are determined for two parameterised halo families with
different inner and outer density slopes. The models are regularised towards maximum
entropy.

\subsection{Luminous and dark matter}
In each galaxy, models with dark matter fit better than models without.
A constant mass-to-light ratio is significantly ruled out in about half of the sample
(eight galaxies where dark matter is detected on at least the 95 percent confidence level). 
In four galaxies the case for dark matter
is weak. The mass distribution in one of these systems (GMP1990) is in fact consistent to
follow the light. Five galaxies are intermediate cases where the formal evidence for dark
matter is low, although fits with and without dark matter differ systematically in either
their radial dispersion profiles or in their outermost LOSVDs. We believe that
the low signal for dark matter in these systems is partly due to our very conservative 
treatment of the error bars.

Our inferences about dark matter are based on the mass decomposition provided 
by equation (\ref{rhorho}). In particular, we have assumed that the stellar mass-to-light
ratio is constant throughout the galaxy. In this context, GMP1990 is not surrounded
by significant amounts of dark matter. However, before finally concluding upon dark
matter in our galaxies a detailed comparison to independent estimates of 
the stellar mass is required (Thomas et al. 2007a, in preparation). For example, GMP1990
has a large mass-to-light ratio $\Upsilon = 10.0$ ($R_C$-band). If the actual stellar mass
can only account for a fraction of it, then our result does not argue against dark
matter in this galaxy, but merely implies that dark matter follows closely the light 
of the system.

Constant mass-to-light ratios have been reported to be consistent with planetary nebulae 
kinematics in the outskirts of three roundish objects (spherical modelling; \citealt{R03}). 
Also some of the round and non-rotating ellipticals of \citet{Kr00} are consistent with
the mass distribution following the light distribution. Many of these latter systems lack of
kinematic data beyond $\reff$, however. In addition to the related uncertainties
for the outer dark halo, spherical modelling of round galaxies generally
suffers from the ambiguity related to the flattening along the line-of-sight. In this
sense, GMP1990 is an interesting case, because its apparent flattening 
implies a viewing-angle close to $i=90\degr$ and the kinematic data extend relatively
far out ($3 \, \reff$). 

Best-fit dark matter halos are in 4 out of 17 cases of the NFW-type and in all other
cases logarithmic. Differences in the goodness-of-fit based on one or the other
halo family are marginal in most cases. 
Central dark matter densities are at least one to two orders
of magnitude lower than the corresponding mass densities in stars. Between
10 and 50 percent of the mass inside the half-light radius $\reff$ is
formally dark in most Coma galaxies. These dark matter fractions 
are in general agreement with earlier results of dark matter modelling in round, non-rotating
ellipticals \citep{G01} as well as with the analysis of cold gas kinematics
\citep{Ber93,Oos02}, hot halo gas \citep{Loe99,Fuk06,Hum06}
and strong lensing studies \citep{Kee01,Tre04}. \citet{Cap05} concluded for similar
dark matter fractions in the SAURON-ellipticals (although in their models it is assumed
that mass follows light).

The combination of luminous and
dark matter results in circular velocity curves of various shapes: some galaxies have outer 
decreasing $\vcirc$ while others
show an indication for a dip in $\vcirc$ around $\approx 10 \, \kpc$ and a subsequent 
increase of $\vcirc$ towards larger radii. 
We cannot easily quantify the significance of
this dip. Its appearance close to the outermost kinematic radius is suspicious to
reflect a modelling artifact. However, in contrast to the orbital
structure, whose reconstruction becomes uncertain around (and beyond) the last 
kinematic data point \citep{Kra05,Tho05}, the mass reconstruction in these regions
is more robust \citep{Tho05}. In addition, the dip does not appear in all galaxies,
suggesting that it is related to some observable property of the corresponding objects.
In any case, it is interesting to note that
similar dips are also indicated in temperature profiles of elliptical galaxy X-ray halos 
\citep{Fuk06} and can also been seen in spherical models of some round galaxies 
\citep{G01}. More extended kinematic data sets may in the future allow to better constrain 
the outer shape of the circular velocity curve.

In rotating systems
the circular velocity is fairly constant over the observationally sampled radial region
(10 percent fractional variation). Similarly flat circular velocity curves have also
been inferred from stellar kinematics of round systems 
(spherical modelling; \citealt{G01,Mag01}) and from strong gravitational 
lensing \citep{Kop06}.

To the resolution of our orbit models all dark halos are supported by at least
one positive definite phase-space distribution function. In case of NFW-halos smooth
DFs can be constructed, but for LOG-halos with large core radii even the maximisation of
orbital entropy does not yield smooth DFs. It is not obvious whether the corresponding
spatial density profiles are stable or not. Further modelling is required to investigate
whether phase-space arguments can be used to rule out logarithmic halos in elliptical
galaxies.

\subsection{Kinematics}
With decreasing total mass the influence of rotation on the stellar 
phase-space distribution increases. In many galaxies rotation arises by an overpopulation
of prograde orbits and a simultaneous underpopulation of retrograde orbits. At least one system
lacks of the depopulation of retrograde orbits, proving that it is not a general
artifact of our modelling approach.
 
Some galaxies show strong tangential anisotropy along the minor-axis. This derives from
low minor-axis $H_4$-measurements, because observed $H_4$ and modelled orbital anisotropy
along the minor-axis turn out to be correlated. Slope and zero-point of 
this correlation are largely independent of regularisation, but -- to some degree -- 
stronger regularisation tightens the relation. Such a relation between an intrinsic 
quantity on the one hand and an observed one on the other cannot originate from the
entropy maximisation alone.

Along the major-axis, a slight tendency of increasing $\betaphi$ with
increasing $H_4$ is noticeable, but with much larger scatter than along the minor-axis.
Only one system shows indication of tangential anisotropy ($\betatheta<0$), 
all other galaxies are isotropic or mildly radially anisotropic ($\betatheta > 0$,
$\betaphi \ga 0$). Radial anisotropy also appears characteristic for spherical models of round
galaxies \citep{Kr00,G01}. A suppression of vertical energy (corresponding to
$\betatheta > 0$) has recently been reported for SAURON ellipticals \citep{Cap07}. We
plan a detailed investigation of the orbital structure of the Coma galaxies for the
future.

\subsection{Regularisation}
Stellar mass-to-light ratios, dark matter fractions and the shape of circular velocity curves
turn out to be robust against different choices of the regularisation parameter $\alpha$.
The strongest effect $\alpha$ has is on the reconstructed anisotropies: their absolute
values tend to
increase if the weight on regularisation constraints is lowered. At a fixed radius the
dependency of anisotropy on $\alpha$ is mostly monotonic, such that the general
quality of a galaxy to be radially or tangentially anisotropic, respectively, is independent
of $\alpha$. What changes instead is the actual amount of anisotropy.

\subsection{Outlook}
Detailed investigations of luminous and dark matter scaling relations, of stellar population
properties and their connection to the phase-space distribution of orbits are in
preparation.

\section*{Acknowledgements}
We thank Eric Emsellem for his constructive referee report that helped 
to improve the presentation. JT acknowledges financial support by the 
Sonderforschungsbereich 375 ``Astro-Teilchenphysik'' of the Deutsche Forschungsgemeinschaft.
EMC receives support from the grant PRIN2005/32 by Istituto Nazionale di Astrofisica 
(INAF) and from the grant CPDA068415/06 by the Padua University.
Support for Program number HST-GO-10884.0-A was provided by NASA through a grant
from the Space Telescope Science Institute which is operated by the Association
of Universities for Research in Astronomy, Incorporated, under NASA contract
NAS5-26555.

\appendix

\section{Data fits}
\label{sec:fits}
\begin{figure}\centering
\includegraphics[width=90mm,angle=0]{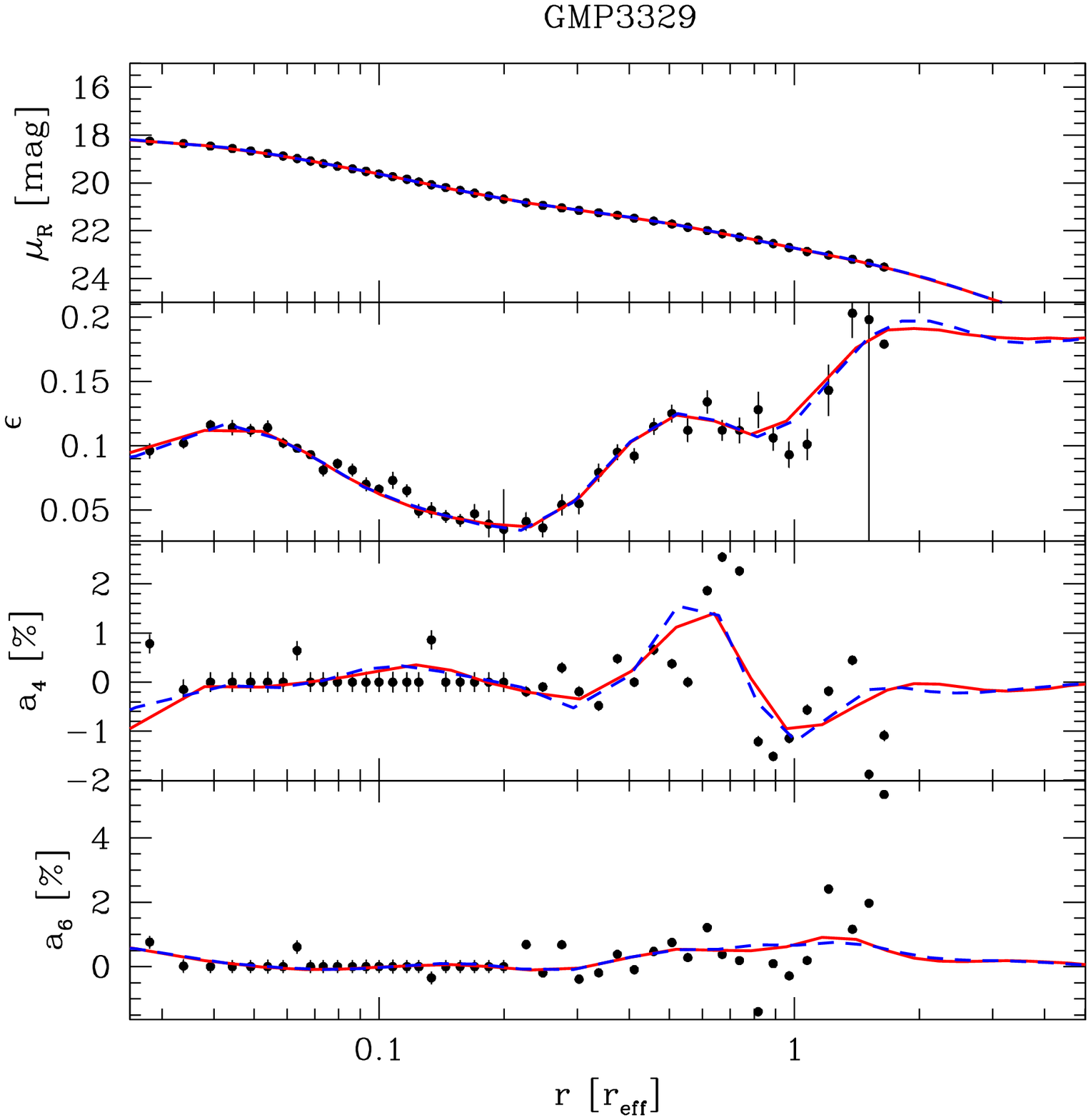}
\includegraphics[width=90mm,angle=0]{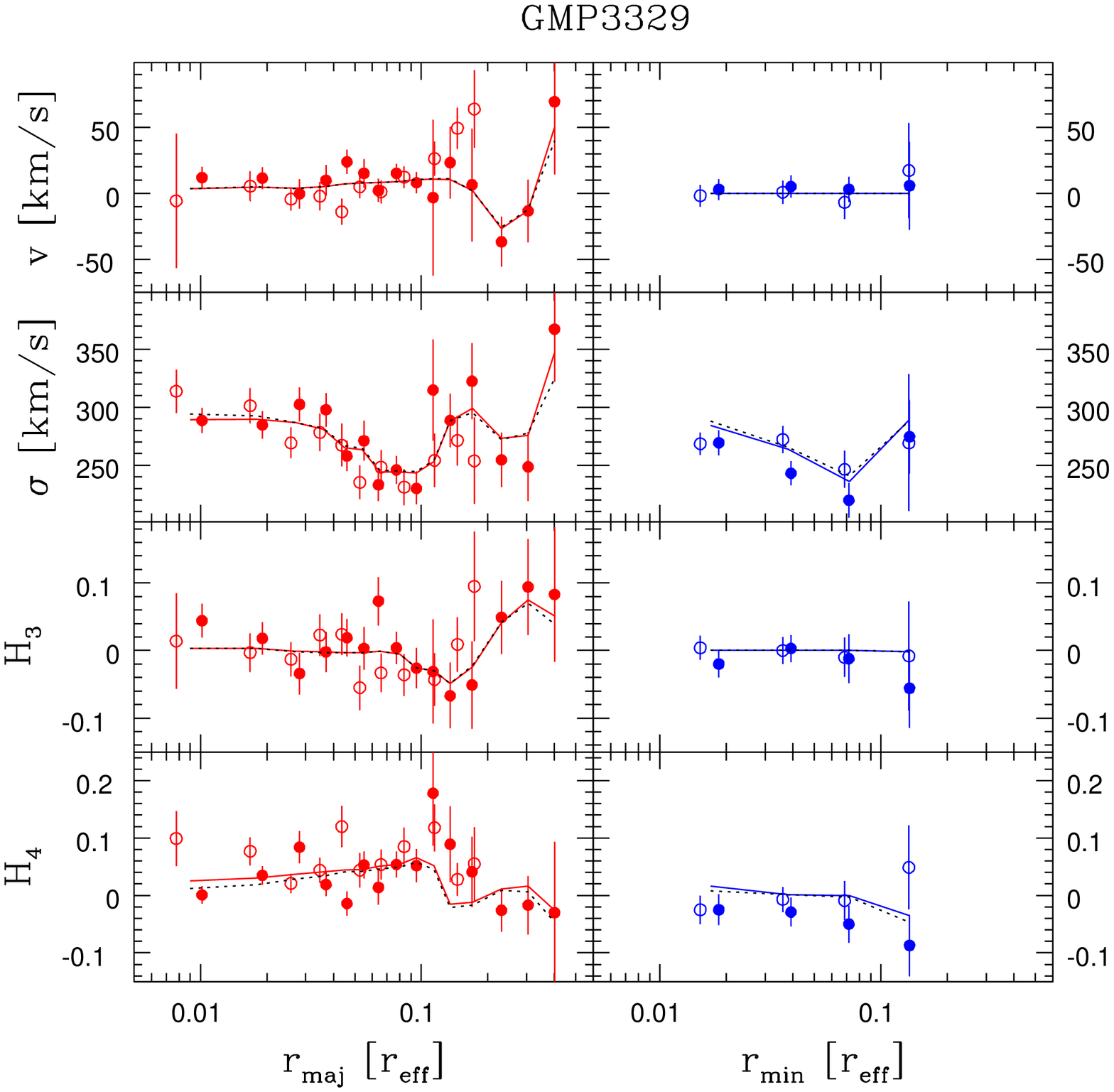}
\caption{Upper panel: Joint ground-based and HST photometry of GMP3329/NGC 4874. Lines:
best-fit deprojection (red) and its edge-on reprojection (blue). Lower panel: 
stellar kinematics along major (left/red) and minor axis (right/blue); 
filled and open circles refer to the two sides of the galaxy; dotted: best-fit model
without dark matter.}
\label{isoplotgh3329}
\end{figure}

\begin{figure}\centering
\includegraphics[width=90mm,angle=0]{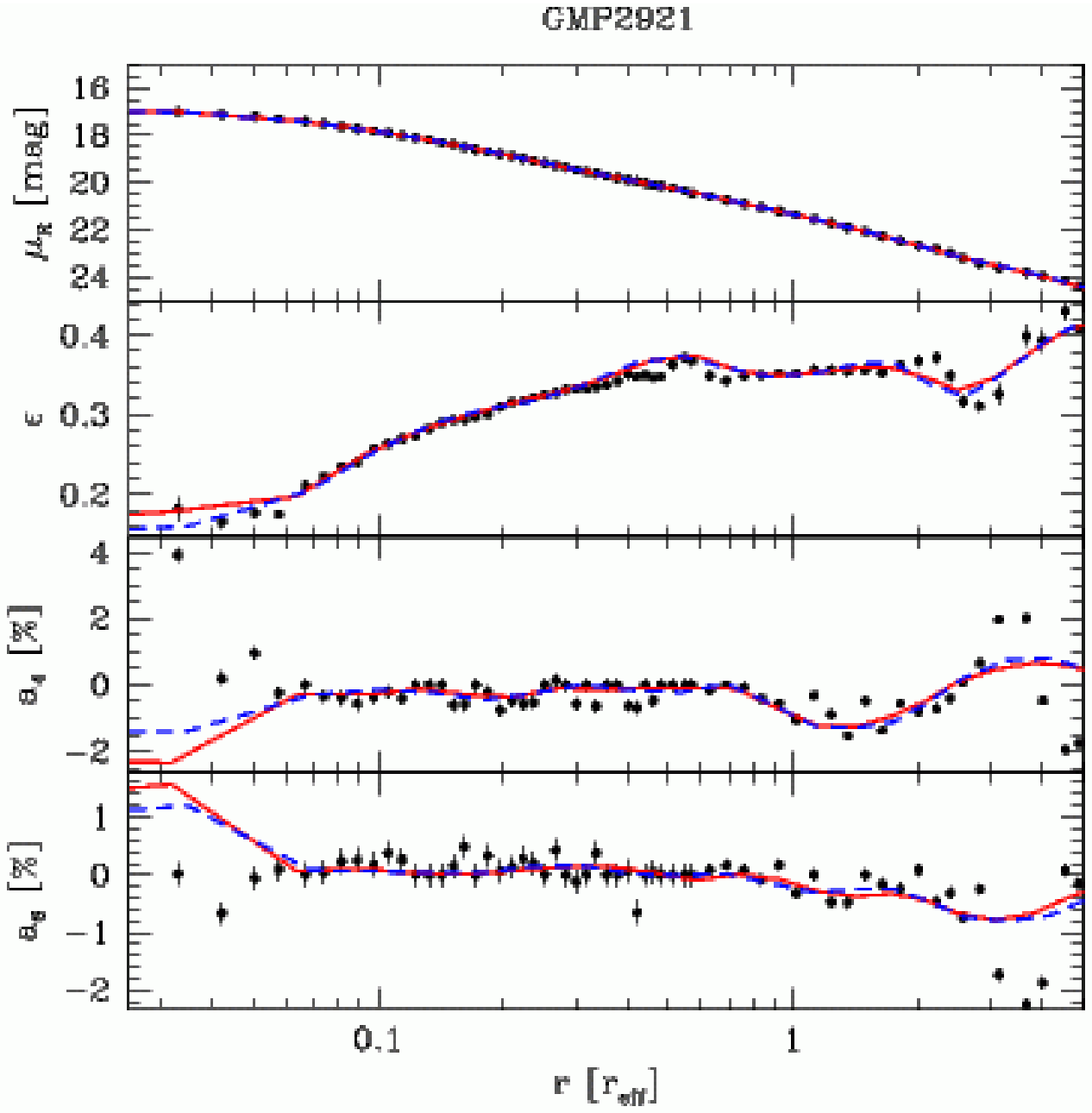}
\includegraphics[width=90mm,angle=0]{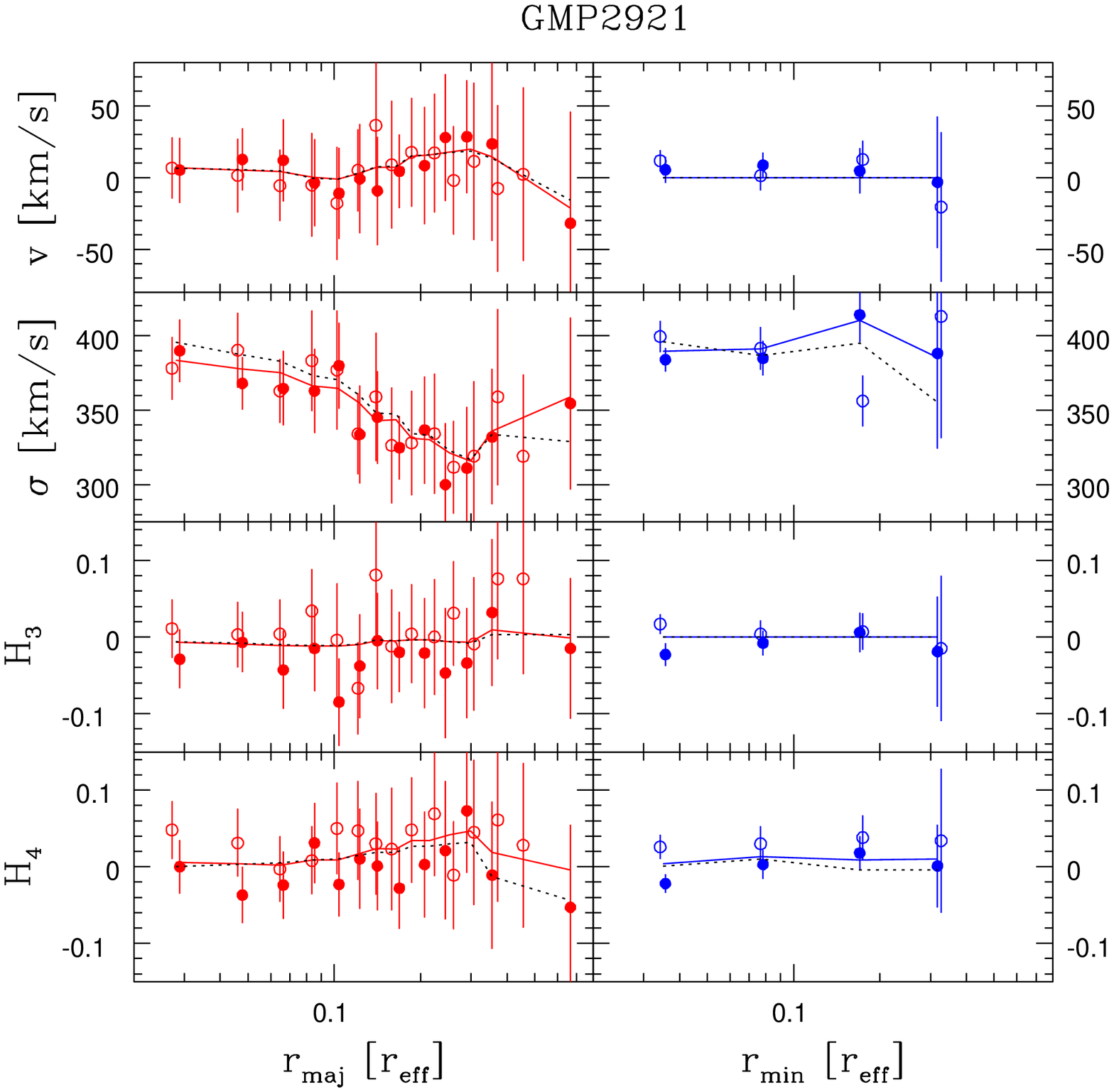}
\caption{As Fig.~\ref{isoplotgh3329}, but for GMP2921/NGC 4889.}
\label{isoplotgh2921}
\end{figure}

\begin{figure}\centering
\includegraphics[width=90mm,angle=0]{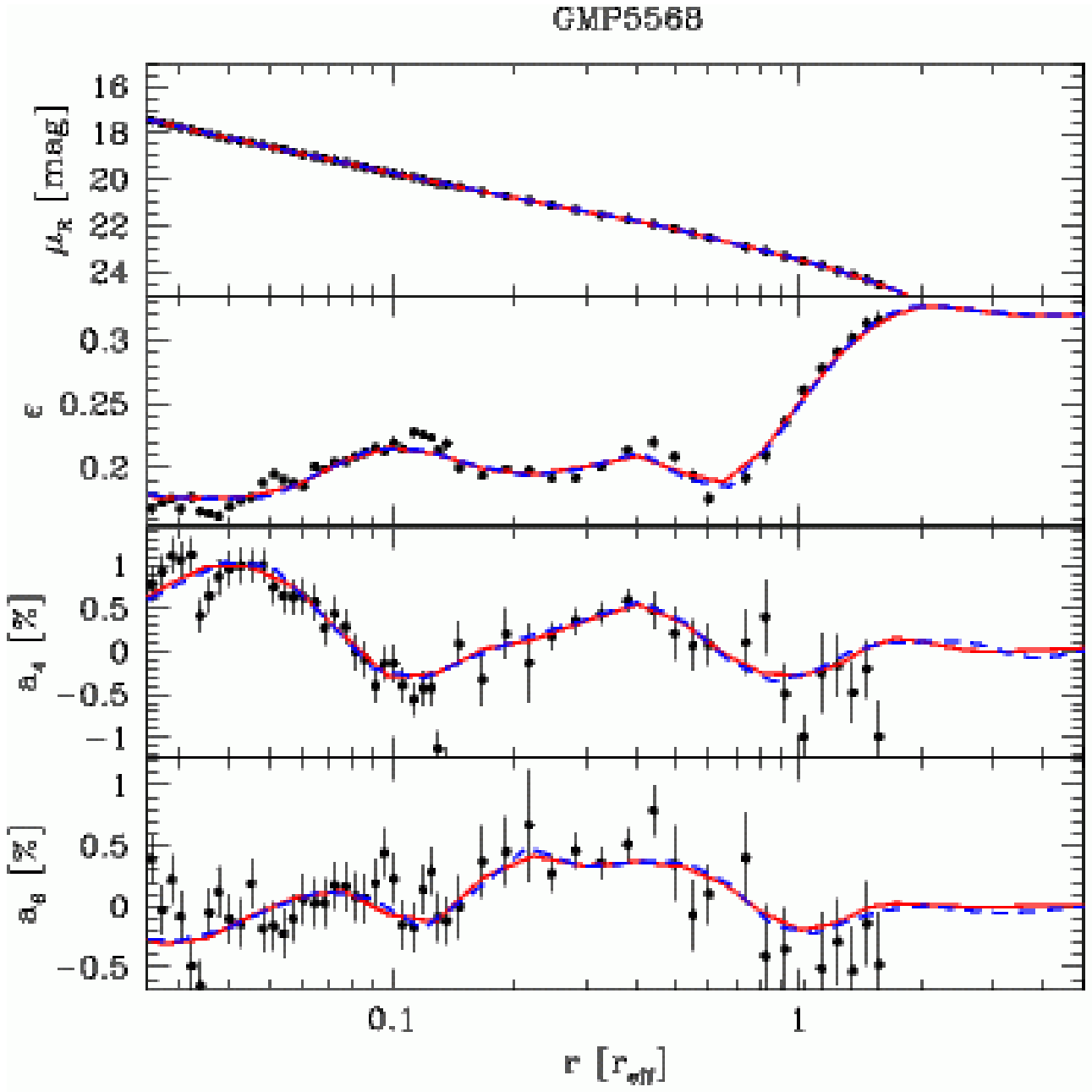}
\includegraphics[width=90mm,angle=0]{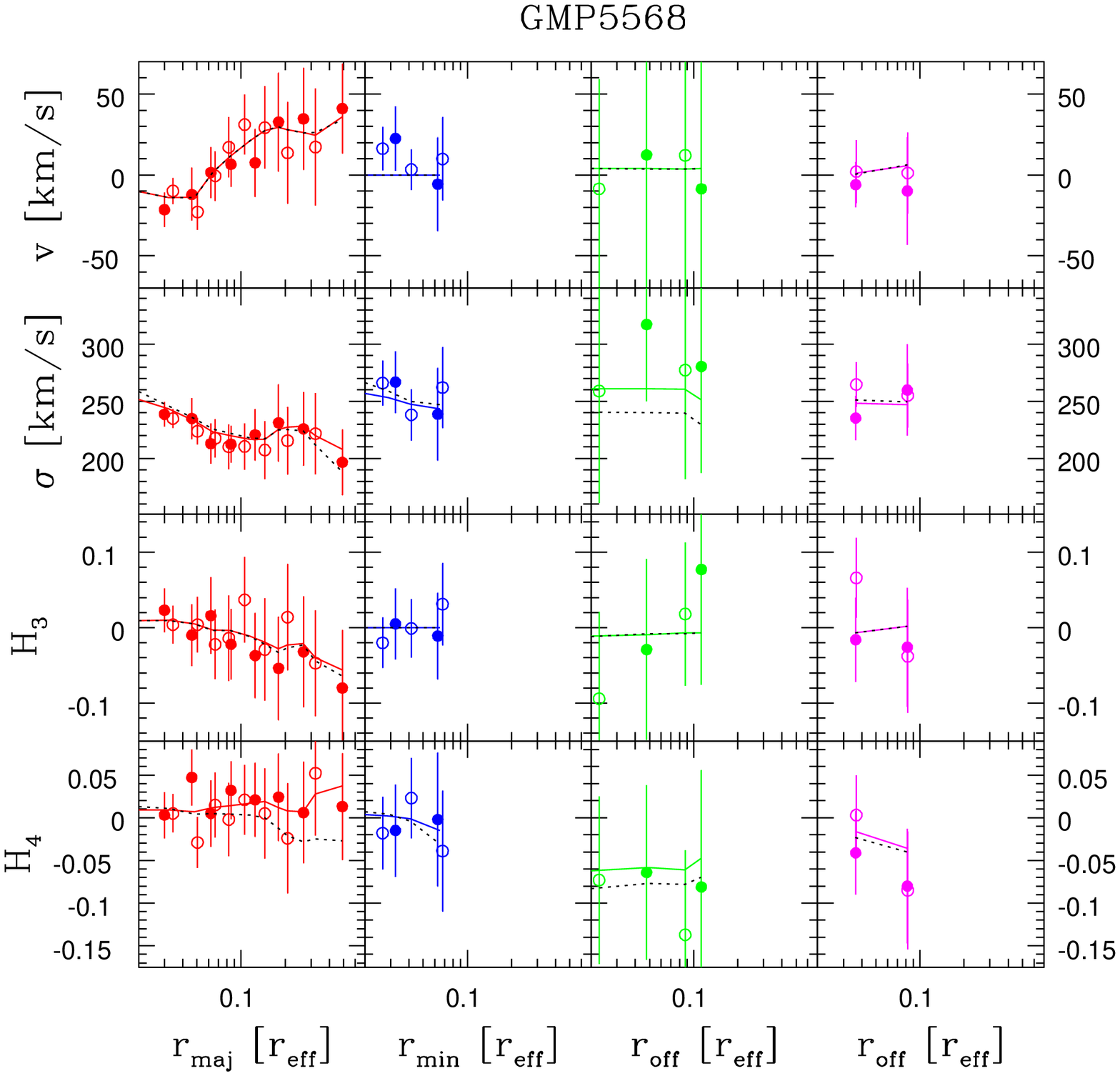}
\caption{As Fig.~\ref{isoplotgh3329}, but for GMP5568/NGC 4816. Green/third column:
offset to major-axis $\reff/4$; magenta/fourth column: offset to major-axis
$\reff/20$.}
\label{isoplotgh5568}
\end{figure}

\begin{figure}\centering
\includegraphics[width=90mm,angle=0]{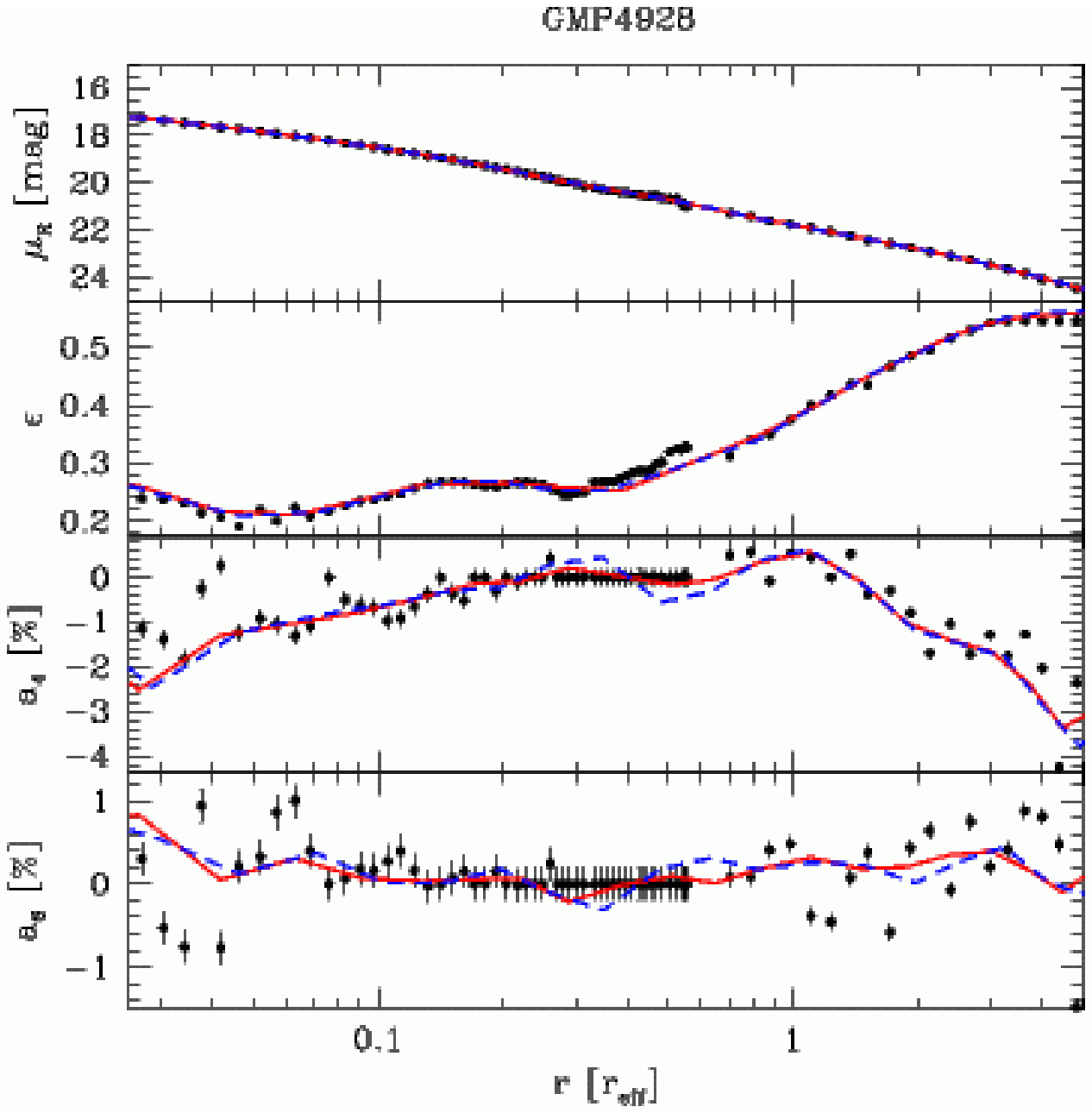}
\includegraphics[width=90mm,angle=0]{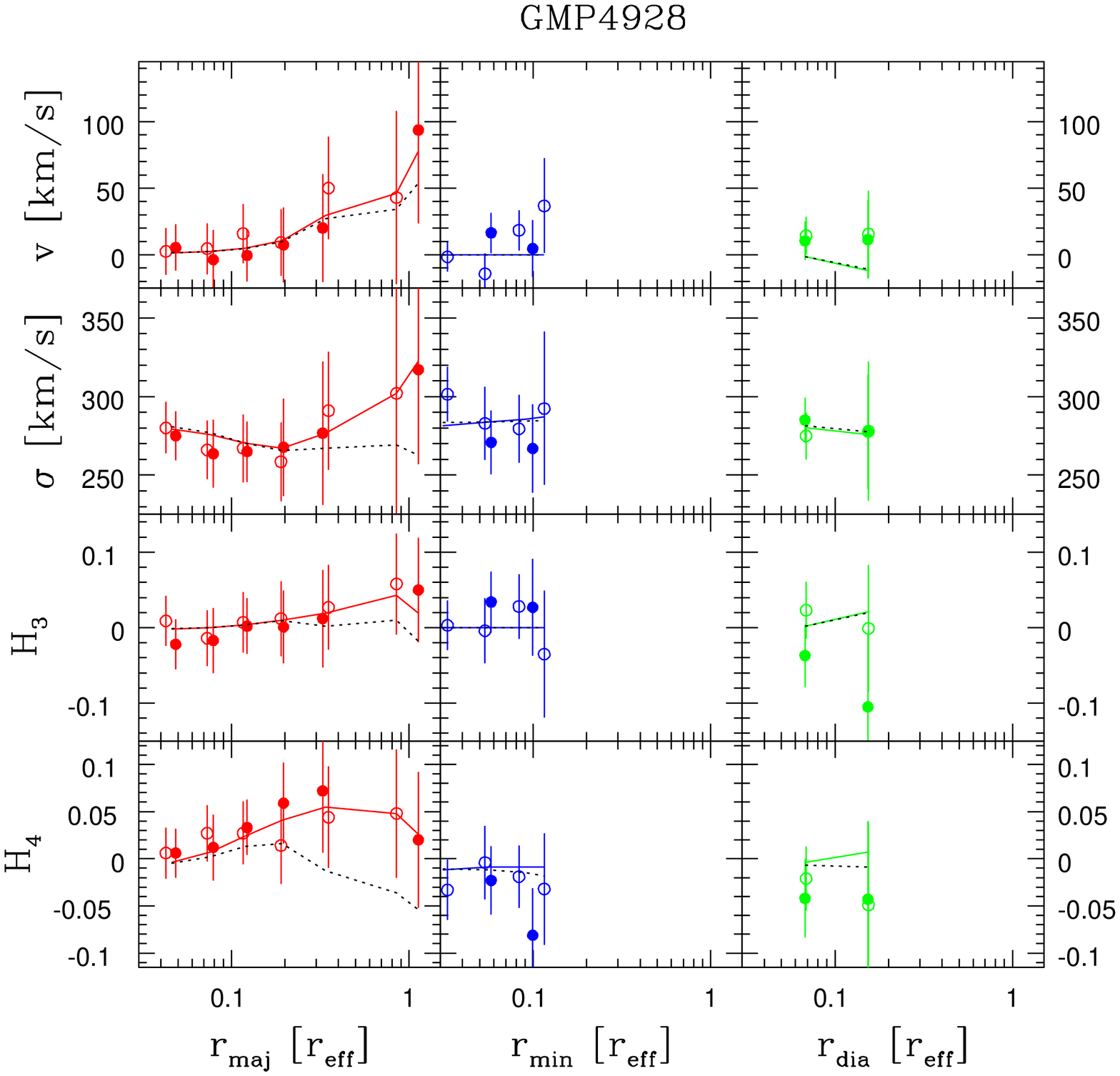}
\caption{As Fig.~\ref{isoplotgh3329}, but for GMP4928/NGC 4839; green/third column:
diagonal axis.}
\label{isoplotgh4928}
\end{figure}

\begin{figure}\centering
\includegraphics[width=90mm,angle=0]{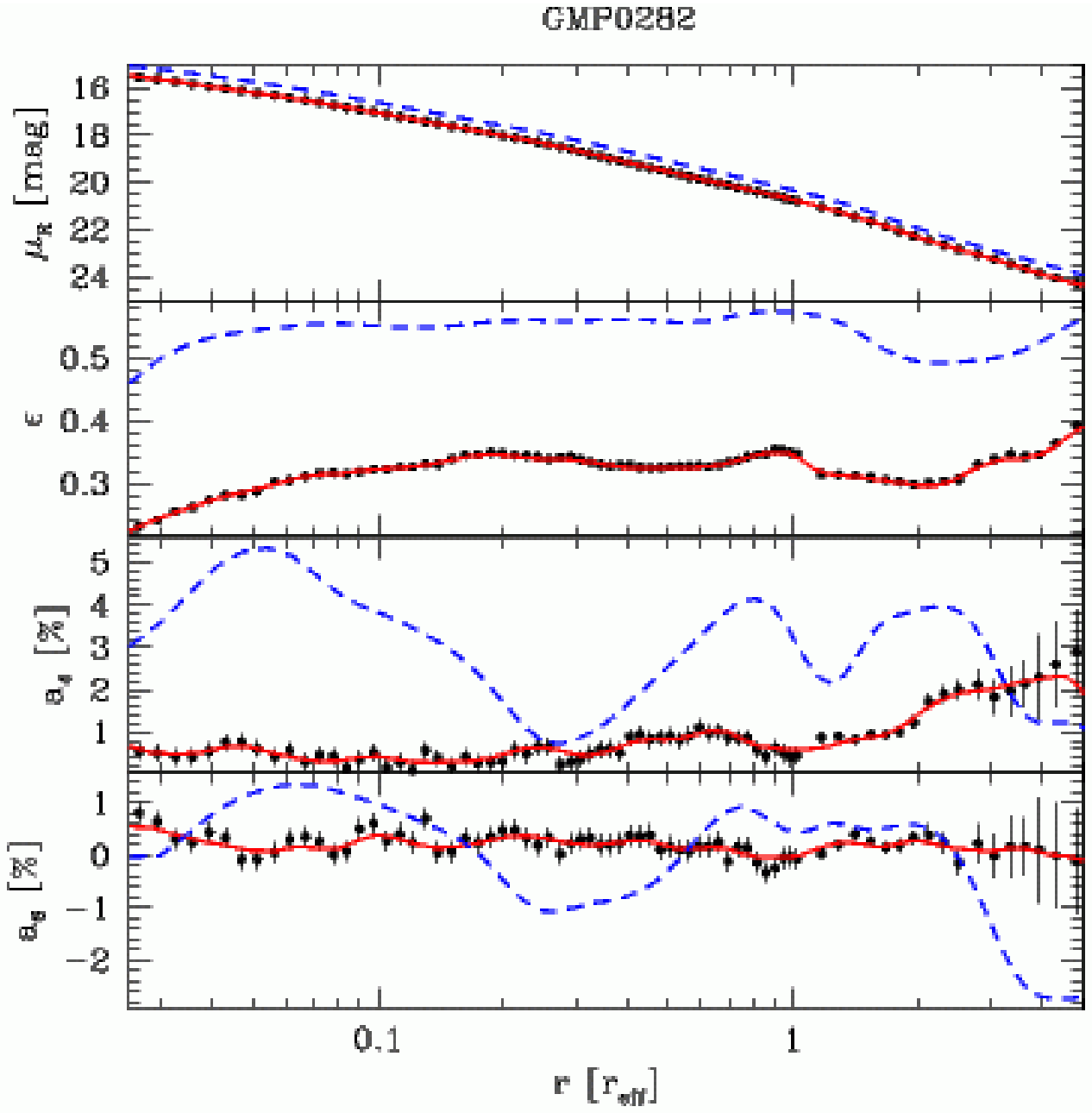}
\includegraphics[width=90mm,angle=0]{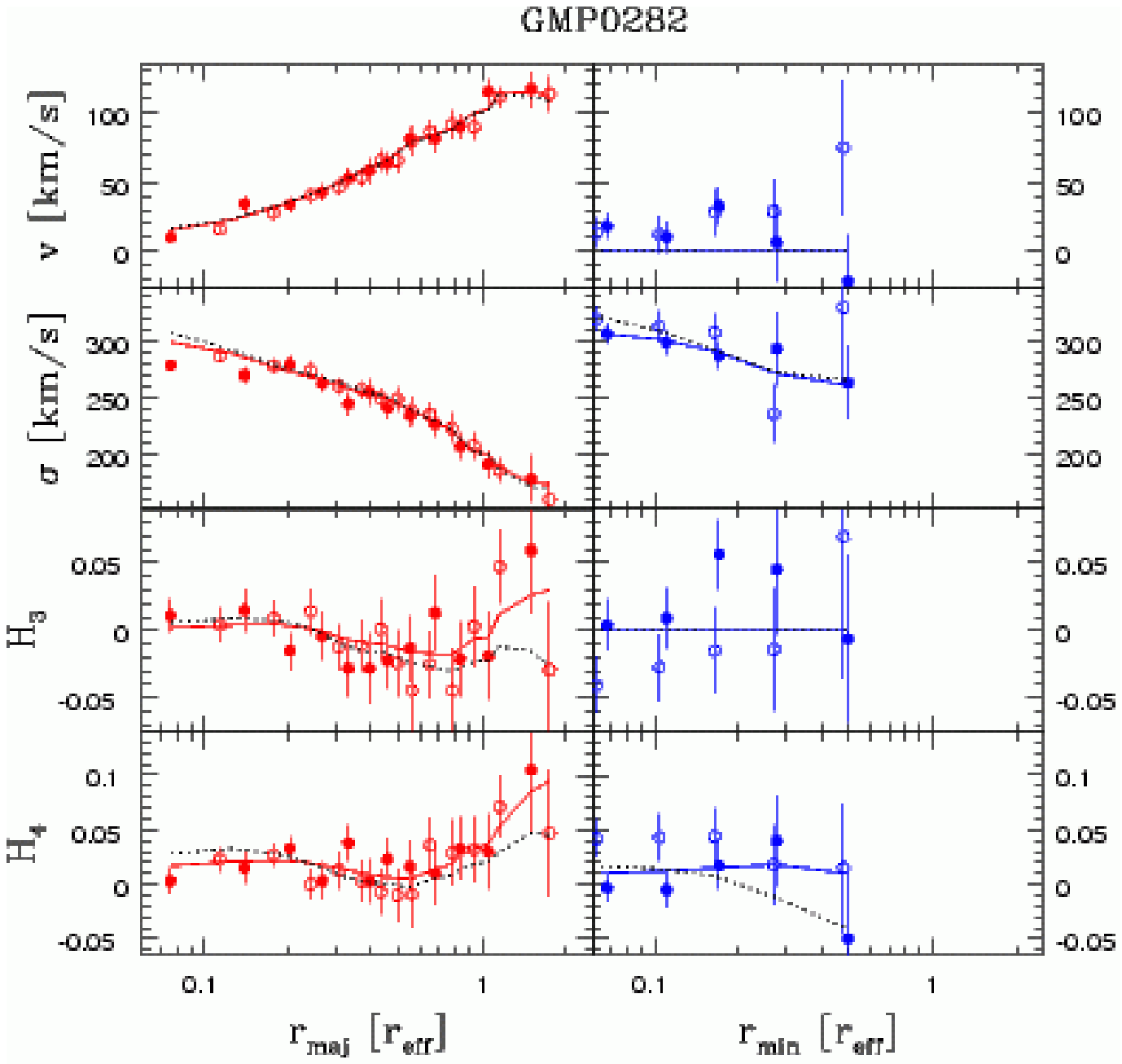}
\caption{As Fig.~\ref{isoplotgh3329}, but for GMP0282/NGC 4952.}
\label{isoplotgh0282}
\end{figure}

\begin{figure}\centering
\includegraphics[width=90mm,angle=0]{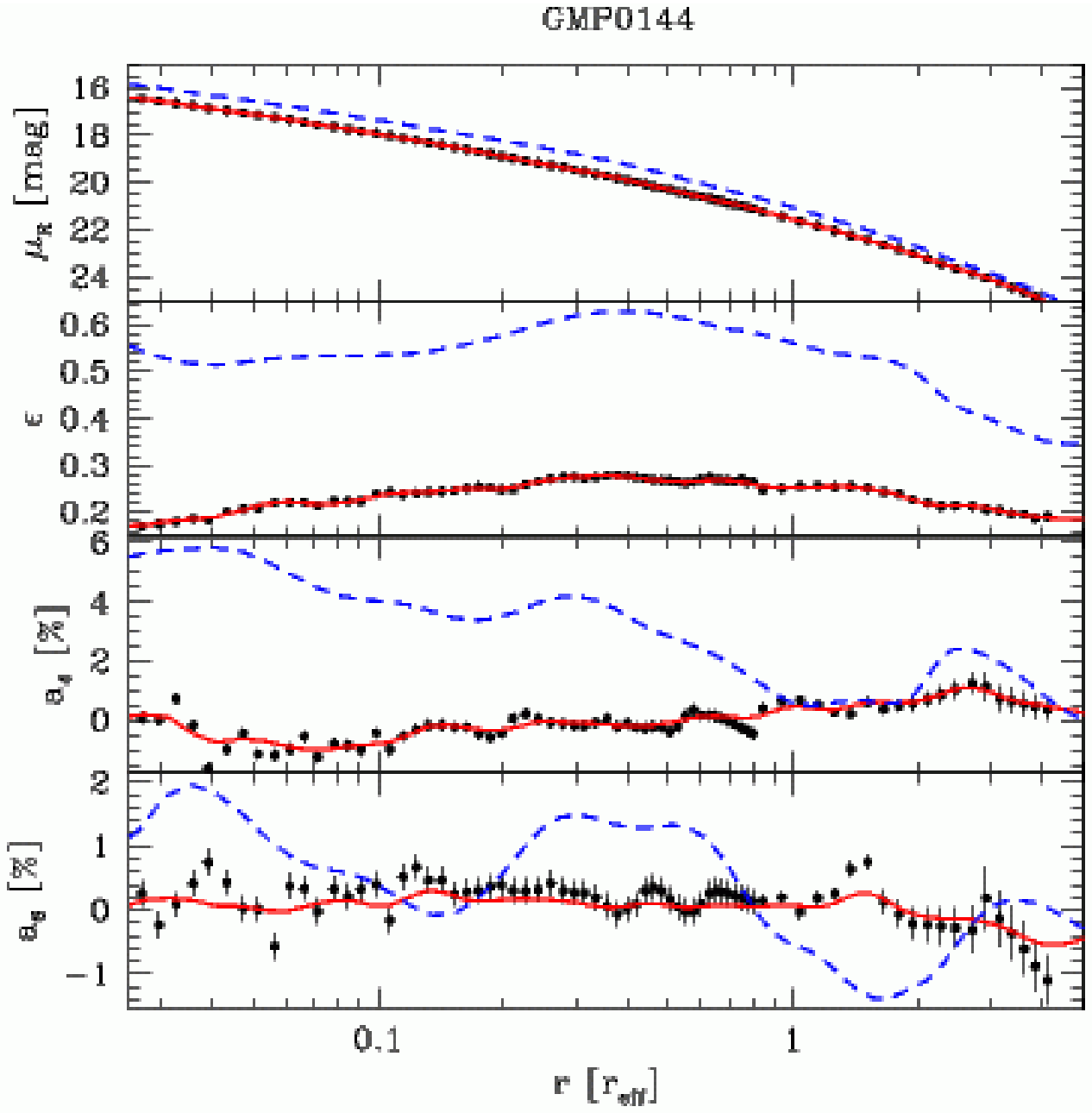}
\includegraphics[width=90mm,angle=0]{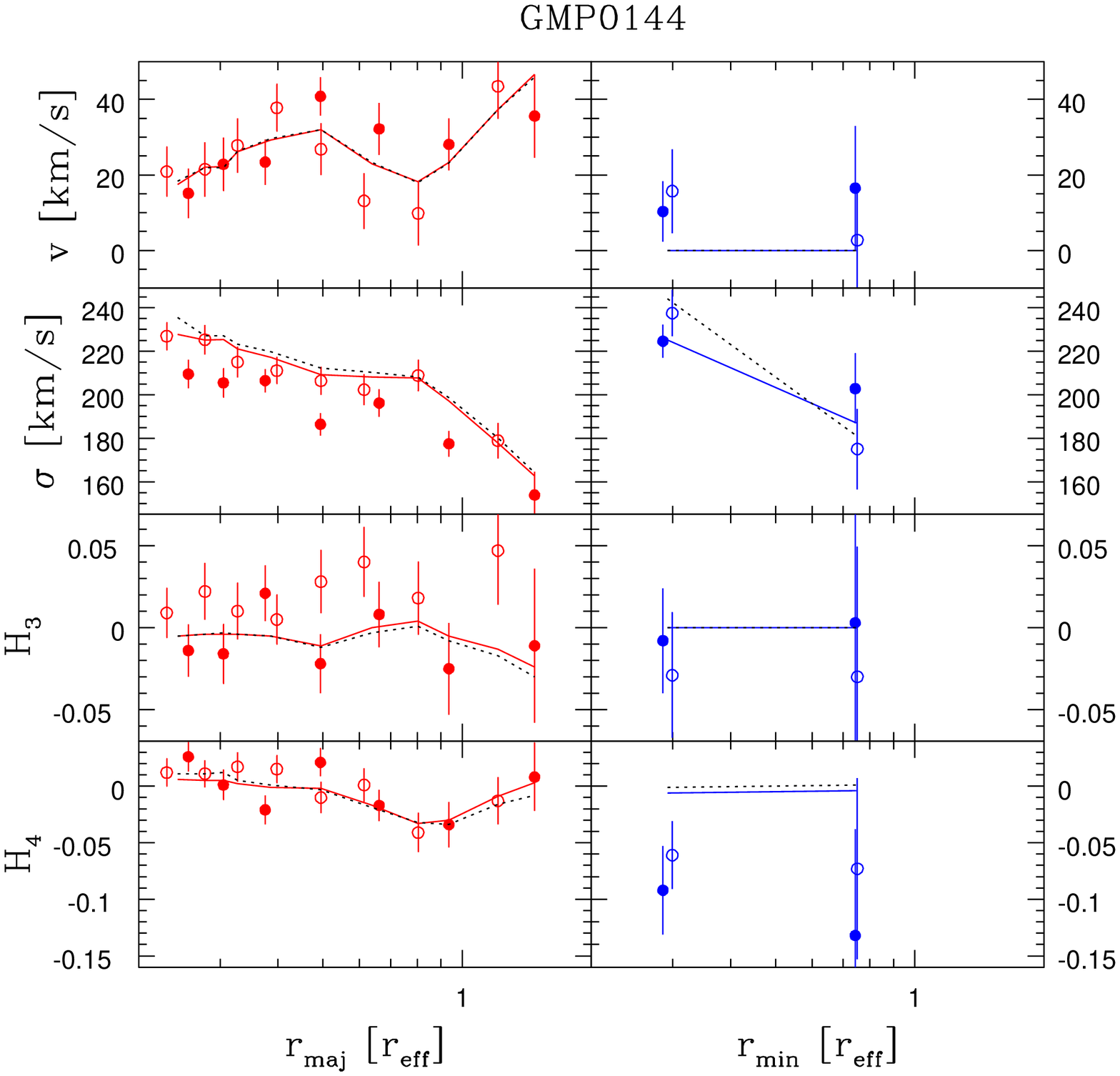}
\caption{As Fig.~\ref{isoplotgh3329}, but for GMP0144/NGC 4957.}
\label{isoplotgh0144}
\end{figure}

\begin{figure}\centering
\includegraphics[width=90mm,angle=0]{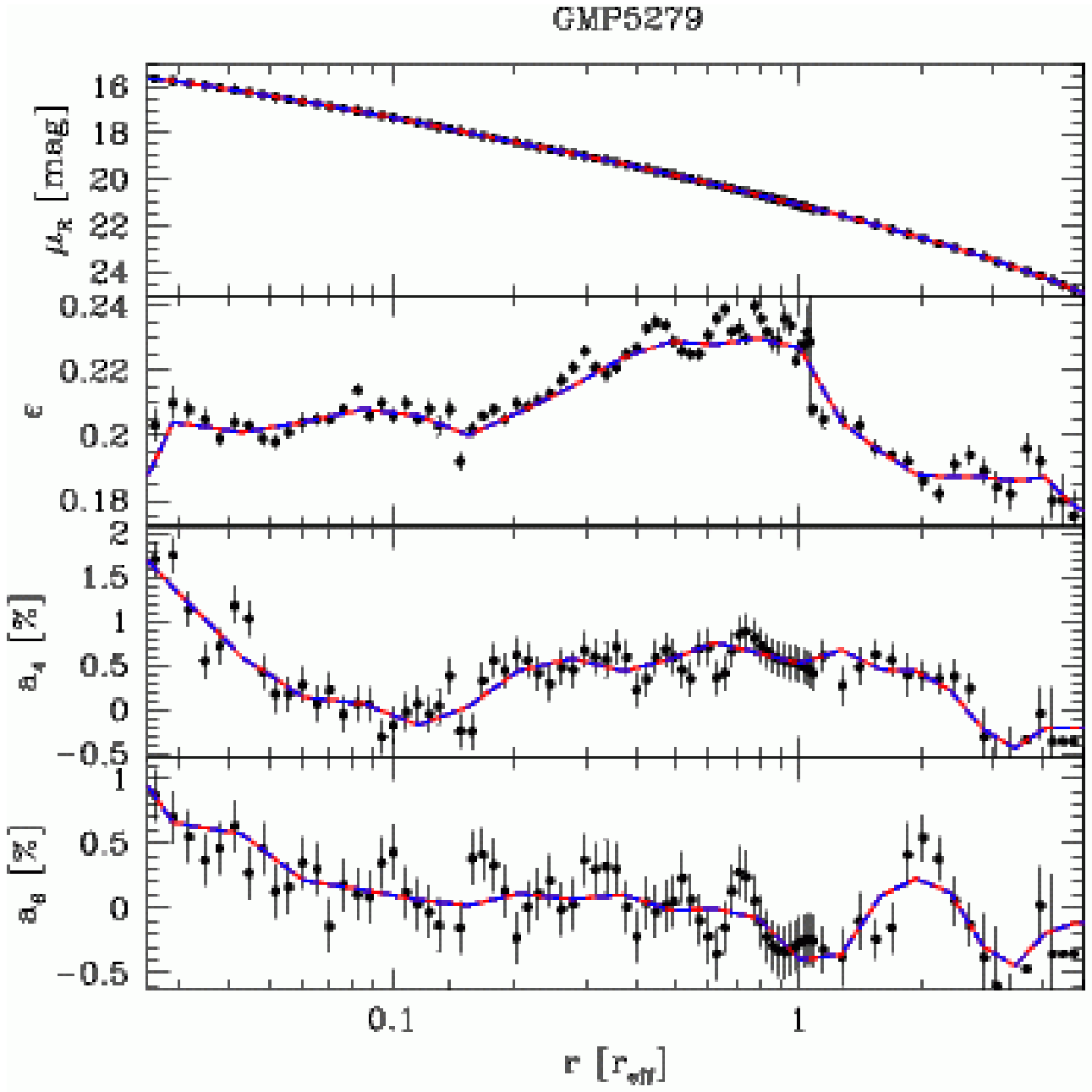}
\includegraphics[width=90mm,angle=0]{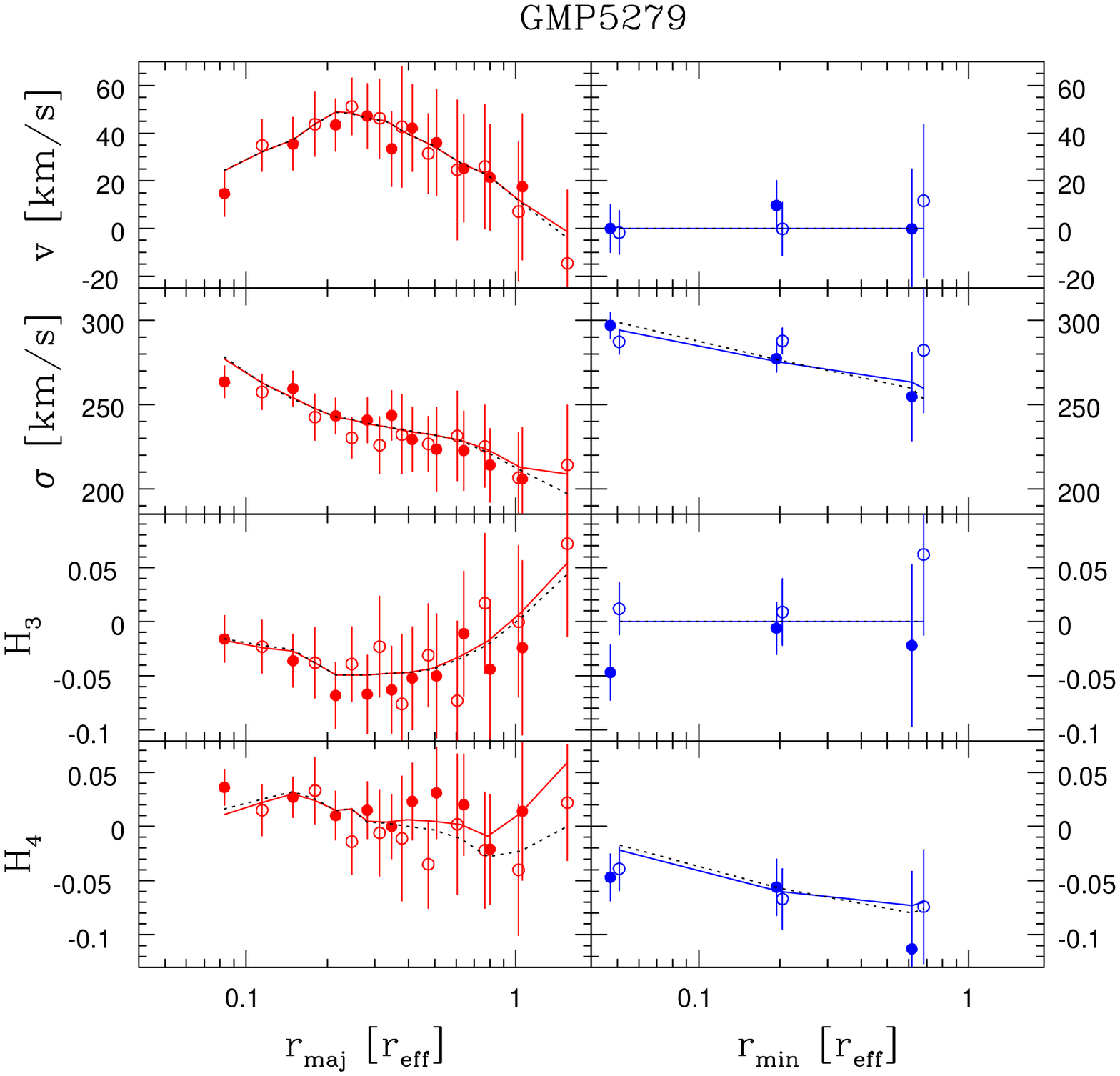}
\caption{As Fig.~\ref{isoplotgh3329}, but for GMP5279/NGC 4827.}
\label{isoplotgh5279}
\end{figure}

\begin{figure}\centering
\includegraphics[width=90mm,angle=0]{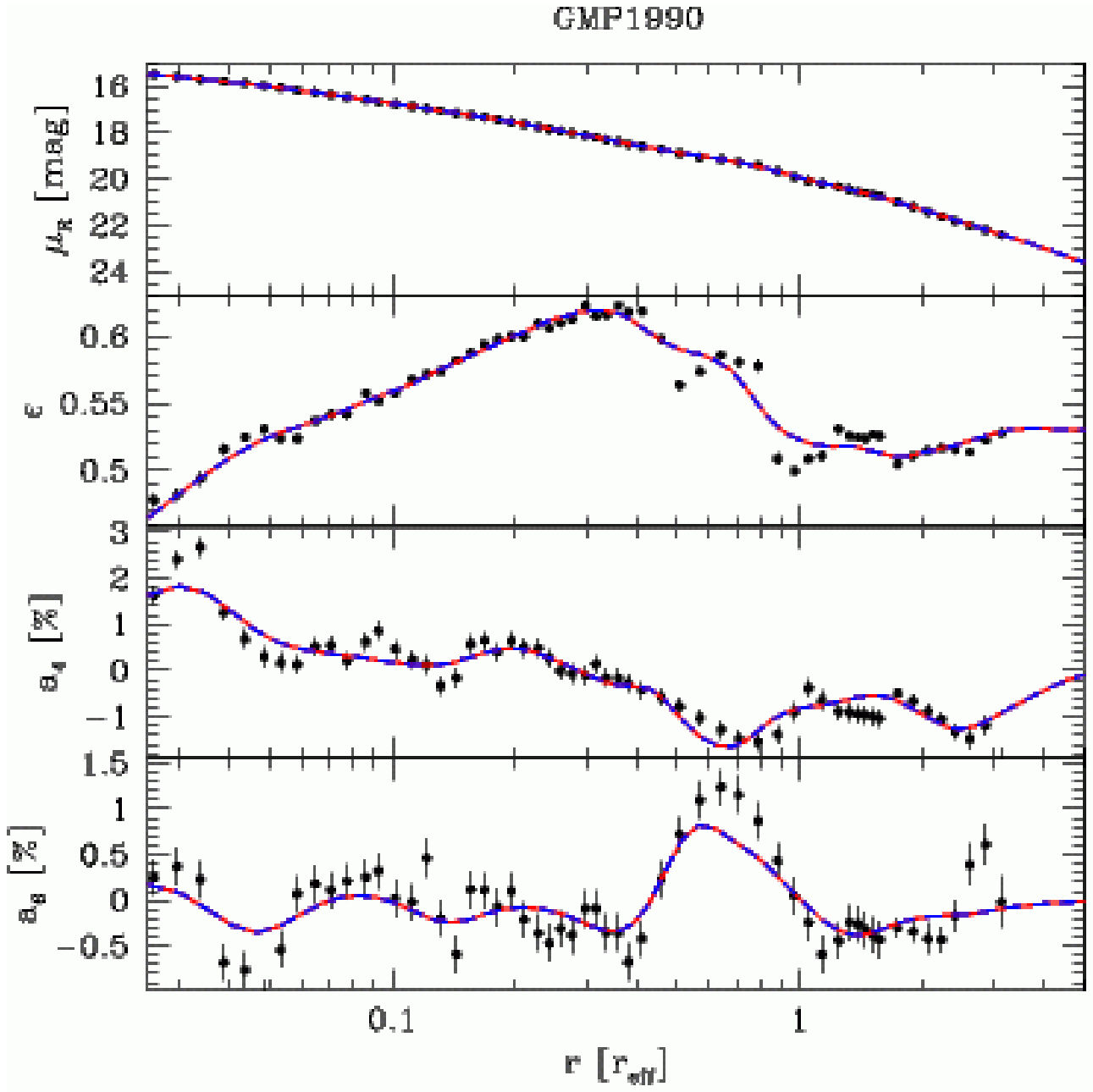}
\includegraphics[width=90mm,angle=0]{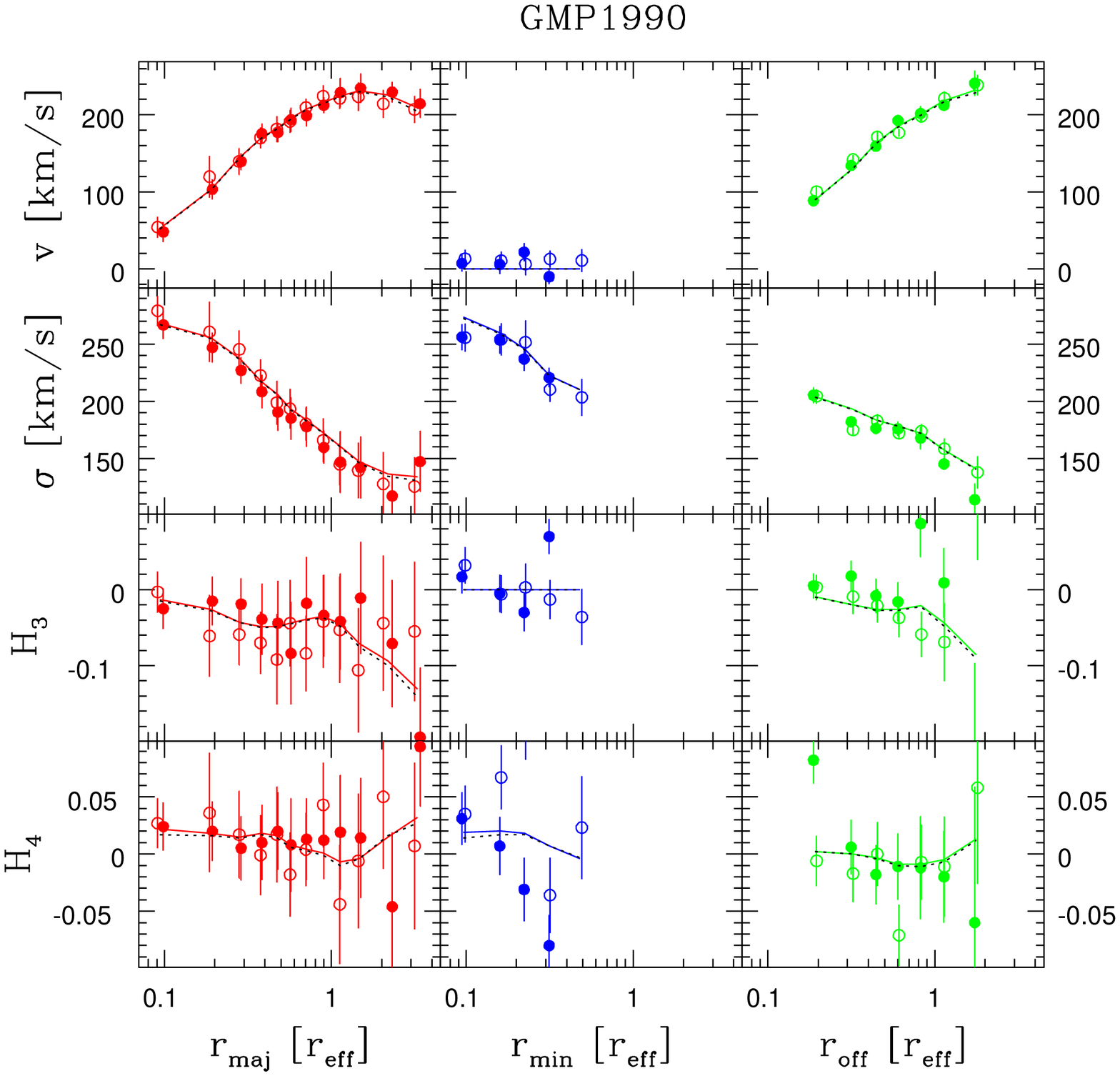}
\caption{As Fig.~\ref{isoplotgh3329}, but for GMP1990/IC 843; green/third column:
offset to major-axis $\reff/3$.}
\label{isoplotgh1990}
\end{figure}

\begin{figure}\centering
\includegraphics[width=90mm,angle=0]{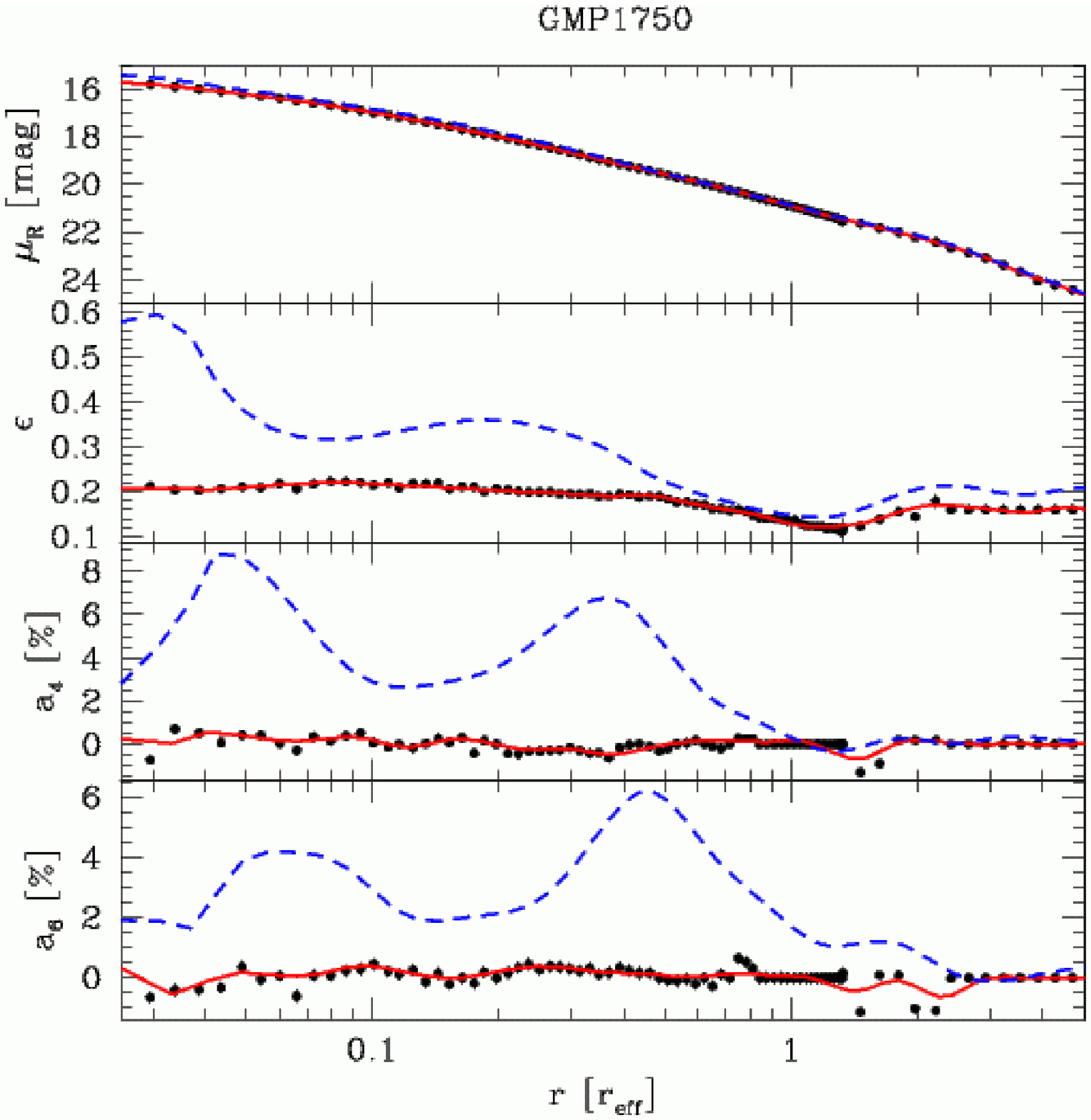}
\includegraphics[width=90mm,angle=0]{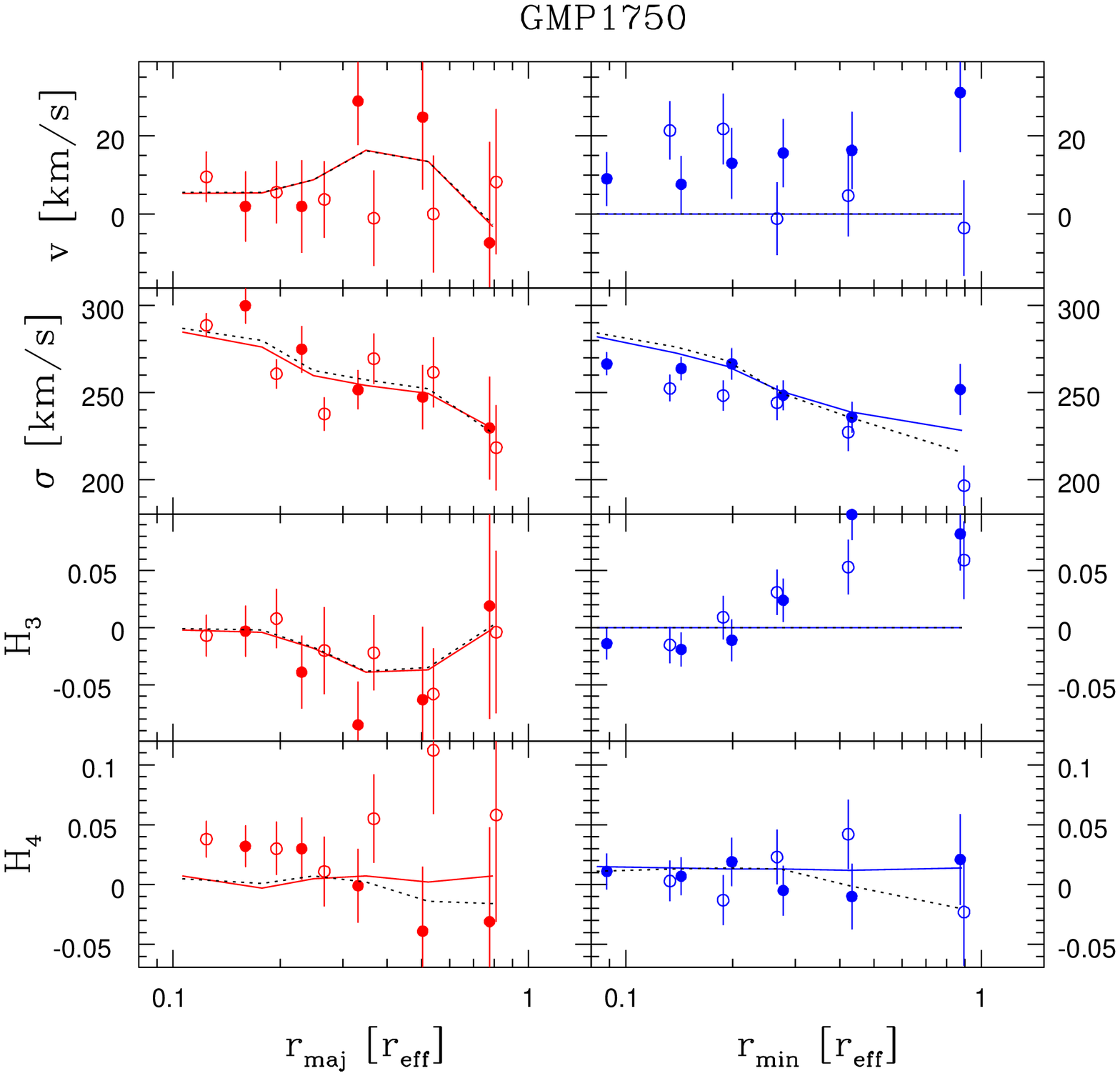}
\caption{As Fig.~\ref{isoplotgh3329}, but for GMP1750/NGC 4926.}
\label{isoplotgh1750}
\end{figure}

\begin{figure}\centering
\includegraphics[width=90mm,angle=0]{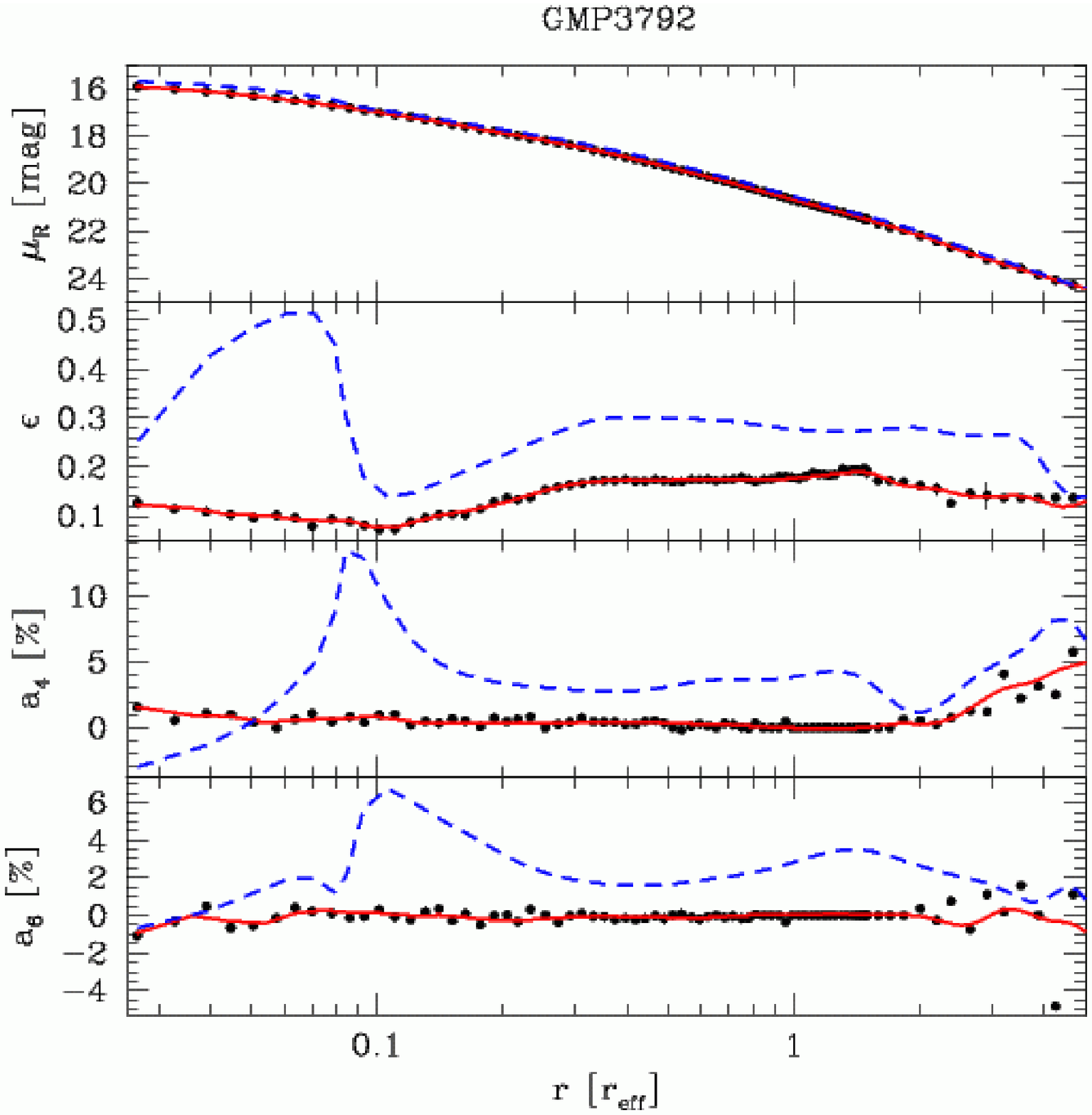}
\includegraphics[width=90mm,angle=0]{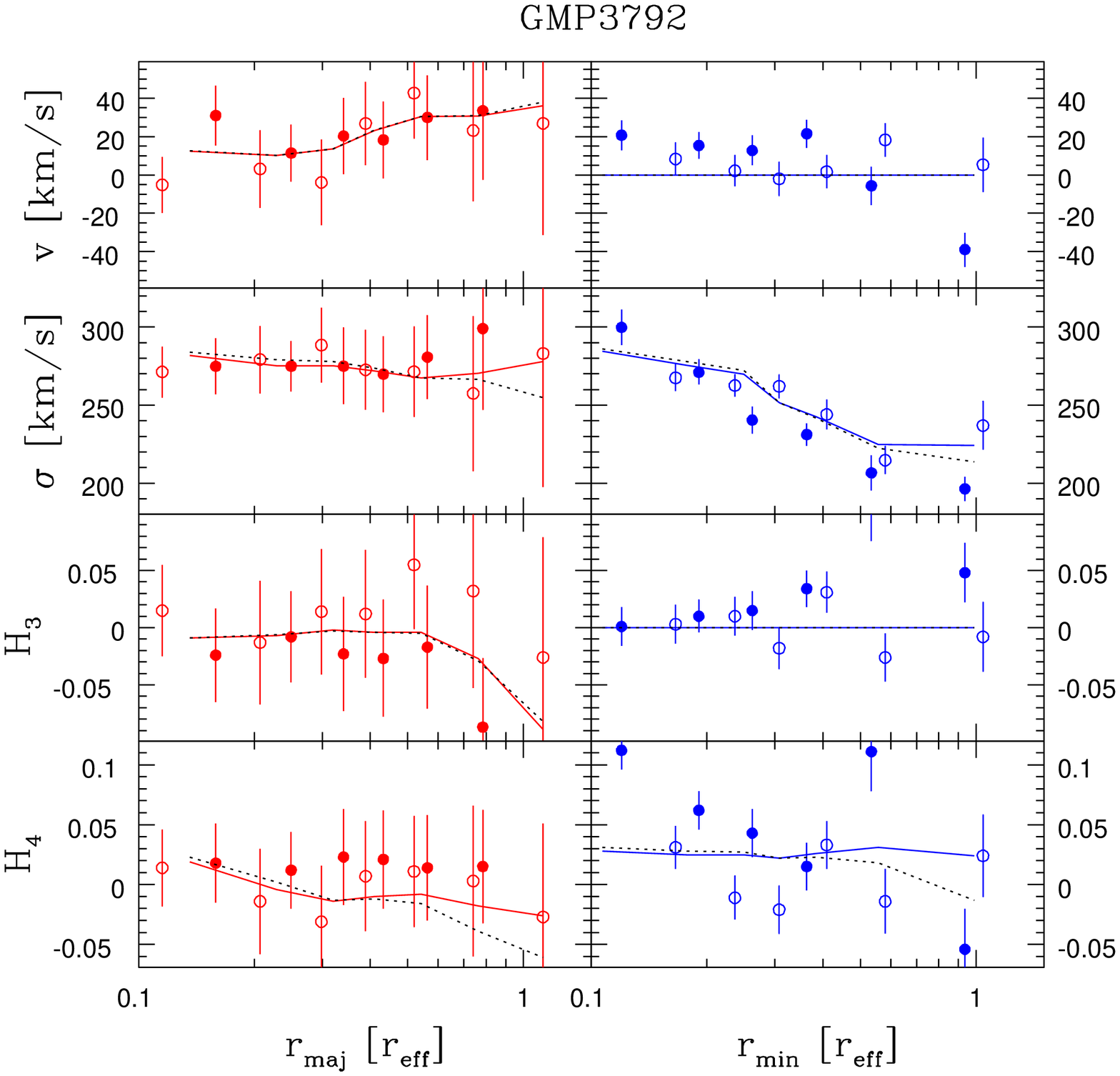}
\caption{As Fig.~\ref{isoplotgh3329}, but for GMP3792/NGC 4860.}
\label{isoplotgh3792}
\end{figure}

\begin{figure}\centering
\includegraphics[width=90mm,angle=0]{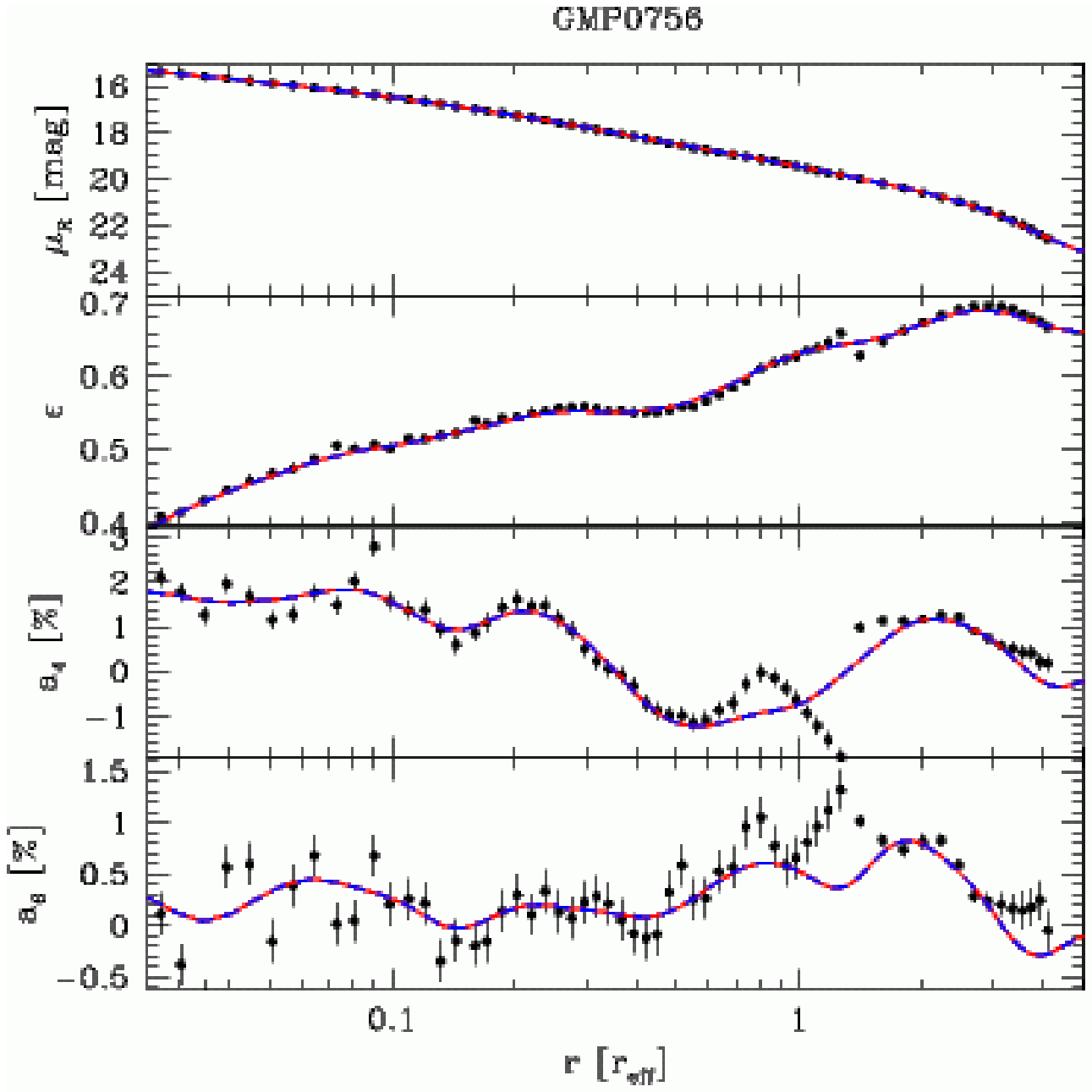}
\includegraphics[width=90mm,angle=0]{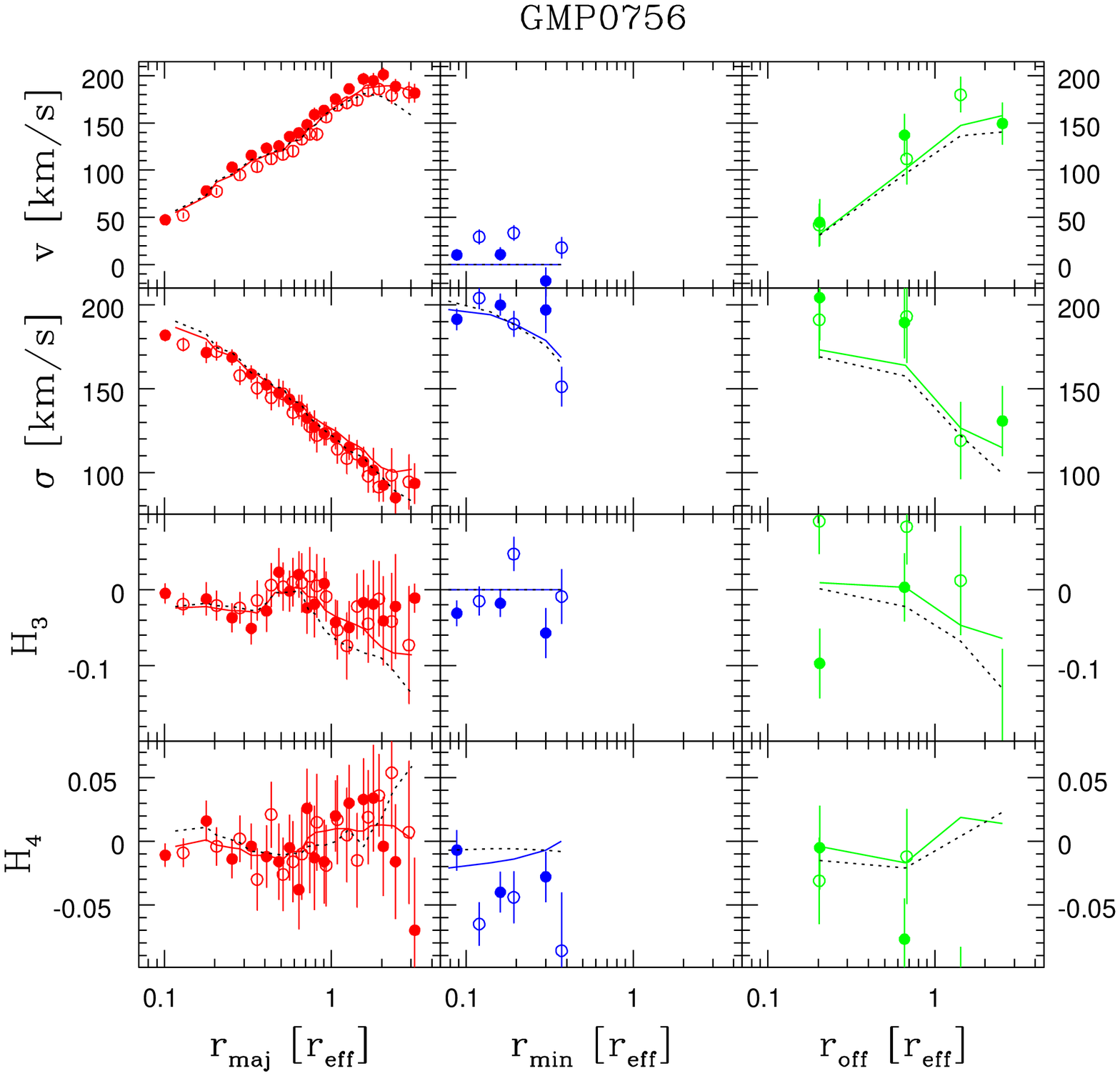}
\caption{As Fig.~\ref{isoplotgh3329}, but for GMP0756/NGC 4944; green/third column:
offset to major-axis $\reff/2$ (the two outermost $H_4<-0.1$ are omitted in the plot).}
\label{isoplotgh0756}
\end{figure}

\begin{figure}\centering
\includegraphics[width=90mm,angle=0]{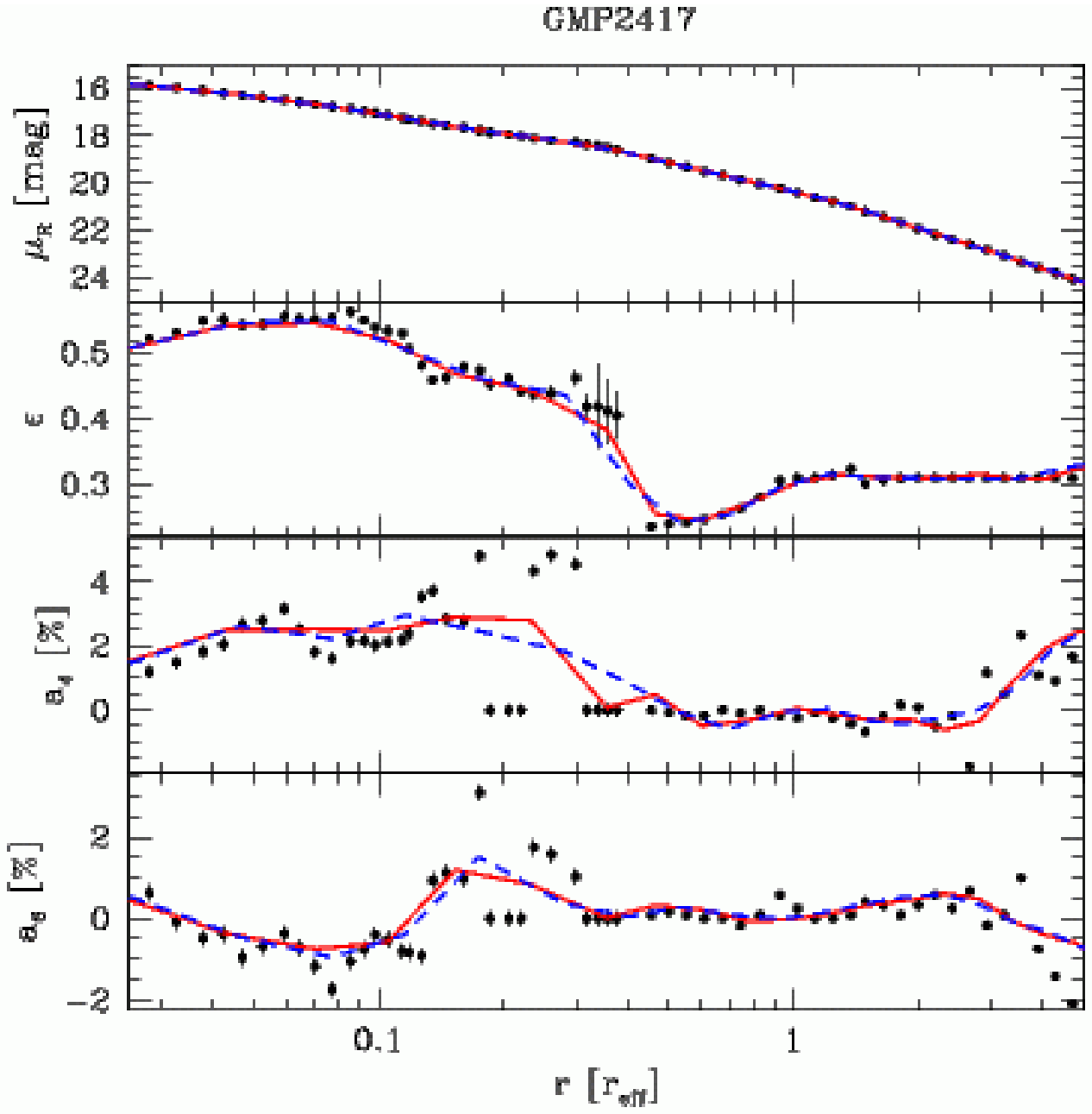}
\includegraphics[width=90mm,angle=0]{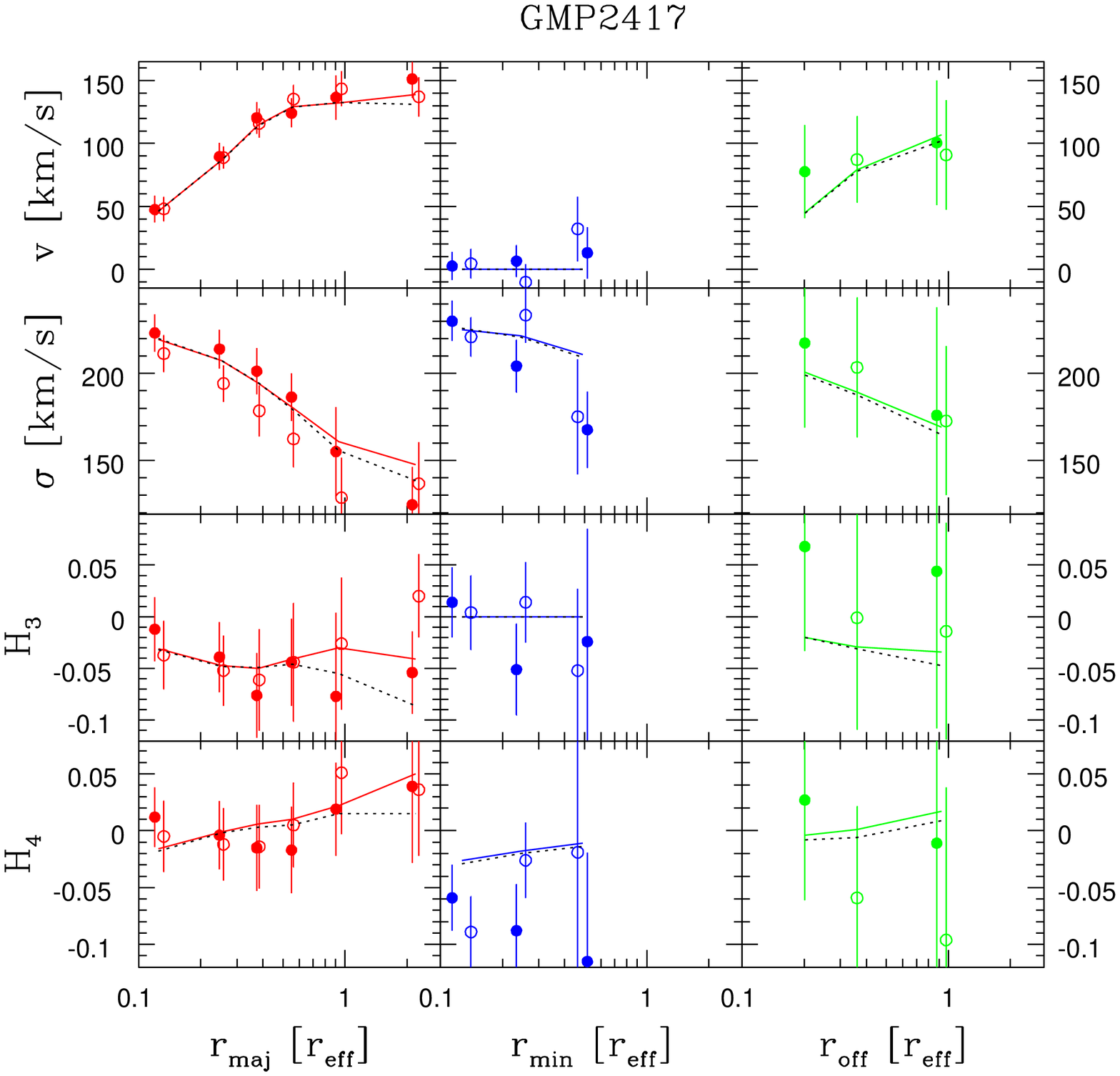}
\caption{As Fig.~\ref{isoplotgh3329}, but for GMP2417/NGC 4908; green/third column:
offset to major-axis $\reff/2$.}
\label{isoplotgh2417}
\end{figure}

\begin{figure}\centering
\includegraphics[width=90mm,angle=0]{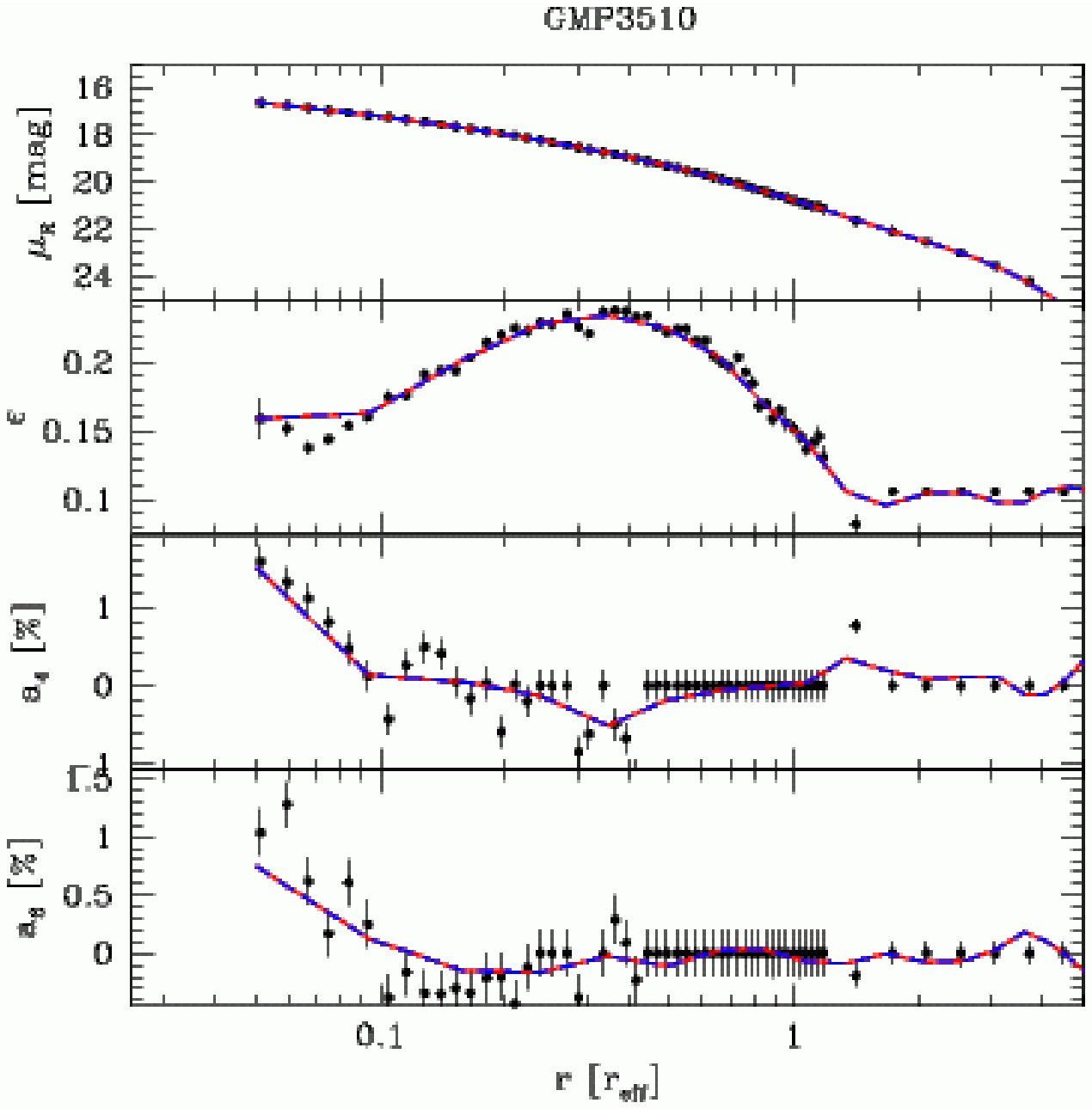}
\includegraphics[width=90mm,angle=0]{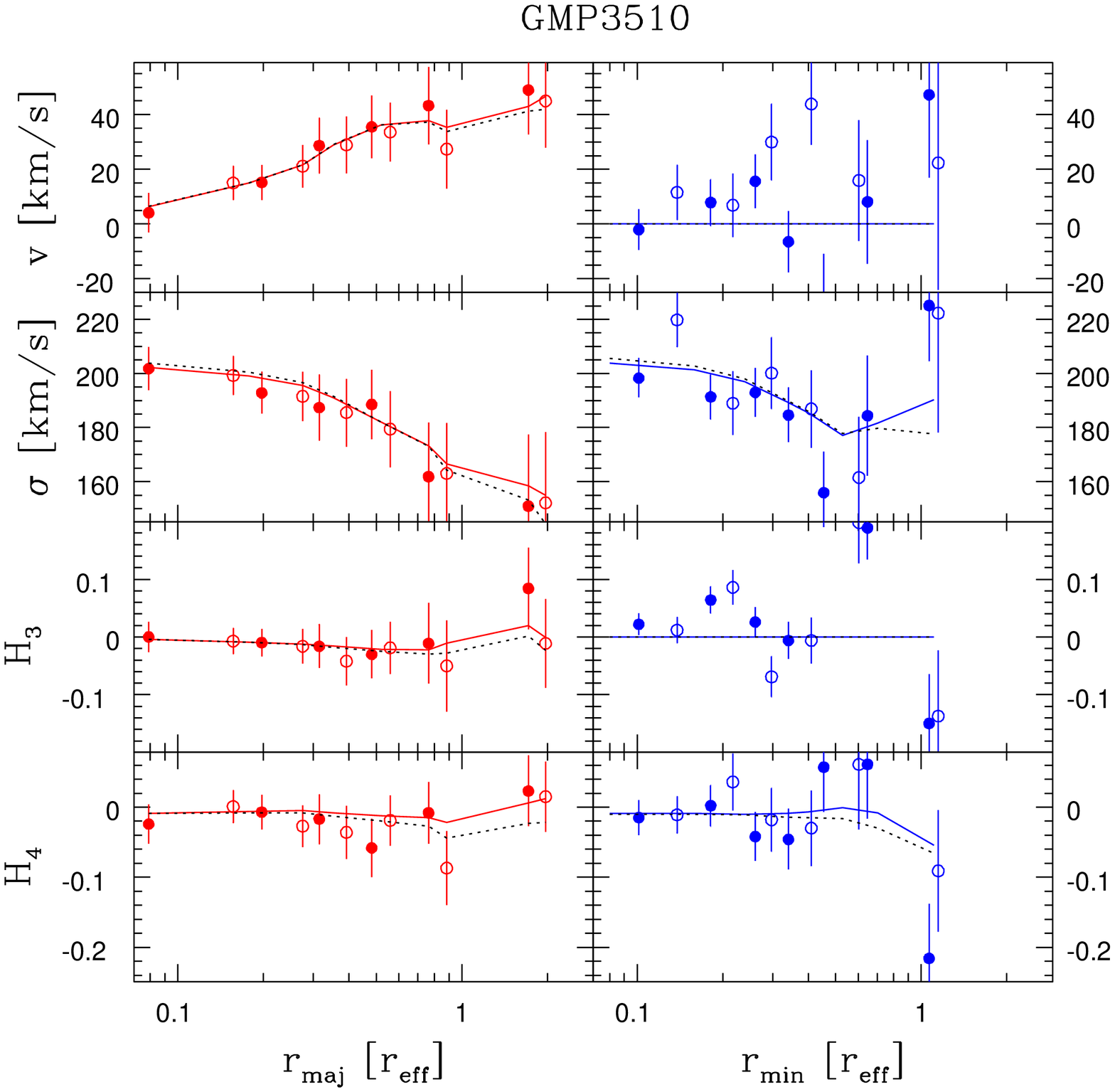}
\caption{As Fig.~\ref{isoplotgh3329}, but for GMP3510/NGC 4869.}
\label{isoplotgh3510}
\end{figure}

\begin{figure}\centering
\includegraphics[width=90mm,angle=0]{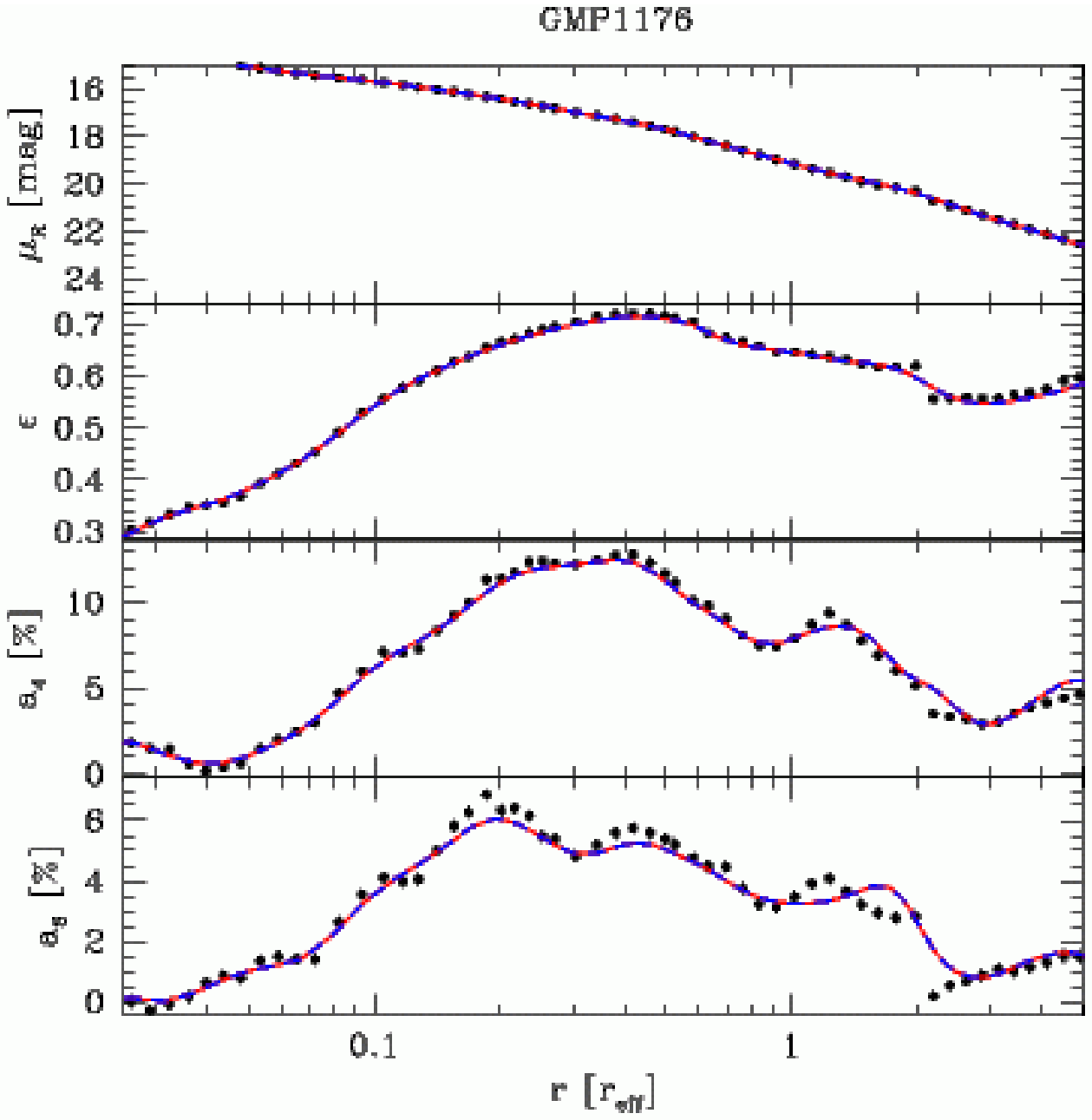}
\includegraphics[width=90mm,angle=0]{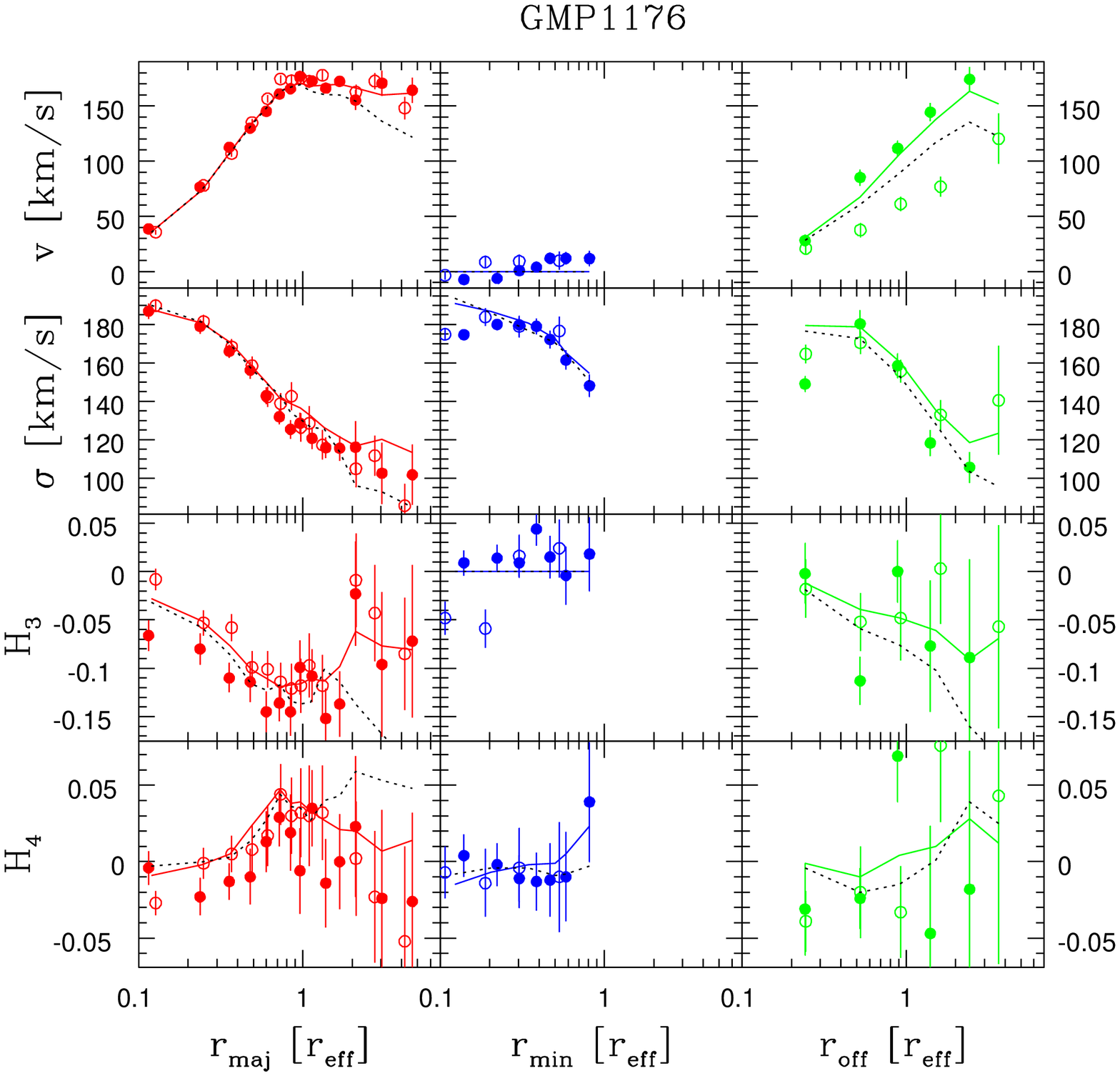}
\caption{As Fig.~\ref{isoplotgh0144}, but for GMP1176/NGC 4931; green/third column:
offset to major-axis $\reff/3$.}
\label{isoplotgh1176}
\end{figure}

\begin{figure}\centering
\includegraphics[width=90mm,angle=0]{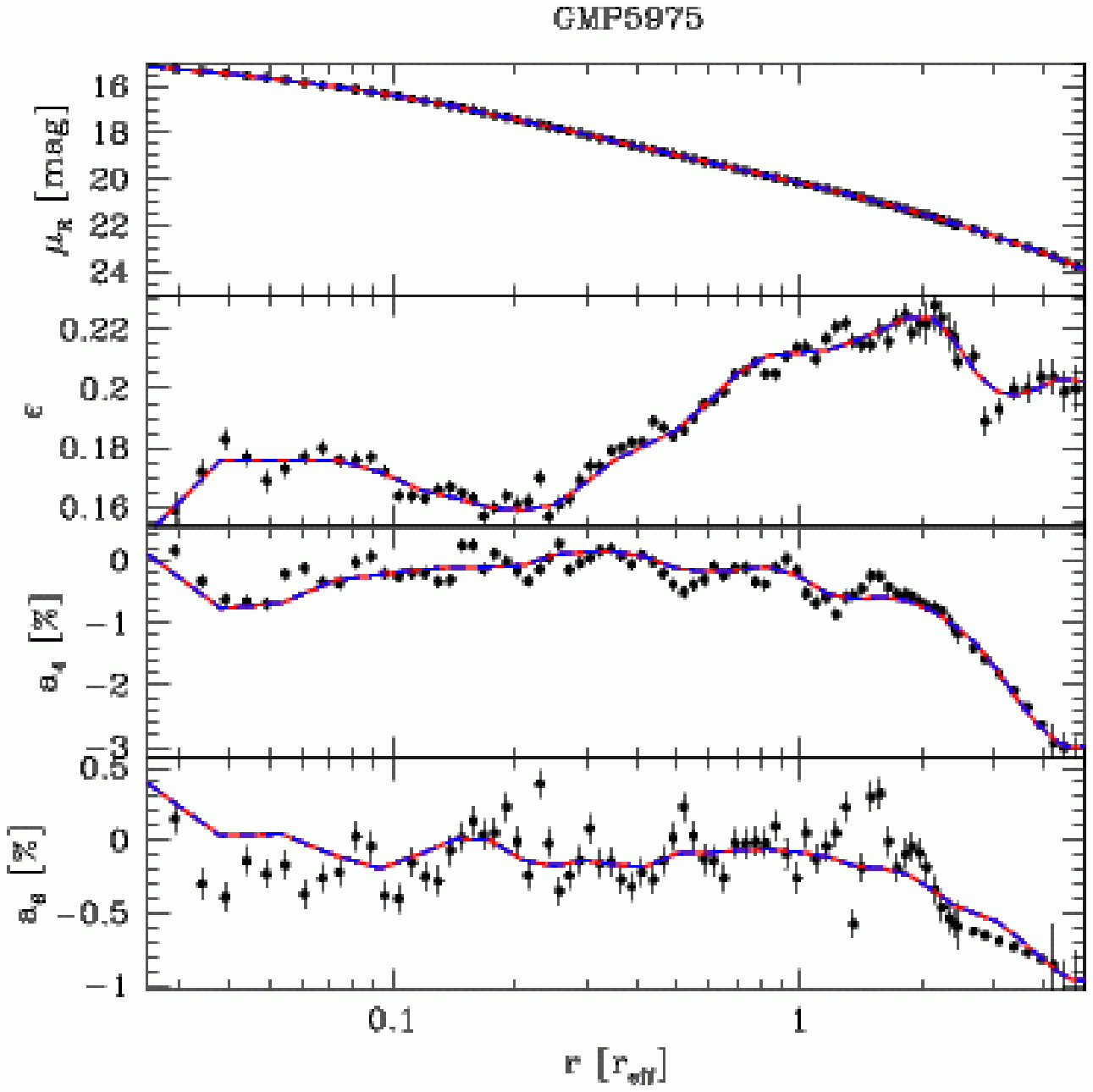}
\includegraphics[width=90mm,angle=0]{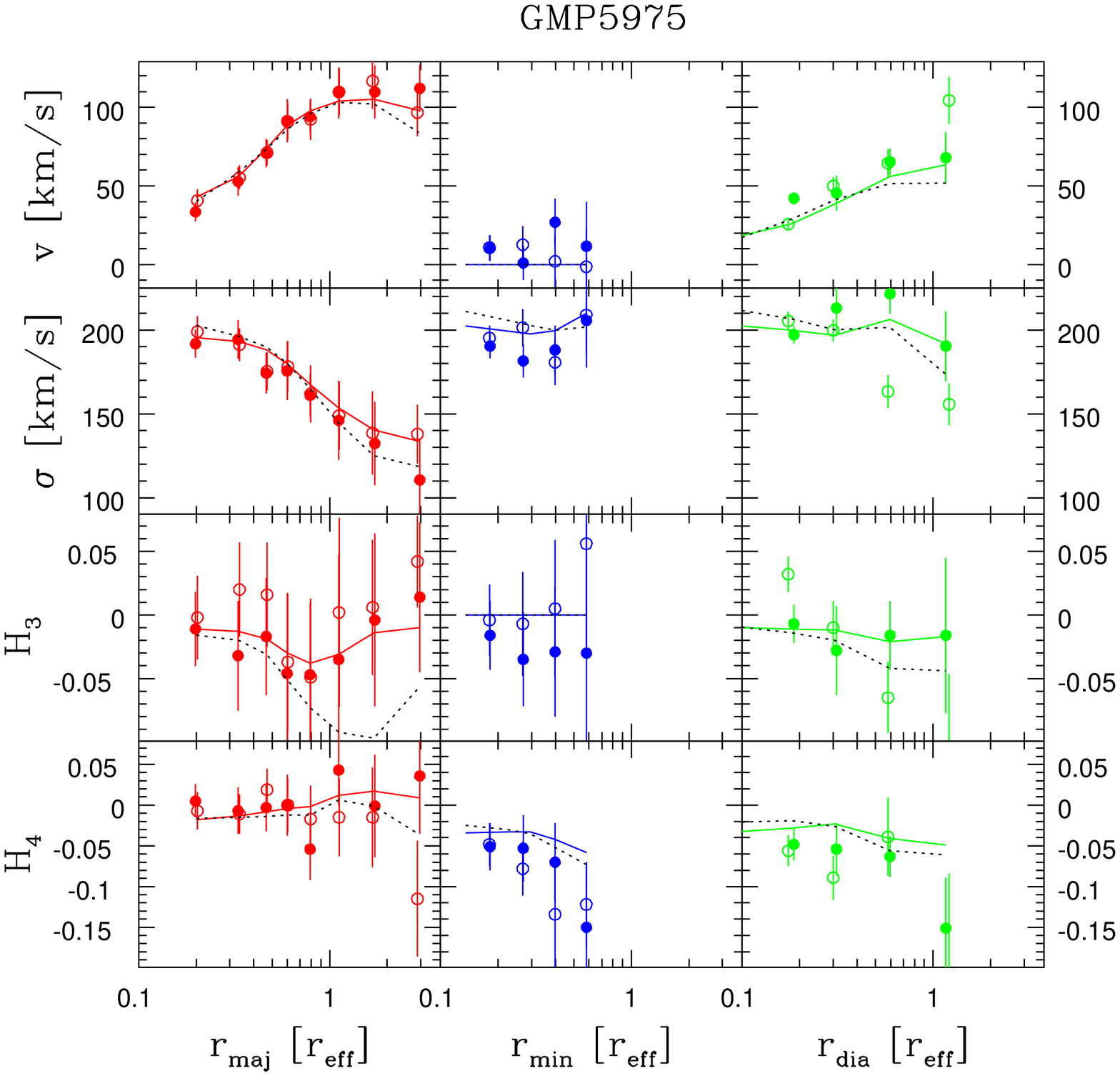}
\caption{As Fig.~\ref{isoplotgh3329}, but for GMP5975/NGC 4807; green/third column:
diagonal axis.}
\label{isoplotgh5975}
\end{figure}

\begin{figure}\centering
\includegraphics[width=90mm,angle=0]{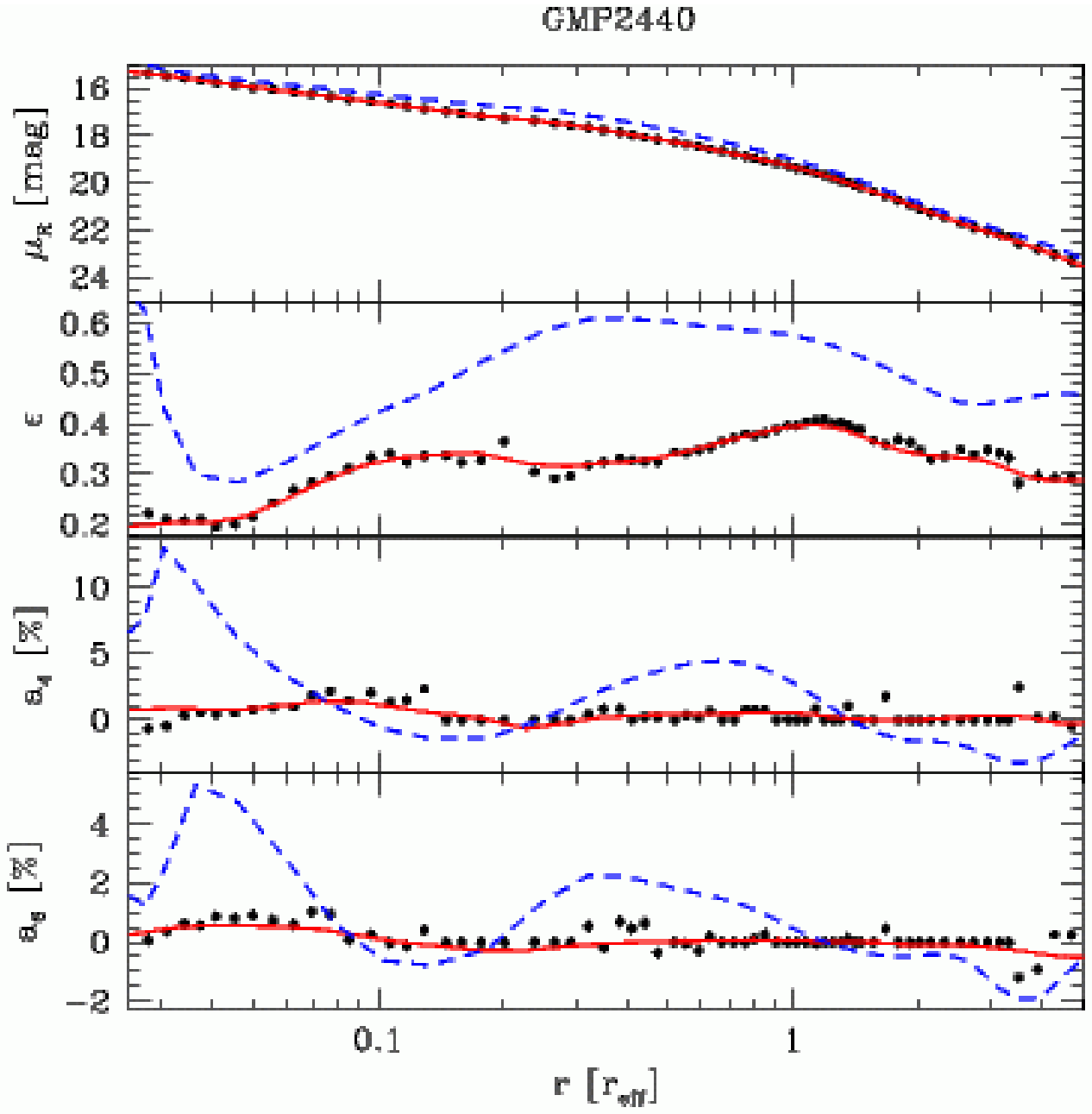}
\includegraphics[width=90mm,angle=0]{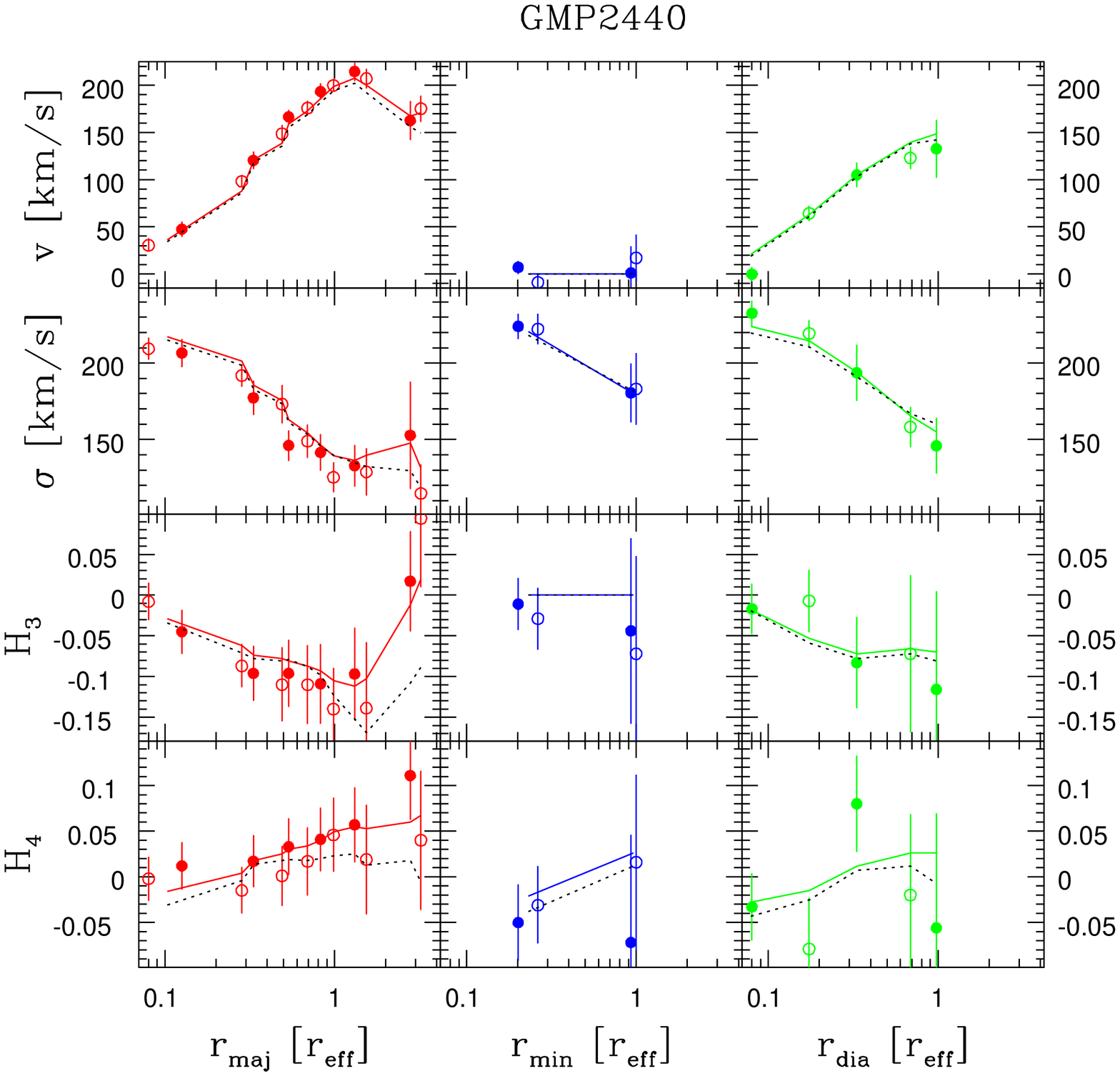}
\caption{As Fig.~\ref{isoplotgh3329}, but for GMP2440/IC 4045; green/third column:
diagonal axis.}
\label{isoplotgh2440}
\end{figure}

\begin{figure}\centering
\includegraphics[width=90mm,angle=0]{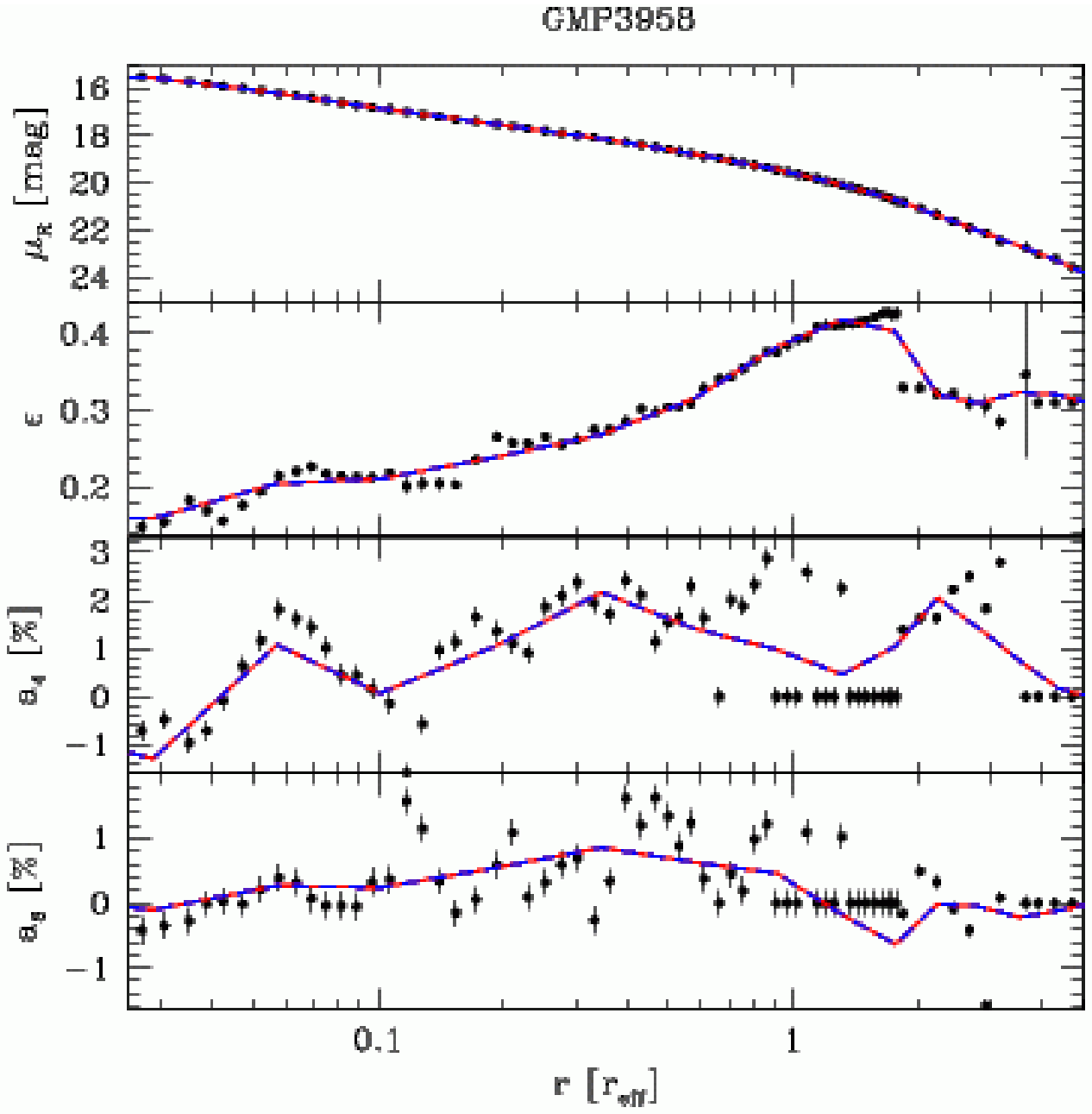}
\includegraphics[width=90mm,angle=0]{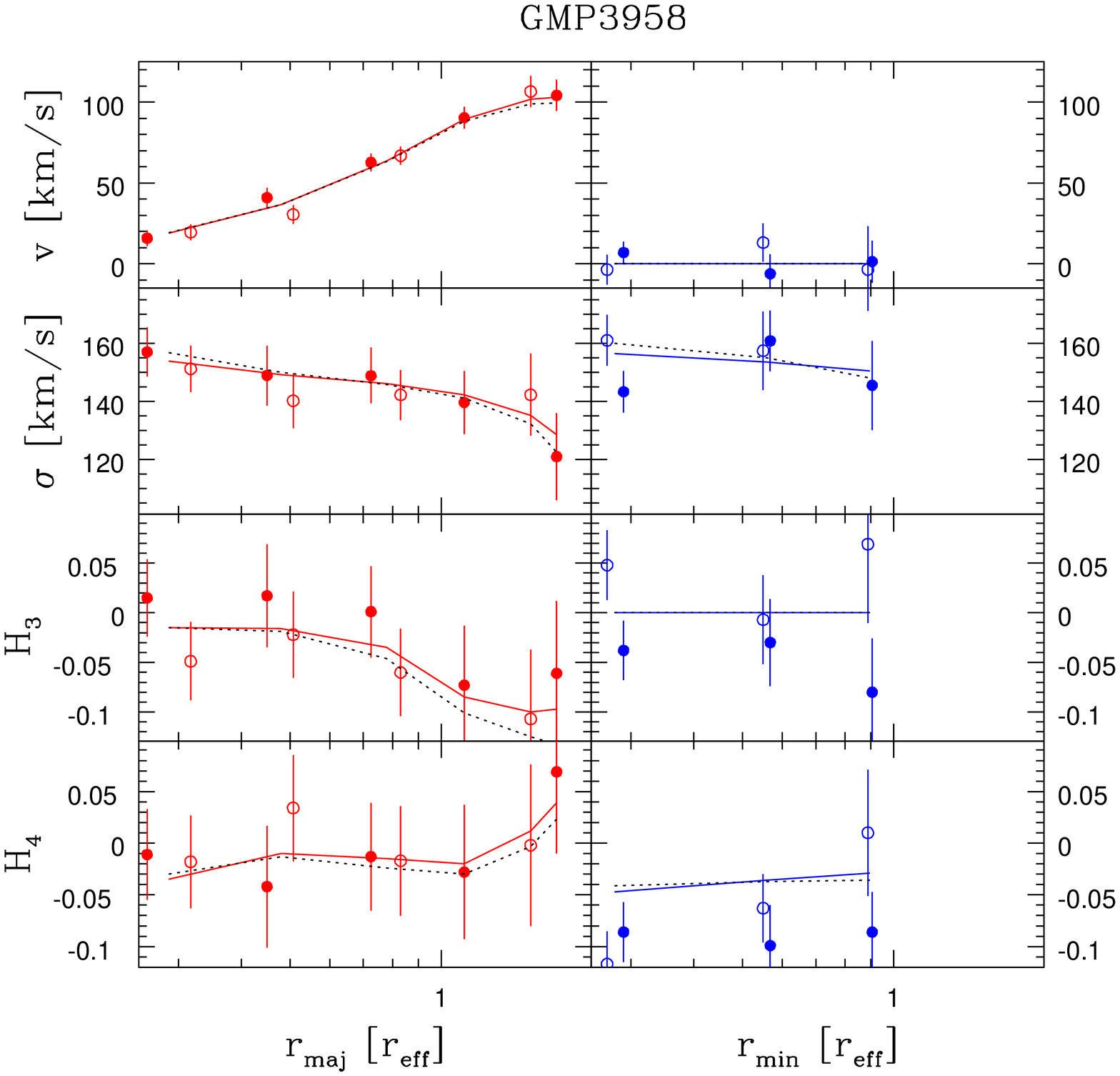}
\caption{As Fig.~\ref{isoplotgh3329}, but for GMP3958/IC 3947.}
\label{isoplotgh3958}
\end{figure}

Figs.~\ref{isoplotgh3329} - \ref{isoplotgh3958} survey the fits to the photometric
and kinematical data for each galaxy (galaxies are arranged in order of decreasing 
total mass inside $\reff$). 

\newpage

\bsp

\label{lastpage}


\begin{thebibliography}{99}
\bibitem[\protect\citeauthoryear{Baes, Dejonghe \& Buyle}{2003}]{Bae03b} Baes M., Dejonghe H., Buyle P.,  2003, A\&A, 432, 411
\bibitem[\protect\citeauthoryear{Bender \& M\"ollenhoff}{1987}]{bm87} Bender R., M\"ollenhoff C., 1987, A\&A, 177, 71
\bibitem[\protect\citeauthoryear{Bertola et al.}{1993}]{Ber93} Bertola F., Pizzella A., Persic M., Salucci P., 1993, ApJL, 416, 45
\bibitem[\protect\citeauthoryear{Binney}{1981}]{Bin81} Binney J., 1981, MNRAS, 196, 455
\bibitem[\protect\citeauthoryear{Binney \& Tremaine}{1987}]{Bin87} Binney J., Tremaine S., 1987, Galactic Dynamics (Princeton: Princeton University Press)
\bibitem[\protect\citeauthoryear{Cappellari et al.}{2006}]{Cap05} Cappellari M.~et al., 2006, MNRAS, 366, 1126
\bibitem[\protect\citeauthoryear{Cappellari et al.}{2007}]{Cap07} Cappellari M.~et al., 2007, MNRAS, in press
\bibitem[\protect\citeauthoryear{Ciotti \& Pellegrini}{1992}]{Cio92} Ciotti L., Pellegrini S., 1992, MNRAS,255, 561
\bibitem[\protect\citeauthoryear{Ciotti}{1996}]{Cio96} Ciotti L., 1996, ApJ, 471, 68
\bibitem[\protect\citeauthoryear{Ciotti}{1999}]{Cio99} Ciotti L., 1999, ApJ, 520, 574
\bibitem[\protect\citeauthoryear{Corsini et al.}{2007}]{Cor07} Corsini E.~M., Wegner G., Saglia R.~P., Thomas J., Bender R., Thomas D., submitted to ApJS
\bibitem[\protect\citeauthoryear{Dubinski}{1998}]{Dub98} Dubinski J., 1998, ApJ, 502, 141
\bibitem[\protect\citeauthoryear{Evans}{1993}]{Eva93} Evans N.~W., 1993, MNRAS, 260, 191
\bibitem[\protect\citeauthoryear{Fukazawa et al.}{2006}]{Fuk06} Fukazawa Y., Botoya-Nonesa J.~G., Pu J., Ohto A., Kawano N., 2006, ApJ, 636, 698
\bibitem[\protect\citeauthoryear{Fukugita, Hogan \& Peebles}{1998}]{Fuk98} Fukugita M., Hogan C.~J., Peebles P.~J.~E., 1998, ApJ, 503, 518
\bibitem[\protect\citeauthoryear{Gebhardt et al.}{2000}]{Geb00} Gebhardt K.~et al., 2000, AJ, 119, 1157  
\bibitem[\protect\citeauthoryear{Gebhardt et al.}{2003}]{Geb03} Gebhardt K.~et al., 2003, ApJ, 583, 92 
\bibitem[\protect\citeauthoryear{Gerhard}{1993}]{Ger93} Gerhard O.~E., 1993, MNRAS, 265, 213
\bibitem[\protect\citeauthoryear{Gerhard et al.}{2001}]{G01} Gerhard O.~E., Kronawitter A., Saglia R.~P., Bender R., 2001, AJ, 121, 1936
\bibitem[\protect\citeauthoryear{Gerhard et al.}{2007}]{Ger07} Gerhard O.~E., Arnaboldi M., Freeman K.~C., Okamura S., Kashikawa N., Yasuda N., 2007, A\&A, 468, 815
\bibitem[\protect\citeauthoryear{Godwin, Metcalfe \& Peach}{1983}]{GMP} Godwin J.~G., Metcalfe N., Peach J.~V., 1983, MNRAS, 202, 113
\bibitem[\protect\citeauthoryear{Hernquist}{1992}]{Her92} Hernquist L., 1992, ApJ, 409, 548 
\bibitem[\protect\citeauthoryear{Hernquist}{1993}]{Her93} Hernquist L., 1993, ApJ, 400, 460 
\bibitem[\protect\citeauthoryear{Hoekstra et al.}{2004}]{Hoe04} Hoekstra H., Yee H.~K.~C., Gladders M.~D., 2004, ApJ, 606, 67
\bibitem[\protect\citeauthoryear{Humphrey et al.}{2006}]{Hum06} Humphrey P.~J., Buote, D.~A., Gastaldello F., Zappacosta L., Bullock J.~S., Brighenti F., Mathews W.~G., 2006, preprint (astro-ph/0601301)
\bibitem[\protect\citeauthoryear{Jesseit, Naab \& Burkert}{2005}]{Jes05} Jesseit R., Naab T., Burkert A., 2005, MNRAS, 360, 1185
\bibitem[\protect\citeauthoryear{Jing \& Suto}{2000}]{J00} Jing Y.~P., Suto Y., 2000, ApJ, 529, L69
\bibitem[\protect\citeauthoryear{J\o rgensen \& Franx}{1994}]{Jor94} J\o rgensen I., Franx M., 1994, ApJ, 433, 553
\bibitem[\protect\citeauthoryear{Keeton}{2001}]{Kee01} Keeton C.~R., 2001, ApJ, 561, 46
\bibitem[\protect\citeauthoryear{Kleinheinrich et al.}{2006}]{Kle06} Kleinheinrich M.~et al., 2006, A\&A, 455, 441
\bibitem[\protect\citeauthoryear{Koopmans et al.}{2006}]{Kop06} Koopmans L.~V.~E., Treu T., Bolton A.~S., Burles S., Moustakas L.~A., 2006, ApJ, 649, 599
\bibitem[\protect\citeauthoryear{Krajnovi\'c et al.}{2005}]{Kra05} Krajnovi\'c D., Cappellari M., Emsellem E., McDermid R.~M., de Zeeuw P.~T., 2005, MNRAS, 357, 1113
\bibitem[\protect\citeauthoryear{Kronawitter et al.}{2000}]{Kr00} Kronawitter A., Saglia R.~P., Gerhard O.~E., Bender R., 2000, A\&AS, 144, 53
\bibitem[\protect\citeauthoryear{Loewenstein \& White}{1999}]{Loe99} Loewenstein M., White R.~E., 1999, ApJ 518, 50
\bibitem[\protect\citeauthoryear{Magorrian}{1999}]{mag99} Magorrian J., 1999, MNRAS, 302, 530
\bibitem[\protect\citeauthoryear{Magorrian \& Ballantyne}{2001}]{Mag01} Magorrian J., Ballantyne D., 2001, MNRAS, 322, 702
\bibitem[\protect\citeauthoryear{Mandelbaum et al.}{2006}]{Man06} Mandelbaum R., Seljak U., Kauffmann G., Hirata C., Brinkmann J., 2006, MNRAS, 368, 715
\bibitem[\protect\citeauthoryear{Mehlert et al.}{1998}]{Meh98} Mehlert D., Saglia R.~P., Bender R., Wegner G., 1998, A\&A, 332, 33
\bibitem[\protect\citeauthoryear{Mehlert et al.}{2000}]{Meh00} Mehlert D., Saglia R.~P., Bender R., Wegner G., 2000, A\&AS, 141, 449
\bibitem[\protect\citeauthoryear{Naab \& Burkert}{2003}]{Nab03} Naab T., Burkert A., 2003, ApJ, 597, 893
\bibitem[\protect\citeauthoryear{Navarro, Frenk \& White}{1996}]{nfw96} Navarro J.~F., Frenk C.~S., White S.~D.~M., 1996, ApJ, 462, 563
\bibitem[\protect\citeauthoryear{Oosterloo et al.}{2002}]{Oos02} Oosterloo T.~A., Morganti R., Sadler E.~M., Vergani D., Caldwell N., 2002, AJ, 123, 729
\bibitem[\protect\citeauthoryear{Pierce et al.}{2006}]{Pie06} Pierce M.~et al., 2006, MNRAS, 366, 1253
\bibitem[\protect\citeauthoryear{Pizzella et al.}{1997}]{Piz97} Pizzella A.~et al., 1997, A\&A, 323, 349
\bibitem[\protect\citeauthoryear{Renzini}{2006}]{Ren06} Renzini A., 2006, ARA\&A, 44, 141
\bibitem[\protect\citeauthoryear{Richstone \& Tremaine}{1988}]{Ric88} Richstone D.~O., Tremaine S., 1988, ApJ, 327, 82 
\bibitem[\protect\citeauthoryear{Rix et al.}{1997}]{R97} Rix H.~W., de Zeeuw P.~T., Cretton N., van der Marel R.~P., Carollo C.~M., 1997, ApJ, 488, 702 
\bibitem[\protect\citeauthoryear{Romanowsky et al.}{2003}]{R03} Romanowsky A.~J., Douglas N.~G., Arnaboldi M., Kuijken K., Merrifield M.~R., Napolitano N.~R., Capaccioli M., Freeman K.~C., 2003, Sci, 301, 1696
\bibitem[\protect\citeauthoryear{Saglia et al.}{2000}]{Sag00} Saglia R.~P., Kronawitter A., Gerhard O.~E., Bender R., 2000, AJ, 119, 153
\bibitem[\protect\citeauthoryear{Schwarzschild}{1979}]{S79} Schwarzschild M., 1979, ApJ, 232, 236 
\bibitem[\protect\citeauthoryear{Thomas et al.}{2004}]{Tho04} Thomas J., Saglia R.~P., Bender R., Thomas D., Gebhardt K., Magorrian J., Richstone D., 2004, MNRAS, 353, 391
\bibitem[\protect\citeauthoryear{Thomas et al.}{2005}]{Tho05} Thomas J., Saglia R.~P., Bender R., Thomas D., Gebhardt K., Magorrian J., Corsini E.~M., Wegner G., 2005, MNRAS, 360, 1355
\bibitem[\protect\citeauthoryear{Thomas et al.}{2007b}]{Tho07} Thomas J., Jesseit R., Naab T., Saglia R.~P., Burkert A., Bender R., MNRAS in press
\bibitem[\protect\citeauthoryear{Tremblay \& Merritt}{1996}]{Tre96} Tremblay B., Merritt D., 1996, AJ, 111, 2243
\bibitem[\protect\citeauthoryear{Treu \& Koopmans}{2004}]{Tre04} Treu T., Koopmans L.~V.~E., 2004, ApJ, 611, 739
\bibitem[\protect\citeauthoryear{van Albada}{1982}]{vanAl82} van Albada T.~S., 1982, MNRAS, 201, 939 
\bibitem[\protect\citeauthoryear{van der Marel \& Franx}{1993}]{vdMF93} van der Marel R.~P., Franx M., 1993, ApJ, 407, 525
\bibitem[\protect\citeauthoryear{Vincent \& Ryden}{2005}]{Vin05} Vincent R.~A., Ryden B.~S., 2005, ApJ, 623, 137
\bibitem[\protect\citeauthoryear{Wechsler et al.}{2002}]{W02} Wechsler R.~H., Bullock J.~S., Primack J.~R., Kravtsov A.~V., Dekel A., 2002, ApJ, 568, 52
\bibitem[\protect\citeauthoryear{Wegner et al.}{2002}]{Weg02} Wegner G., Corsini E.~M., Saglia R.~P., Bender R., Merkl D., Thomas D., Thomas J., Mehlert D., 2002, A\&A, 395, 753
\bibitem[\protect\citeauthoryear{Weil \& Hernquist}{1996}]{Wei96} Weil M., Hernquist L., 1996, ApJ, 457, 51
\end{thebibliography}
\end{document}